\documentclass[12pt]{article}
\pdfoutput=1

\usepackage{putex}
\usepackage{autobreak}
\usepackage{graphicx}
\usepackage{caption}
\usepackage{amsmath}
\usepackage{array}
\usepackage{subcaption}
\usepackage{epstopdf}
\usepackage{enumerate}
\usepackage{cite}
\usepackage{youngtab}
\usepackage{tensor}
\usepackage{slashed}
\usepackage[aligntableaux=center]{ytableau}
\usepackage[utf8]{inputenc}
\usepackage{rotating}
\usepackage[
      colorlinks=true,
      linkcolor=blue,
      urlcolor=blue,
      filecolor=black,
      citecolor=red,
      ]{hyperref}

\newcommand{\abs}[1]{\left\lvert #1 \right\rvert}

\newcommand {\be} {\begin {equation}}
\newcommand {\ee} {\end {equation}}

\newcommand {\bes} {\begin {equation*}}
\newcommand {\ees} {\end {equation*}}

\newcommand{\es}[2] {\begin{equation} \label{#1} \begin{split} #2 \end{split} \end{equation}}

\newcommand{\CP}{\mathbb{CP}}
\newcommand{\Z}{\mathbb{Z}}

\newcommand{\R}{\mathbb{R}}

\newcommand{\cA}{{\mathcal A}}

\newcommand{\cC}{{\mathcal C}}

\newcommand{\cE}{{\mathcal E}}
\newcommand{\cF}{{\mathcal F}}
\newcommand{\cG}{{\mathcal G}}

\newcommand{\cN}{{\mathcal N}}

\newcommand{\cP}{{\mathcal P}}

\newcommand{\cR}{{\mathcal R}}
\newcommand{\cS}{{\mathcal S}}
\newcommand{\cT}{{\mathcal T}}
\newcommand{\cW}{{\mathcal W}}

\newcommand{\cZ}{{\mathcal Z}}

\newcommand{\beq}{\begin{equation}}
\newcommand{\eeq}{\end{equation}}

\def\ie{\begin{equation}\begin{aligned}}
\def\fe{\end{aligned}\end{equation}}

\newcommand{\ra}{\rangle}

\newcommand{\A}{{\alpha}}
\newcommand{\B}{{\beta}}

\newcommand{\mf}{\mathfrak }

\numberwithin{equation}{section}

\def\<{\langle}
\def\>{\rangle}

\begin{document}

\preprint{PUPT-2591}

\institution{PU}{Joseph Henry Laboratories, Princeton University, Princeton, NJ 08544, USA}
\institution{Exile}{Department of Particle Physics and Astrophysics, Weizmann Institute of Science, Rehovot, Israel}

\title{
AdS$_4$/CFT$_3$ from Weak to Strong String Coupling
}

\authors{Damon J.~Binder,\worksat{\PU} Shai M.~Chester,\worksat{\Exile} and Silviu S.~Pufu\worksat{\PU}}

\abstract{
We consider the four-point function of operators in the stress tensor multiplet of the $U(N)_k\times U(N)_{-k}$ ABJM theory, in the limit where $N$ is taken to infinity while $N/k^{5}$ is held fixed.  In this limit, ABJM theory is holographically dual to type IIA string theory on $AdS_4\times \mathbb{CP}^3$ at finite string coupling $g_s \sim (N/k^5)^{1/4}$.  While at leading order in $1/N$, the stress tensor multiplet four-point function can be computed from type IIA supergravity, in this work we focus on the first subleading correction, which comes from tree level Witten diagrams with an $R^4$ interaction vertex.  Using superconformal Ward identities, bulk locality, and the mass deformed sphere free energy previously computed to all orders in $1/N$ from supersymmetric localization, we determine this $R^4$ correction as a function of $N/k^5$.  Taking its flat space limit, we recover the known $R^4$ contribution to the type IIA S-matrix and reproduce the fact that it only receives perturbative contributions in $g_s$ from genus zero and genus one string worldsheets. This is the first check of AdS/CFT at finite $g_s$ for local operators. Our result for the four-point correlator interpolates between the large $N$, large 't Hooft coupling limit and the large $N$ finite $k$ limit.  From the bulk perspective, this is an interpolation between type IIA string theory on $AdS_4\times \mathbb{CP}^3$ at small string coupling and M-theory on $AdS_4\times S^7/\mathbb{Z}_k$.
}
\date{}

\maketitle

\tableofcontents

\section{Introduction and Summary}

Even though holographic correlators have been a subject of study since the early days of the AdS/CFT correspondence \cite{Maldacena:1997re,Witten:1998qj,Gubser:1998bc} (see for example \cite{DHoker:1998vkc,Freedman:1998bj,DHoker:1998bqu,DHoker:1998ecp,DHoker:1999bve,DHoker:1999kzh,DHoker:1999mqo,DHoker:1999jke,Freedman:1998tz} for early work on four-point functions), they are in many cases hard or even impossible to compute directly.  For instance, in the case of higher derivative contact interactions in string theory or M-theory, where the full supersymmetric completion of the first correction to the supergravity action is not completely known (see however \cite{deHaro:2002vk,Policastro:2006vt,Paulos:2008tn,Liu:2013dna}), one cannot even write down the full set of relevant Witten diagrams.   In the past few years, however, it has become clear that in certain cases one can essentially `bootstrap' the answer using various consistency conditions \cite{Goncalves:2014ffa,Rastelli:2017ymc,Rastelli:2017udc,Rastelli:2016nze,Binder:2019jwn,Zhou:2017zaw,Chester:2018aca,Binder:2018yvd,Chester:2018dga}.  These consistency conditions include crossing symmetry, the analytic properties of the correlators in Mellin space, and supersymmetry.  In particular, for tree level Witten diagrams with supergravity and/or higher derivative vertices in 2d \cite{Giusto:2019pxc,Rastelli:2019gtj,Giusto:2018ovt}, 3d \cite{Zhou:2017zaw,Chester:2018aca,Binder:2018yvd}, 4d \cite{Rastelli:2017udc,Rastelli:2016nze,Binder:2019jwn}, 5d \cite{Zhou:2018ofp}, and 6d \cite{Rastelli:2017ymc,Chester:2018dga} maximally supersymmetric theories, these consistency conditions determine the Witten diagrams contributing to the 4-point functions\footnote{See also \cite{Goncalves:2019znr} for recent work on holographic five-point functions in the 4d $\cN = 4$ super-Yang-Mills theory in the supergravity approximation.} of $1/2$-BPS operators up to a finite number of coefficients.  For low orders in the derivative expansion, one can further determine these coefficients using other methods, such as supersymmetric localization \cite{Pestun:2007rz,Kapustin:2009kz} or the relation between the Mellin amplitudes and flat space scattering amplitudes in 10d or 11d \cite{Polchinski:1999ry,Susskind:1998vk,Giddings:1999jq,Penedones:2010ue,Fitzpatrick:2011hu,Fitzpatrick:2011jn}.  In particular, Refs.~\cite{Chester:2018aca,Chester:2018dga,Binder:2019jwn} showed that the tree-level Witten diagram corresponding to an $R^4$ contact interaction, which is the first correction to supergravity in both 10d and 11d, can be completely determined using either supersymmetric localization or the flat space scattering amplitudes.  The agreement between the two methods of fixing the undetermined coefficients in this case provides a precision test of AdS/CFT beyond supergravity.

The goal of this work is to move away both from maximal supersymmetry and from $1/2$-BPS multiplets and to study the stress tensor multiplet tree level Witten diagrams in the 3d $\cN = 6$ $U(N)_k \times U(N)_{-k}$ gauge theory of Aharony, Bergman, Jafferis, and Maldacena (ABJM theory) \cite{Aharony:2008ug}, at large $N$.\footnote{One could also consider the $U(N)_k\times U(M)_{-k}$ and $N\neq M$ theory due to Aharony, Bergman, and Jafferis \cite{Aharony:2008gk}, but we will not do so here.}  The reason for pursuing this generalization is that it offers the possibility of an unprecedented test of AdS/CFT at finite string coupling $g_s$.  Indeed, if in ABJM theory we take $N$ to be large and of the same order as $k^5$, then the holographic dual is a weakly curved $AdS_4 \times \CP^3$ background of type IIA string theory with finite $g_s$ \cite{Aharony:2008ug}.   Using the consistency conditions mentioned above supplemented by supersymmetric localization results, we will be able to fully determine the contribution of the $R^4$ contact diagrams to the four-point function of the lowest dimension operator in the same super-multiplet as the stress tensor.  The flat space limit of the Mellin amplitude then reproduces precisely the $R^4$ contribution to the four-point scattering of super-gravitons in type IIA string theory as a function of $g_s$.  This function receives contributions from genus zero and genus one string worldsheets \cite{Green:1997di}.  The reason why such a finite $g_s$ test of AdS/CFT was not available in the maximally supersymmetric cases is that in 3d and 6d the bulk dual was an M-theory as opposed to string theory background, while in the 4d case, whose dual is type IIB string theory on $AdS_5 \times S^5$, the required supersymmetric localization result in the limit of large $N$ and finite $g_s \propto g_\text{YM}^2$  is hard to evaluate due to the contribution of instantons in the localized $S^4$ partition function \cite{Pestun:2007rz,Nekrasov:2002qd,Nekrasov:2003rj,Losev:1997tp,Moore:1997dj}.

In more detail, in this work we consider the four-point function of the scalar superconformal primary of the $\cN = 6$ stress tensor multiplet, which is a $1/3$-BPS operator that can be represented as a traceless tensor $S_a{}^b$, with $a, b = 1, \ldots, 4$, transforming in the $\bf 15$ of the $SU(4)_R$ R-symmetry \cite{Dolan:2008vc,Liendo:2015cgi,Cordova:2016emh}.  In addition to the large $N$, fixed $N/k^5$ limit mentioned above where ABJM theory is dual to type IIA string theory at finite $g_s$, we will also consider the M-theory limit where $N$ is taken to infinity while $k$ is kept fixed, as well as the 't Hooft strong coupling limit where $N$ is taken to infinity while $N/k$ is fixed and large and where ABJM theory is dual to weakly coupled type IIA strings on $AdS_4 \times \CP^3$.  The latter two limits can be obtained from the first:  for small values of $N / k^5$, one recovers the weakly coupled type IIA limit, while for large $N/k^5$ one recovers the M-theory limit.  In all these limits, we focus on the first few tree-level Witten diagrams that compute the $\langle SSSS \rangle$ correlator.  Our results will be expressed in terms of the following Mellin amplitudes (whose definition will be made precise in the next section):
 \es{MellinAmps}{
  M_\text{SG}(s, t)&: \qquad 
   \text{meromorphic Mellin amplitude with linear growth at large $s, t$} \\
  M_3(s, t)&: \qquad
   \text{polynomial Mellin amplitude of degree $3$} \\
  M_4(s, t)&: \qquad
   \text{polynomial Mellin amplitude of degree $4$}
 } 
Each of these Mellin amplitudes gives rise to correlation functions that are crossing-invariant and solve the superconformal Ward identities.  The first one, $M_\text{SG}(s, t)$ corresponds to the sum of the contact and exchange diagrams using supergravity vertices.  The other two correspond to six-derivative and eight-derivative interaction vertices, respectively.  

With these ingredients and the definitions $\mu \equiv N/ k^5$ and $\lambda \approx N/k$ (see Eq.~\eqref{thooft} for the precise definition), we find
 \es{MellinSummary}{
  \text{M-theory}: \quad M(s, t)&= \frac{1}{c_T}\frac{32}{\pi^2} M_\text{SG}(s, t) +\frac{1}{c_T^{\frac53}}\frac{1120}{3\pi^3}\left(\frac{6\pi}{k^2}\right)^{1/3} M_4(s, t)+O\left(c_T^{-2}\right) \,, \\
  \text{'t Hooft}: \quad M(s, t) &= \frac{1}{c_T}\left(\frac{32}{\pi^2}M_\text{SG}(s, t)+ \frac{3\sqrt{2}\zeta(3)}{4\pi^5}  \bigl[
 35 M_4(s, t)  -72 M_3(s, t) \bigr] \lambda^{
-\frac32} +O(\lambda^{-\frac52})\right) \\
&{}+ \frac{1}{c_T^{2}}\left(\frac{4480\sqrt{2}}{3\pi^3} M_4(s, t)  \lambda^{\frac12}+O(\lambda^0)\right)+O(c_T^{-3}) \,, \\
\text{fixed $\mu$:} \quad M(s, t)&= \frac{1}{c_T}\frac{32}{\pi^2} M_\text{SG}(s, t) \\
  &{}+\frac{1}{c_T^{\frac74}}\left(-\frac{576\ 2^{3/8} 3^{\frac14}\zeta (3)}{\pi ^{23/4} \mu^{3/8}}M_3(s,t)+\frac{2^{\frac38}280}{3^{3/4} \pi ^{23/4} } \left(4\sqrt{2} \pi ^3\mu^{\frac18}+3 \zeta
   (3)\mu^{-\frac38}\right)M_4(s,t)\right) \\
   &{}+O\left(c_T^{-2}\right) \,,
 }
where we expanded the Mellin amplitudes in $1/c_T$ instead of $1/N$, with $c_T$ being the theory-dependent constant that appears in the two-point function of the canonically-normalized stress tensor $T_{\mu\nu}$:
  \es{CanStress}{
  \langle T_{\mu\nu}(\vec{x}) T_{\rho \sigma}(0) \rangle = \frac{c_T}{64} \left(P_{\mu\rho} P_{\nu \sigma} + P_{\nu \rho} P_{\mu \sigma} - P_{\mu\nu} P_{\rho\sigma} \right) \frac{1}{16 \pi^2 \vec{x}^2} \,, \qquad P_{\mu\nu} \equiv \eta_{\mu\nu} \nabla^2 - \partial_\mu \partial_\nu \,.
 }
(As shown in \cite{Chester:2014fya}, $c_T$ is exactly calculable in ABJM theory using the supersymmetric localization results of \cite{Jafferis:2010un} and \cite{Closset:2012vg}.  It behaves as $c_T \propto k^{1/2} N^{3/2}$ at large $N$.)  The Mellin amplitudes in \eqref{MellinSummary} can then be related to the 4-point scattering amplitudes of super-gravitons in 11d and 10d flat space using the relation proposed in \cite{Penedones:2010ue}:
 \es{ScattSummary}{
 \text{M-theory:} \quad \mathcal{A}^{11} &=\mathcal{A}^{11}_\text{SG}\left[  1 + \ell_{11}^6 \frac{1}{3\cdot 2^7}stu+ O(\ell_{11}^9) \right]\,, \\
 \text{type IIA, small $g_s$:} \quad {\mathcal{A}}^{10} &=\mathcal{A}^{10}_\text{SG}\biggl[  1 + \ell_{s}^6  \left(  \frac{\zeta(3)}{32}stu +O(\ell_s^{10}) \right) \\
 &{}+g_s^2 \left(\ell_s^6  \frac{\pi^2}{96}stu +O(\ell_s^8)\right)+O(g_s^4)\biggr] \,, \\
  \text{type IIA, finite $g_s$:} \quad {\mathcal{A}}^{10} &=\mathcal{A}^{10}_\text{SG}\left[  1 + \ell_{s}^6  stu \left(  \frac{\zeta(3)}{32} 
   + g_s^2  \frac{\pi^2}{96}  \right) +O(\ell_s^8)\right] \,,
 }
where $\cA^{11}_\text{SG}$ and $\cA^{10}_\text{SG}$ are the scattering amplitudes in 11d and 10d supergravity, respectively, $\ell_{11}$ is the 11d Planck length, $\ell_s$ is the 10d string length, and $s$, $t$, $u = - s - t$ are the Mandelstam invariants. 

Eqs.~\eqref{ScattSummary} are the well-known formulas describing the scattering of massless states in M-theory and string theory in the small momentum expansion.  They have the following structure.   The leading term in each equation is the supergravity scattering amplitude, and it contains information about the polarization and the type of massless particles being scattered.  In each case, the corrections to the supergravity amplitude are captured by a single function of $s$ and $t$ that can be expanded at small $s$ and $t$.  Besides the supergravity terms, the only other terms written down in \eqref{ScattSummary} are proportional to $s t u$ and correspond to an $R^4$ correction.\footnote{It would be interesting to study the next few terms not written down in \eqref{ScattSummary} in future work. In particular, in M-theory the next correction not written down in \eqref{ScattSummary} is at order $\ell_{11}^9$, and it comes from the 11d supergravity amplitude that can be found in \cite{Green:1997as,Russo:1997mk}.  The term after that, at order $\ell_{11}^{12}$, comes from a $D^6 R^4$ interaction, it is protected, and it can be related to the $D^6 R^4$ term in the type IIA string theory amplitude at order $g_s^4$ \cite{Green:1998by,Green:2005ba,Pioline:2015yea}.   In the string theory case, all terms at order $g_s^0$ can be resummed into an expression involving Gamma functions that can be found, for instance, in \cite{Green:1997ud,Polchinski:1998rr}.  Starting at order $g_s^2$, the scattering amplitude contains both analytic and non-analytic terms that can be derived from the tree level terms using unitarity \cite{Green:2008uj}.   While the type II string theory S-matrix is known to order $g_s^4$ for finite $\ell_s$ \cite{Gomez:2010ad,DHoker:2005jhf}, the lowest few protected terms in the small $\ell_s$ expansion are also known to order $g_s^6$ \cite{Gomez:2013sla}.  For work on the Mellin amplitudes corresponding to the non-analytic terms, see \cite{Alday:2018kkw,Alday:2017vkk,Alday:2018pdi}.}   The various supergravity and $R^4$ terms in the three equations are not independent.  Indeed, if in the first equation, one makes the replacement $\ell_{11} = \ell_s (2 \pi g_s)^{1/3}$, then the supergravity term in the first equation matches the supergravity terms in the other two, and the $R^4$ term in the first equation matches the $g_s^2 s t u$ terms in the other two.   Consequently, the first term in the first equation of \eqref{MellinSummary} is identical to the first term in the second equation of \eqref{MellinSummary} and the second term in the first equation of \eqref{MellinSummary} is identical to the first term on the second line of the second equation of \eqref{MellinSummary}.

The terms given in \eqref{MellinSummary} are derived solely using supersymmetric localization \cite{Jafferis:2010un,Kapustin:2009kz}, as was originally done in \cite{Binder:2018yvd} for $k=1,2$ when the theory has enhanced $\mathcal{N}=8$ supersymmetry.   Supersymmetric localization can be used to compute the $S^3$ free energy in the presence of real mass deformations of Lagrangian theories with at least $\mathcal{N}=2$ supersymmetry. When viewed as an $\mathcal{N}=2$ SCFT, ABJM theory has an $SO(4)\times U(1)$ flavor symmetry,\footnote{The $U(1)$ is a flavor symmetry whose current lies in the $\mathcal{N}=6$ stress tensor multiplet, and so exists for all $\mathcal{N}=6$ SCFTs \cite{Bashkirov:2011fr}. If the theory has $\mathcal{N}=8$ supersymmetry, then $SO(6)_R\times U(1)$ is enhanced to $SO(8)_R$.} and it can be deformed by three real mass parameters corresponding to the Cartan of $SO(4)\times U(1)$.  We will focus on two of the three masses, which we denote by $m_+$ and $m_-$. The $S^3$ free energy $F(m_+,m_-)$ was computed to all orders in $1/N$ for any ${k\leq N}$ in \cite{Nosaka:2015iiw} using the Fermi gas formalism developed in \cite{Marino:2011eh}. The two independent choices of four derivatives $\frac{\partial^4 F}{\partial m_\pm^4}\big|_{m_\pm =0}$ and $\frac{\partial^4 F}{\partial m_+^2 \partial m_-^2}\big|_{m_\pm =0}$ can be related to integrated four-point functions of the stress tensor multiplet, which can in turn be related to $\langle SSSS\rangle$ using Ward identities to fix all the coefficients shown in \eqref{MellinSummary}.  
In the $m_{\pm}\to0$ limit, the non-perturbative corrections to $F(m_+,m_-)$ are expected to take the form $e^{-\sqrt{Nk}}$ and $e^{-\sqrt{N/k}}$, so this expansion also holds to all order in the finite 't Hooft coupling $\lambda\sim N/k$ and finite $\mu=N/k^5$ expansions, with no non-perturbative in $\mu$ terms. 

The rest of this paper is organized as follows. In Section~\ref{4POINT}, we set up the computation of the $\langle SSSS \rangle$ correlator in terms of tree-level Mellin amplitudes.  In particular, we determine $M_3$ and $M_4$ using the consistency conditions mentioned above.  Implementing these constraints is much trickier than in the maximal SUSY, $1/2$-BPS case, and we get guidance from solving a similar problem for flat space scattering amplitudes.  Section~\ref{locloc} contains a derivation of the supersymmetric localization constraints in ABJM theory.  In Section~\ref{CORRELATORS}, we combine the localization constraints with the general setup developed in Section~\ref{4POINT}.  We end with a discussion of our results in Section~\ref{disc}.  Many technical details are relegated to the Appendices.

\section{The $\langle SSSS\rangle$ correlator at strong coupling}
\label{4POINT}

We will begin by discussing the $\langle SSSS\rangle$ four-point function at strong coupling. In any of the strong coupling limits mentioned in the Introduction, the correlator $\langle SSSS\rangle$ can be written in terms of tree-level and loop Witten diagrams, although in this paper we focus only on the tree-level contributions.  The leading tree-level contribution comes from supergravity exchange diagrams. These are corrected by higher derivative contact interactions, suppressed by the ratio $\ell_p / L$ in 11d or $\ell_s / L$ in 10d, depending on the limit being taken.  Beyond the supergravity term, the tree-level Witten diagrams take a particularly simple form in Mellin space: at each order in the perturbative expansion only a finite number of Mellin amplitudes $M^i(s,t)$ contribute, each of which is polynomial in $s, t$.  In this section our task is to determine the first few such amplitudes, using the flat space limit, crossing symmetry, the supersymmetric Ward identities, and locality.

\subsection{Setup}
As mentioned in the Introduction, the $S$ operator is the superconformal primary of the stress tensor multiplet, and transforms in the $\bf 15$ of the $\mathfrak{so}(6)$ R-symmetry. In index notation we write the operator as $S_b{}^a(\vec{x})$, where the raised index $a=1,\ldots ,4$ transforms in the $\bf 4$ of $\mf{su}(4) \cong \mf{so}(6)$ and the lowered indices in the $\overline{\bf 4}.$ We will find it more convenient however to use an index-free notation by defining
 \es{SNorm}{
  S(\vec x,X) = X_a{}^b S_b{}^a(\vec x) \,,
 }
where $X$ is an arbitrary traceless ${\bf 4}\otimes\overline{\bf 4}$ matrix. We normalize this operator so that
\begin{equation}\label{sNorm}
  \langle S(\vec x_1,X_1)S(\vec x_2,X_2)\rangle = \frac{\text{Tr}(X_1X_2)}{x_{12}^2} \,.
\end{equation}
Using both conformal and $\mathfrak{so}(6)$ symmetry, we can expand
\begin{equation}\label{SSSScor}\begin{aligned}
\langle S(\vec x_1,X_1) \cdots &S(\vec x_4,X_4)\rangle = \frac 1{x_{12}^2 x_{34}^2}\Bigg[\cS^1(U,V)A_{12}A_{34}+\cS^2(U,V)A_{13}A_{24}+\cS^3(U,V)A_{14}A_{23} \\
& + \cS^4(U,V)B_{1423} + \cS^5(U,V)B_{1234}+\cS^6(U,V)B_{1342}\Bigg] \,,
\end{aligned}\end{equation}
where we define the R-symmetry structures
\begin{equation}\label{ABRDef}
A_{ij} = \tr(X_iX_j) \,, \qquad B_{ijkl} = \tr(X_i X_j X_k X_l) + \tr(X_l X_k X_j X_i)  \,, 
\end{equation}
and where $\cS^i$ are functions of the conformal cross-ratios
\begin{equation} U \equiv \frac{x_{12}^2x_{34}^2}{x_{13}^2x_{24}^2} \,, \qquad  V \equiv \frac{x_{14}^2x_{23}^2}{x_{13}^2x_{24}^2} \,.
\end{equation}

For future reference, we note that it is sometimes useful to write the four-point function in a conformal block expansion, which reads\footnote{We could reorganize this block expansion into superconformal blocks (as opposed to conformal blocks) for each supermultiplet, but it is unnecessary to do so for our purposes.}
\es{blockExp}{
\langle S(\vec x_1,X_1)S(\vec x_2,X_2)S(\vec x_3,X_3)S(\vec x_4,X_4)\rangle&=\frac{1}{x_{12}^2x^2_{34}} \sum_R T_R(X_i) \cS_R(U, V) \\
\cS_R(U, V) &\equiv \sum_{\Delta,\ell}\lambda^2_{\Delta,\ell,R}G_{\Delta,\ell}(U,V)\,,
}
where $G_{\Delta, \ell}(U, V)$ are the 3d conformal blocks normalized as in \cite{Kos:2013tga}, $T_R(X_i)$ are the $SU(4)$ invariants corresponding to the $s$-channel exchange of an operator in the irrep $R$, and $\lambda^2_{\Delta,\ell,R}$ are squared OPE coefficients. The $SU(4)$ irreps $R$ of the operators that appear in $S\times S$ are
\es{irreps}{
{\bf15}\otimes{\bf15}&={\bf1}_s\oplus{\bf15}_a\oplus{\bf15}_s\oplus{\bf20'}_s\oplus{\bf45}_a\oplus{\bf\overline{45}}_a\oplus{\bf84}_s\,,
}
where $s/a$ denotes the symmetric/antisymmetric product.   As explained in Appendix~\ref{SU4STRUCTURES}, we find 
\es{TR}{
\begin{pmatrix}T_{{\bf 1}_s} \\ T_{{\bf 15}_a} \\T_{{\bf 15}_s} \\T_{{\bf 20}'_s} \\T_{{\bf 45}_a \oplus {\overline{\bf 45}}_a} \\T_{{\bf 84}_s} \\\end{pmatrix}=
\begin{pmatrix}
 1 & 0 & 0 & 0 & 0 & 0 \\
 0 & 0 & 0 & 0 & 1 & -1 \\
 -1 & 0 & 0 & 0 & 1 & 1 \\
 2 & 6 & 6 & -6 & -3 & -3 \\
 0 & 4 & -4 & 0 & 1 & -1 \\
 \frac{4}{15} & 4 & 4 & 4 & -\frac{2}{3} & -\frac{2}{3} 
\end{pmatrix}\begin{pmatrix}A_{12}A_{34}\\A_{13}A_{24}\\A_{14}A_{23}\\B_{1423}\\B_{1234}\\B_{1342} \end{pmatrix}\,.
}
We can distinguish between ${\bf15}_s$ and ${\bf{15}}_a$ by (anti)symmetrizing appropriately, and we should only consider the real combination ${\bf45}\oplus\overline{\bf45}$.

Holographic correlators are simpler in Mellin space. To compute the Mellin transform of $\cS^i(U,V)$, we first compute the connected correlator by subtracting the disconnected part
\begin{equation}\cS^i_{\text{conn}}(U,V)\equiv \cS^i(U,V)-\cS^i_{\text{disc}}(U,V) \,,  \qquad
 \cS^i_{\text{disc}} = \begin{pmatrix} 1 & U & \frac UV &0&0&0\end{pmatrix} \,,
 \end{equation}
and then we define $M^i(s,t)$ through
\begin{equation}\begin{split}\label{melDef}
\cS^i_{\text{conn}}(U,V)=&\int_{-i\infty}^{i\infty}\frac{ds\, dt}{(4\pi i)^2}\ U^{\frac s2}V^{\frac u2-1} \Gamma^2\left[1-\frac s2\right]\Gamma^{2}\left[1-\frac t2\right]\Gamma^{2}\left[1-\frac u2\right] M^i(s,t)\,,
\end{split}\end{equation}
where $u = 4 - s - t$. The Mellin transform \eqref{melDef} is defined such that a bulk contact Witten diagrams coming from a vertex with $n = 2m$ derivatives gives rise to a polynomial $M^i(s, t)$ of degree $m$ \cite{Penedones:2010ue}.  (This property holds both for scalars and for operators with spin, provided that the Mellin amplitudes for operators with spin are defined appropriately.)  The two integration contours in \eqref{melDef} are chosen such that\footnote{This is the correct choice of contour provided that $M^i(s, t)$ does not have any poles with $\Re(s) <2$ or $\Re(t) < 2$ or $\Re(u) < 2$.  If this is not the case (such as for the supergravity Mellin amplitude), the integration contour will have to be modified in such a way that the extra poles are on the same side of the contour as the other poles in $s$, $t$, $u$, respectively.}
\begin{equation}\label{contour}
\text{Re}(s) < 2\,,\quad \text{Re}(t) < 2\,,\quad \text{Re}(u) = 4-\text{Re}(s) - \text{Re}(t) < 2\,,
\end{equation}
which include all poles of the Gamma functions on one side or the other of the contour.   These poles naturally incorporate the effect of double trace operators \cite{Mack:2009mi}.   



In this paper we focus on tree-level Witten diagrams, and in the rest of this section we aim to determine a basis of Mellin amplitudes that can be used to write the contribution from contact Witten diagrams with small numbers of derivatives.  These Mellin amplitudes obey three constraints:
 \begin{enumerate}
  \item They obey the crossing symmetry requirements
\es{CrossingM}{
M^1(s,t) &= M^1(s,u)\,, \qquad M^2(s,t) = M^1(t,s)\,, \qquad M^3(s,t) = M^1(u,t)  \,, \\
M^4(s,t) &= M^4(s,u)\,, \qquad M^5(s,t) = M^4(t,s)\,, \qquad M^6(s,t) = M^4(u,t) 
}
coming from the crossing symmetry of the full $\langle SSSS \rangle$ correlator.
\item They obey the SUSY Ward identities following from ${\cal N} = 6$ superconformal symmetry.  The SUSY Ward identities not only constrain $M^i(s, t)$, but they also allow us to determine the Mellin amplitudes corresponding to correlators of other operators in the stress-tensor multiplet.
\item The $M^i(s, t)$ and all other Mellin amplitudes related to them by SUSY are polynomials in $s$, $t$.  We call the collection of Mellin amplitudes corresponding to four-point functions of operators in the same super-multiplet a super-Mellin amplitude, and we define the degree of a polynomial super-Mellin amplitude $n$ to be the highest degree of any component Mellin amplitude.
\end{enumerate}
For fixed $m$, we will label the Mellin amplitudes obeying these requirements as $M^i_{m}(s,t)$ in case there is a unique such amplitude for a given $m$ or by $M^i_{m,k}(s,t)$ in the case that there are multiple such amplitudes indexed by $k$.  These Mellin amplitudes represent a basis for contact Witten diagrams, with the number of derivatives in the interaction vertex being bounded from below by $2m$.  In Section~\ref{CORRELATORS}, we will use these Mellin amplitudes and the constraints coming from supersymmetric localization explored in the next section in order to determine the first few terms in the strong coupling expansion of the $\langle SSSS \rangle$ correlator.

Note that, in general, supersymmetry relates the contact interactions for bulk fields with various spins, and in flat space SUSY preserves the number of derivatives of the interaction vertices it relates.  In AdS however, the number of derivatives within a given super-vertex may vary, with the change in the number of derivatives being compensated by an appropriate power of the AdS radius $L$.  Thus, it may happen that a four-scalar vertex with a given number of derivatives is part of a supervertex containing other vertices with more derivatives.  The corresponding Mellin amplitudes $M^i(s,t)$ will then have lower degree than those of some four-point function of superconformal descendants of $S$, and so $M_n^i(s, t)$ may have degree less than $n$.  This fact will be very important in the analysis that follows.


\subsection{The flat-space limit and a toy problem}
\label{FLATLIM}

Finding the Mellin amplitudes $M^i_n(s, t)$ that obey the conditions listed above is a difficult task, as satisfying the third condition requires us to calculate Ward identities for many different correlators and then examine the locality properties of their Mellin amplitudes. We can simplify matters by first solving an analogous problem for flat space scattering amplitudes. 

At large AdS radius, we can recover flat space scattering amplitudes for scalars using the Penedones formula \cite{Fitzpatrick:2011hu}. Applied to the superconformal primary $S$ the relationship is (up to an overall normalization $\cN(L)$)
\es{FlatLimit}{\cA^i(s,t) &= \lim_{L \to \infty}{\cN(L)}\sqrt{\pi} \int_{\kappa-i\infty}^{\kappa+ i \infty} \frac{d\alpha}{2 \pi i} \, e^\alpha \alpha^{-\frac12} M^i\left(\frac{L^2}{2 \alpha} s,  \frac{L^2}{2 \alpha} t \right) \,.
}
Here, $\kappa > 0$, and $\cA^i(s,t)$ is the corresponding 4d flat space scattering amplitude of graviscalars (or more precisely a scattering amplitude in 10d string theory or 11d M-theory with the momenta restricted to lie within 4d and polarizations transverse to this 4d space), computed in the limit where the AdS radius $L$ is taken to infinity while keeping some other dimensionful length scale $\ell_\text{UV}$ fixed.  For string or M-theory duals we can take $\ell_\text{UV}$ to be either the 10d string length or 11d Planck length, as we will do in Section~\ref{CORRELATORS}. 

From \eqref{FlatLimit} we expect that each Mellin amplitude $M^i_{m,k}(s,t)$ must give rise to a local $\cN = 6$ scattering amplitude $\cA^i_{m,k}(s,t)$. This mapping should furthermore be one-to-one, since if two amplitudes $M^i_{m,k_1}$ and $M^i_{m,k_2}$ have the same large $s, t$ limit, then their difference $M^i_{m,k_1}-M^i_{m,k_2}$ will be a local Mellin amplitude with degree at most $m-1$. Thus, if we can find all of the number of local scattering amplitudes of a given degree in $s, t$, then this will also tell us the number of Mellin amplitudes which occur at this degree:\footnote{At a more abstract level, we can justify the correspondence \eqref{AmplitudeCFT} as follows. Local Mellin amplitudes correspond to bulk contact Witten diagrams, which are themselves in one-to-one correspondence with local counterterms in AdS\@.  But since AdS is maximally symmetric, local counterterms in AdS are equivalent to local counterterms in flat-space. Since local counterterms in flat-space correspond exactly to scattering amplitudes, we find that Mellin amplitudes and scattering amplitudes are in one-to-one correspondence.}
 \es{AmplitudeCFT}{
  \text{\# of degree $m$ scattering amplitudes in 4d SUGRA}& \quad \\ 
   &\hspace{-2in}= \quad \text{\# of degree $m$ Mellin amplitudes in 3d SCFT} \,.
 }
Because the flat space scattering amplitudes are obtained as the large $s$, $t$ limits of Mellin amplitudes, finding all crossing-invariant, supersymmetric, and local $\cN = 6$ flat space scattering amplitudes is a strictly simpler problem than finding all Mellin amplitudes with the same properties.

\subsection{Counterterms in $\cN = 6$ supergravity}
\label{COUNTERN6}
The toy problem described in the previous section is that of finding four-point scattering amplitudes corresponding to counterterms in 4d $\cN = 6$ supergravity. Spinor helicity and on-shell supersymmetric methods provide an efficient means to classify allowed counterterms in a theory. They were first applied to 4d $\cN = 8$ in \cite{Elvang:2010jv,Elvang:2010xn}, and have subsequently been generalized to other maximally supersymmetric theories in \cite{Wang:2015jna,Wang:2015aua}. In the context of $\cN = 6$ supergravity these methods have been applied to study amplitudes involving bulk graviton exchange \cite{Elvang:2011fx,Freedman:2018mrv}. 

\begin{table}
\begin{center}
\hspace{-.4in}
{\renewcommand{\arraystretch}{1.2}
\begin{tabular}{|l||r|r|r|r|r|r|r|}\hline
$\Phi$ Particles & $h^+$ & $\psi^+$ & $g^+$ & $F^+$ & $\phi$ & $\chi^-$ & $a^-$ \\ 
Helicity & $+2$ & $+3/2$ & $+1$ & $+1/2$ & $0$ & $-1/2$  & $-1$\\
$SU(6)_R$ & $\bf 1$ & $\bf 6$ & $\bf 15$ & $\bf 20$ & $\overline{\bf 15}$ & $\overline{\bf 6}$ & $\bf 1$\\ \hline
$\Psi$ Particles & $a^+$ & $\chi^+$ & $\overline \phi$ & $F^-$ & $g^-$ & $\psi^-$ & $h^-$\\
Helicity & $+1$ & $+1/2$ & $0$ & $-1/2$ & $-1$ & $-3/2$  & $-2$\\
$SU(6)_R$ & $\bf 1$ & $\bf 6$ & $\bf 15$ & $\bf 20$ & $\overline{\bf 15}$ & $\overline{\bf 6}$ & $\bf 1$ \\ \hline
\end{tabular}}
\caption{Massless particles in $\cN = 6$ supergravity.}
\label{sugraMult}
\end{center}
\end{table}

Let us begin with a quick review of on-shell superspace (see also Appendix~\ref{REVIEWSPINOR}); for a detailed textbook treatment of the subject we recommend \cite{Elvang:2015rqa}.
In $\cN = 6$ supergravity, the massless particles split into two supermultiplets: a multiplet we denote by $\Phi$ that contains the positive helicity graviton $h^+$, and its $\cC\cP\mathcal T$ conjugate multiplet we denote by $\Psi$ that contains the negative helicity graviton $h^-$.  In addition to the graviton $h^\pm$, these multiplets also contain the gravitino $\psi^\pm$, the gauginos $g^\pm$, fermions $F^\pm$, scalars $\phi$, and the graviphoton $a^\pm$.  Table \ref{sugraMult} lists the particles in these multiplets, along with their transformation properties under the $SU(6)$ R-symmetry of $\cN = 6$ supergravity.  In the on-shell superspace formalism, the $\Phi$ and $\Psi$ superfields are polynomials in the Grassmann variables $\eta^I$, with $I = 1, \ldots 6$ transforming in the $\overline{\bf 6}$ of $SU(6)$:\footnote{Upper $I, J, K, \ldots$ indices transform in the $\overline{\bf 6}$ of $SU(6)$ while lower $I, J, K, \ldots$ indices transform in the ${\bf 6}$ of $SU(6)$.}
\begin{equation}\begin{split}
\Phi &\equiv h^+ + \eta^I \psi^+_I + \frac 1 {2!}\eta^I \eta^J g_{IJ}^++ \frac 1 {3!}\eta^I\eta^J\eta^K F_{IJK}^+ + \frac1{4!2} \eta^I\eta^J\eta^K\eta^L \epsilon_{IJKLMN}\phi^{MN} \\ &+ \frac 1 {5!}\eta^I\eta^J\eta^K\eta^L\eta^M\epsilon_{IJKLMN}\chi^{N-} + \frac 1 {6!}\eta^I\eta^J\eta^K\eta^L\eta^M\eta^N\epsilon_{IJKLMN} a^- \\
\Psi &\equiv a^+ + \eta^I \chi^+_I + \frac 1 {2!}\eta^I\eta^J \overline\phi_{IJ}+ \frac 1 {3!}\eta^I\eta^J\eta^K F_{IJK}^+ + \frac1{4!2} \eta^I\eta^J\eta^K\eta^L \epsilon_{IJKLMN}g^{MN} \\ &+ \frac 1 {5!}\eta^I\eta^J\eta^K\eta^L\eta^M \epsilon_{IJKLMN}\psi^{N-} + \frac 1 {6!}\eta^I\eta^J\eta^K\eta^L\eta^M\eta^N \epsilon_{IJKLMN} h^-\,.\end{split}\end{equation}
In a four-point superamplitude, such as $\cA[\Phi\Phi\Psi\Psi]$, each particle $i = 1, \ldots, 4$ is associated to some Grassmannian variable $\eta^I_i$. To compute a component scattering amplitude we simply differentiate with respect to some of the Grassmannian variables while setting all others to zero. For instance:
\begin{equation}\label{exampleSA}\begin{split}
A[h^+h^+h^+h^+] &= \cA[\Phi\Phi\Psi\Psi]\Bigg|_{\eta_i^I = 0} \,, \\
A[h^+h^+h^-h^-] &= \left(\prod_{J = 1}^6\frac{\partial}{\partial\eta_3^J}\right)\left(\prod_{K = 1}^6\frac{\partial}{\partial\eta_4^K}\right)\cA[\Phi\Phi\Psi\Psi]\Bigg|_{\eta_i^I = 0} \,, \\
A[\phi^{56}\phi^{56}\overline\phi_{12}\overline\phi_{12}] &= \left(\prod_{J = 1}^4\frac{\partial}{\partial\eta_1^J}\right)\left(\prod_{K = 1}^4\frac{\partial}{\partial\eta_2^K}\right)\left(\prod_{L = 1}^2\frac{\partial}{\partial\eta_3^L}\right)\left(\prod_{M = 1}^2\frac{\partial}{\partial\eta_4^M}\right)\cA[\Phi\Phi\Psi\Psi]\Bigg|_{\eta_i^I = 0}.\\
\end{split}\end{equation}
In this way a superamplitude ${\cal A}$ encodes all the amplitudes of its component particles.

Up to crossing there are five possible 4 particle superamplitudes we can construct from $\Phi$ and $\Psi$. However, under $\cC\cP\mathcal T$ the two supermultiplets $\Phi$ and $\Psi$ are conjugates, and their scattering amplitudes are related by complex conjugation (see Appendix~\ref{DISCRETEAMPLITUDES} for a description of how discrete space-time symmetries act on the scattering amplitudes):\footnote{Note that $[ij]^* = \langle ji \rangle$ in terms of the spinor-helicity angle and square brackets.}
\begin{equation}
\cA[\Psi\Psi\Psi\Psi] = (\cA[\Phi\Phi\Phi\Phi])^*\,,\qquad \cA[\Psi\Psi\Psi\Phi] = (\cA[\Phi\Phi\Phi\Psi])^* \,.
\end{equation}
This leaves us only three independent superamplitudes, $\cA[\Phi\Phi\Psi\Psi]$, $\cA[\Phi\Phi\Phi\Psi]$, and $\cA[\Phi\Phi\Phi\Phi]$. Our task now is to constrain the forms of these superamplitudes, beginning with invariance under supersymmetry.

As explained in~\cite{Elvang:2015rqa}, for a given particle $i$ the supermomentum is defined to be
\ie
q^I_i=|i\ra \eta^I_i \,,\qquad   \tilde q_{Ii}=|i] {\frac\partial {\partial \eta^I_i}} \,,
\fe
and it satisfies the on-shell SUSY algebra by construction.  For a given amplitude the total supermomentum is thus:
\ie
Q^I=\sum_i q^I_i \,, \qquad \tilde Q_I=\sum_i \tilde q_{Ii} \,.
\fe
Superamplitudes must be annihiliated by these supercharges. For a four-point amplitude such as $\cA[\Phi\Phi\Psi\Psi]$ this implies that
\begin{equation}
Q^I\cA[\Phi\Phi\Psi\Psi] = 0 \,, \qquad  \tilde Q_I\cA[\Phi\Phi\Psi\Psi] = 0 \,.
\end{equation}
Imposing these conditions uniquely fixes any four-point superamplitudes up to an arbitrary function of $s$ and $t$: 
\begin{equation}\label{FampSol}\begin{split}
\cA[\Phi\Phi\Psi\Psi] &= \delta^{12}(Q)\frac{[12]^4}{\langle 34\rangle^2} f_1(s,t) \,, \\
\cA[\Phi\Phi\Phi\Psi] &= \delta^{12}(Q) \frac{[12]^5\langle 14\rangle\langle 24\rangle}{\langle 34\rangle^4}f_2(s,t) \,, \\
\cA[\Phi\Phi\Phi\Phi] &= \delta^{12}(Q)\frac{[12]^4}{\langle 34\rangle^4} f_3(s,t) \,, 
\end{split}\end{equation}
where the first factor is the Grassmann delta function
 \begin{equation}
  \delta^{12}(Q) = \frac 1 {2^4}\prod_{I = 1}^6\sum_{i,j=1}^4 \langle ij\rangle\eta_i^I\eta_j^I \,, 
 \end{equation}
which is annihilated by both $Q^I$ and $\tilde Q_I$, and $f_i(s, t)$ are arbitrary functions\footnote{Since $\cA[\Phi\Phi\Psi\Psi]$ is self-conjugate under $\cC\cP\mathcal T$ we find that $f_1(s,t)$ is real, as we show in Appendix \ref{DISCRETE}\@. We furthermore show that for $\cC \cT$-invariant theories $f_{2,3}(s,t)$ are also real.} of $s$ and $t$.   The delta function $\delta^{12}(Q)$ is automatically invariant under $SU(6)_R$, even if the full theory does not preserve $SU(6)_R$ \cite{Elvang:2010xn}.\footnote{In flat space $\cN = 6$ the supersymmetry algebra does not require there to be an R-symmetry; it is an accidental symmetry of the supergravity action. On the other hand, the superconformal algebra does require that at least an $SO(6)_R$ symmetry be present in order for an AdS solution to preserve all supersymmetries of the theory.} Note that every term in each superamplitude contains exactly $12$ Grassmannian variables, and, as a result, many component amplitudes vanish, including $A[h^+h^+h^+h^+] =A[\phi\phi\phi\phi] = 0$.  See Table~\ref{ampsTab} for a list of component amplitudes that do not vanish.  The angle and square brackets in \eqref{FampSol} are required so that the $\Phi$ and $\Psi$ components have the correct helicity, which for instance can be fixed by considering
\begin{equation}\label{topcomp}\begin{split}
A[h^+h^+h^-h^-] &= [12]^4\langle 34\rangle^4f_1(s,t) \,, \\
A[h^+h^+h^-a^-] &= [12]^5\langle 34\rangle^2\langle 14\rangle\langle 24\rangle f_2(s,t) \,, \\
A[h^+h^+a^-a^-] &= [12]^4\langle 34\rangle^2f_3(s,t) \,. 
\end{split}\end{equation}

\begin{table}
\begin{center}
{\renewcommand{\arraystretch}{1.2}
\begin{tabular}{c|l|l|l}
 $\sum_i |h_i|$ & $\cA[\Phi\Phi\Psi\Psi]$     & $\cA[\Phi\Phi\Phi\Psi]$  & $\cA[\Phi\Phi\Phi\Phi]$ \\
\hline \hline
0 & $A[\phi\phi\overline\phi\overline\phi]$  & None                             & None \\ \hline
1 & $A[F^+\chi^-\overline\phi\overline\phi]$ & $A[\phi\phi\phi a^+]$           & None\\
  & $A[\phi\phi \chi^+F^-]$                  & $A[\phi\phi F^+ \chi^+]$        & \\
  & $A[\phi\chi^-\chi^+\overline\phi]$       & $A[\phi\phi g^+ \overline\phi]$ & \\
  & $A[\phi F^+F^-\overline\phi]$            &                                 & \\ \hline
2 & $A[F^+F^+F^-F^-]$                        & $A[F^+F^+F^+F^-]$               & $A[F^+F^+F^+F^+]$  \\ 
  & $A[\chi^-\chi^-\chi^+\chi^+]$            & $A[F^+F^+\psi^-\chi^+]$         & $A[g^+F^+F^+\phi]$ \\
  & $A[\phi g^+g^-\overline\phi]$            & $A[\phi\phi\psi^+F^-]$          & $A[g^+g^+\phi\phi]$\\
  & $A[\phi a^-a^+\overline\phi]$            & $A[\psi^+\chi^-\phi\overline\phi]$         & $A[\psi^+F^+\phi\phi]$ \\
  & $A[\phi g^+F^-F^-]$                      & $A[g^+F^+\chi^-\overline\phi]$           & $A[h^+\phi\phi\phi]$ \\
  &   \ldots                                    &  \ldots  &    \\ \hline
\ldots  &   \ldots                                &  \ldots  & \ldots \\ \hline
6 & \ldots                                      & \ldots               & $A[h^+h^+a^-a^-]$ \\ \hline  
7 & \ldots                                      & $A[h^+h^+a^-h^-]$ & None              \\ \hline  
8 & $A[h^+h^+h^-h^-]$                        & None              & None              
\end{tabular}}
\caption{Component amplitudes of each superamplitude, organised by total helicity $\sum_i|h_i|$. Here $h_i$ is the helicity of the $i^{\text{th}}$ particle. We have not included amplitudes equivalent to the ones listed here under crossing. }
\label{ampsTab}
\end{center}
\end{table}

We are now left to constrain the forms of $f_i(s,t)$ using locality and crossing symmetry. A tree-level scattering amplitude is local if and only if it can be written as a polynomial in the spinor helicity variables $[ij]$ and $\langle ij\rangle$; note that
  \begin{equation}
    \label{MandDef} 
     s = [12]\langle12\rangle = [34]\langle 34\rangle\,,\qquad 
     t = [13]\langle13\rangle = [24]\langle 24\rangle\,,\qquad 
     u = [14]\langle14\rangle = [23]\langle 23\rangle \,.
  \end{equation}
From \eqref{topcomp} we immediately see that it is not possible for $f_i(s,t)$ to contain poles in $s, t$ or $u$, or else the amplitudes in \eqref{topcomp} would lead to non-polynomial expressions. Hence $f_i(s,t)$ are necessarily polynomials for tree-level amplitudes. This is also sufficient, as when $f_i(s,t) = 1$ one can check that all amplitudes in the superamplitude are local.

Crossing symmetry imposes a series of further constraints. For instance, in \eqref{topcomp} the amplitudes must be invariant under interchanging the first and second particles. This gives us the relations
\begin{equation}\label{ampCross1}
f_{1,3}(s,t) = f_{1,3}(s,u)\,,\qquad f_2(s,t) = -f_2(s,u)\,,
\end{equation}
where $u = -s-t$ is the third Mandelstam variable. The superamplitudes $\cA[\Phi\Phi\Phi\Phi]$ and $\cA[\Phi\Phi\Phi\Psi]$ are furthermore invariant under crossing which exchange the first and third particles, giving rise to the further conditions:
\begin{equation}\label{ampCross2}
f_2(s,t) = -f_2(u,t)\,,\qquad f_3(s,t) = f_3(u,t)\,.
\end{equation}
Together, Eqs.~\eqref{ampCross1} and \eqref{ampCross2} suffice to guarantee crossing under all possible permutations.

Having determined the allowed forms of $f_i(s,t)$, we can now determine the number of derivatives in each interaction vertex.  To this count each angle and square bracket contribute $1$, $\delta^{12}(Q)$ contributes $6$, and each power of $s$, $t$, $u$ contributes $2$. For instance, if we set $f_2(s,t) = s^k$ and consider the amplitude $\cA[\Phi\Phi\Psi\Psi] = s^k \delta^{12}(Q)\frac{[12]^4}{\langle34\rangle^2}$, it follows that this amplitude comes from an interaction vertex with $8 + 2k$ derivatives, namely from an $D^{2k} R^4$ term. 

With this in mind, we can now systematically find all local counterterms up to a certain number of derivatives. In Table \ref{lowOrderTab} we list all local counterterms up to $15$ derivatives, corresponding to Mellin amplitudes up to degree $7.5$.\footnote{A Mellin amplitude of degree $7.5$ would seem to require non-polynomial contributions to $M^i(s,t)$. In Appendix~\ref{DISCRETE} however we show that due to discrete symmetries the Mellin amplitudes corresponding to $\cA[\Phi\Phi\Phi\Psi]$ never contribute to $\langle SSSS\rangle$, so $M^i(s,t)$ remains a polynomial in $s$ and $t$.}  In particular, the first local counterterm has $6$ derivatives, is unique, and contributes only to ${\cal A}[\Phi \Phi \Phi \Phi]$.  The next local counterterm has $8$ derivatives and is also unique and contributes only to $\cA[\Phi\Phi\Psi\Psi]$.  There are two $10$ derivative counterterms, one contributing to ${\cal A}[\Phi \Phi \Phi \Phi]$ and one to ${\cal A}[\Phi \Phi \Phi \Phi]$, and so on.  The counterterm with the lowest number of derivatives that contributes to $\cA[\Phi\Phi\Phi\Psi]$ has $15$ derivatives and will not be important in this work.
\begin{table}
\begin{center}\hspace{-.2in}
{\renewcommand{\arraystretch}{1.2}
\begin{tabular}{c|c|c|c|c|c}
Mellin deg.       & $f_1(s, t)$  & $f_2(s, t)$ & $f_3(s, t)$ & Counterterms & $\#$ sols. \\ \hline
3                & ---                      & ---                     &           1             & $F^2R^2$                  & 1 \\
4                & 1                        & ---                     & ---                     & $R^4$                     & 1 \\
5                & $s$                      & ---                     & $s^2+t^2+u^2$           & $D^4F^2R^2$\,,\ $D^2R^4$  & 2 \\
6                & $s^2$, $t^2+u^2$         & ---                     & $stu$                   & $D^6F^2R^2$\,,\ $D^4R^4$  & 3 \\
7                & $s^3$, $s(t^2+u^2)$      & ---                     & $(s^2+t^2+u^2)^2$       & $D^8F^2R^2$\,,\ $D^6R^4$  & 3 \\ 
7.5              & ---                      & $(s-t)(t-u)(u-s)$       & ---                     & $D^8FR^3$                 & 1 \\ \hline
\end{tabular}}
\caption{Counterterms in $\cN = 6$ supergravity, up to $15$ derivatives.}
\label{lowOrderTab}
\end{center}
\end{table}

\subsection{Implications for $\cN = 6$ SCFTs}
Having systematically computed the local amplitudes in $\cN = 6$ supergravity, we will now discuss the implications for holographic $\cN = 6$ SCFTs.   First, we can deduce that there are five independent superconformal invariants in the four point function of four stress tensor multiplets. This counting follows from the number of unknown real functions needed to fully determine the scattering amplitudes of supergravitons, one for $f_1(s,t)$ and two each for $f_{2}(s,t)$ and $f_3(s, t)$, as these latter two functions are in general complex.

Second, from Table~\ref{lowOrderTab} we can immediately deduce how many polynomial Mellin super-amplitudes exist for a given degree in $s, t$.  For instance, at third degree we have a single polynomial super-Mellin amplitude with scalar component $M_3^i(s,t)$, and at fourth degree we additionally have another polynomial super-Mellin amplitude with scalar component $M^i_4(s,t)$.  Here, by third and fourth degree we mean that the super-amplitudes that $M_3^i(s, t)$ and $M_4^i(s, t)$ have degree $3$ or $4$ for some of the components of the amplitude, but not necessarily for the scalar components $M_3^i(s, t)$ and $M_4^i(s, t)$ themselves.  These scalar components may be of less than third and fourth degree, respectively.

In fact, it can be argued that while the scalar component $M^i_4(s, t)$ is of degree $4$ in $s, t$, the scalar component $M^i_3(s, t)$ is actually at most quadratic.  This is because the leading order behavior of the super-Mellin amplitudes that $M^i_3(s,t)$ and $M^i_4(s,t)$ are part of at large $s$ and $t$ must match the corresponding super-scattering amplitude.  Since the $M^i_3(s,t)$ amplitude contributes only to the superamplitude $\cA[\Phi\Phi\Phi\Phi]$ (as can be seen from Table~\ref{ampsTab}), it does not give rise to a scalar scattering amplitude.  Therefore $M_3^i(s,t)$ must be at most quadratic, rather than cubic, in $s$ and $t$.  On the other had, $M^i_4(s,t)$ contributes to the superamplitude $\cA[\Phi\Phi\Psi\Psi]$, and this superamplitude does include a scalar scattering amplitude, $A[\phi\phi\overline \phi\overline \phi]$.  Thus, $M^i_4(s, t)$ must have degree $4$.

We can be more precise and also find the leading large $s$, $t$ behavior of all $\langle SSSS \rangle$ Mellin amplitudes $M^i(s, t)$ for which $M^i(s, t)$ is of highest degree in the super-Mellin amplitude.  (This means we will be able to find the leading large $s, t$ behavior of $M_4^i(s, t)$ but not of $M_3^i(s, t)$.)
As per \eqref{FlatLimit}, the leading large $s, t$ behavior of $M^i(s, t)$ comes from the flat space amplitude ${\cal A}^i(s, t)$. The only scattering amplitude with a scalar component is ${\cal A}[\Phi \Phi \Psi \Psi]$ which is fixed in terms of $f_1(s,t)$, and so the leading large $s, t$ behavior of $M^i(s, t)$ depends only on $f_1(s, t)$.  To perform this calculation, we must first extract the scalar $A[\phi \phi \overline{\phi} \overline{\phi}]$ component of $A[\Phi \Phi \Psi \Psi]$, and then must relate $\phi$ and $\overline \phi$ to the superconformal primary $S$. We perform both computations in Appendix~\ref{AMPTOCFT} and find that 
\begin{equation}\label{flatSSSS}\begin{split}
\cA^1(s,t) &= -\frac{1}{2} tu\left(-s^2f_1(s,t)+u^2f_1(u,s)+t^2f_1(t,s)\right) \,, \\
\cA^2(s,t) &= -\frac{1}{2} su\left(s^2f_1(s,t)+u^2f_1(u,s)-t^2f_1(t,s)\right) \,, \\
\cA^3(s,t) &= -\frac{1}{2} ts\left(s^2f_1(s,t) -u^2f_1(u,s)+t^2f_1(t,s)\right) \,, \\
\cA^4(s,t) &= -\frac{1}{2} s t u\left(u f_1(u, s) + t f_1(t, s)\right)\,, \\
\cA^5(s,t) &= -\frac{1}{2} s t u\left(u f_1(u, s) + s f_1(s, t)\right)\,, \\
\cA^6(s,t) &= -\frac{1}{2} s t u\left(s f_1(s, t) + t f_1(t, s)\right)\,.
\end{split}\end{equation}
From \eqref{flatSSSS}, we can also determine $f_1(s, t)$ in terms of $\cA^i(s, t)$:
\begin{equation}
 \label{ATof}
f_1(s,t) = -\frac1{s^3} \left(\frac{\cA^2(s,t)} u + \frac{\cA^3(s,t)} t\right) \,.
\end{equation}

We can then apply \eqref{flatSSSS} to $M_4^i(s, t)$, which at large $s, t$ should asymptote to $\cA^i(s, t)$ with $f_1(s, t) = 1$ (see Table~\ref{lowOrderTab}).  We hence find 
\begin{equation}\begin{split}\label{largeA4}
 M^i_4(s,t) = \begin{pmatrix} 
 t^2u^2  & s^2u^2 & s^2t^2 & \frac{s^2tu}{2} & \frac{st^2u}{2} & \frac{stu^2}{2}
\end{pmatrix} + \text{subleading in $s, t$} \,.
\end{split}\end{equation}

\subsection{Exchange amplitudes}
So far we have considered local contact amplitudes. The only other tree-level diagrams which appear in four point functions consist of exchange diagrams. These can be built up from the on-shell three point amplitudes using on-shell recursion relations (see for instance chapter 3 of \cite{Elvang:2015rqa}), and so our first task is to find the allowed three point amplitudes.

Three point amplitudes are subtle due to special kinematics; conservation of momentum implies that either
\begin{equation} [12] = [13] = [23] = 0 \text{ or } \langle 12\rangle = \langle 13\rangle  = \langle 23\rangle = 0\,.
\end{equation}
For real momenta $[ij]^* = \langle ji\rangle$ so this would seem to rule out any interesting amplitudes. This issue is however resolved by analytically continuing to complex momenta. Locality and little-group scaling then uniquely fix three-point functions to take the form:
\begin{equation}
A[1^{h_1}2^{h_2}3^{h_3}] = 
\begin{cases}
c[12]^{h_1+h_2-h_3}[13]^{h_1+h_3-h_2}[23]^{h_2+h_3-h_1} & \text{ if } h_1 + h_2 + h_3 > 0 \\
c\langle 12\rangle ^{h_3-h_1-h_2}\langle 13\rangle ^{h_2-h_1-h_3}\langle 23\rangle ^{h_1-h_2-h_3}& \text{ if } h_1 + h_2 + h_3 < 0 \\
c & \text{ if } h_1 = h_2 = h_3 = 0 \\
0 & \text{ otherwise} \\
\end{cases}
\end{equation}
where $c$ is an arbitrary constant \cite{McGady:2013sga,Elvang:2015rqa}. Superamplitudes must furthermore satisfy the supersymmetric Ward identities, and this uniquely fixes them to take the form:
\begin{equation}\begin{split}
\cA[\Phi\Phi\Psi] &= \frac {g_1} {[13]^2[23]^2} \delta^{(6)}([12]\eta_3+[23]\eta_1+[31]\eta_2)+\frac {g_2\langle12\rangle^3} {\langle13\rangle^7\langle23\rangle^7} \delta^{(12)}(\langle12\rangle\eta_3+\langle23\rangle\eta_1+\langle31\rangle\eta_2)\,, \\
\cA[\Phi\Phi\Phi] &= \frac {g_3}{[12][13][23]} \delta^{(6)}([12]\eta_3+[23]\eta_1+[31]\eta_2)\,,
\end{split}\end{equation}
where 
\begin{equation}
\begin{split}
\delta^{(6)}([12]\eta_3+[23]\eta_1+[31]\eta_2) &= \prod_{I = 1}^6([12]\eta_{3I}+[23]\eta_{1I}+[31]\eta_{2I})\,,  \\
\delta^{(12)}(\langle12\rangle\eta_3+\langle23\rangle\eta_1+\langle31\rangle\eta_2) &= \prod_{I = 1}^6(\langle12\rangle\eta_{3I}+\langle23\rangle\eta_{1I}+\langle31\rangle\eta_{2I})^2\,.
\end{split}\end{equation}
The $g_1$ term in the $\cA[\Phi\Phi\Psi]$ superamplitude corresponds to the usual supergravity three-point function, and in particular gives rise to a graviton scattering amplitude
\begin{equation}
  \cA[h^+h^+h^-] = g_1\frac{[12]^6}{[13]^2[23]^2} \,.
\end{equation}
The $g_2$ and $g_3$ terms both vanish due to crossing symmetry; if we exchange $1\leftrightarrow2$ then $\cA[\Phi\Phi\Phi]$ and $\cA[\Phi\Phi\Psi]$ must be even, but this is only possible if $g_2 = g_3 = 0$.

Since there is only one supergravity three-point function, we can now determine the corresponding unique four point exchange amplitude. Because the tree-level graviton amplitudes in pure supergravity are identical to those in pure gravity \cite{Elvang:2015rqa}, we can simply use the pure gravity result to deduce that
\begin{equation}
f_1^\text{SG}(s,t) =  \frac {g_1^2} {stu} \,,\qquad f_2^\text{SG}(s,t) = f_3^\text{SG}(s,t) = 0 \,.
\end{equation}
We can then substitute this into \eqref{flatSSSS} to find that the $A[\phi\phi\overline\phi\overline\phi]$ amplitude at large $s, t$ is expected to be
 \es{SGLimit}{
  M^i_\text{SG}(s, t) = g_1^2\begin{pmatrix} 
\frac{tu}{s} &  \frac{su}{t}  &\frac{st}{u} & \frac{s}{2} & \frac{t}{2} & \frac{u}{2}
\end{pmatrix} +  \text{subleading in $s, t$} \,.
 }

\subsection{Supersymmetric Ward identities}
\label{WARDSMAIN}
Our task now is to determine $M_3^i(s,t)$ and $M_4^i(s,t)$. In order to do so we will need to compute the superconformal Ward identities relating the $\cS^i(U,V)$ both to one another and to the correlators of the superconformal descendants of $S_a{}^b$.

The operators in the $\cN = 6$ stress tensor multiplet are shown in Table \ref{stressTable}. There are three fermions with dimension $3/2$, the $\chi_\alpha$, the $F_\alpha$, and its Hermitian conjugate the $\overline F_\alpha$. In addition to the pseudoscalar $P$, at dimension $2$ there are two conserved currents; the R-symmetry current $J_\mu$ in the $\bf 15$, and the $U(1)$ flavour current $j_\mu$ which is an $SO(6)$ singlet. Completing the multiplet are the supercurrent $\psi_{\mu\alpha}$ in the $\bf 6$ and finally the stress tensor itself, $T_{\mu\nu}.$ In Table~\ref{stressTable}, we also list which particles these operators correspond to in the flat space limit.

\begin{table}
\begin{center}
\hspace{-.4in}
{\renewcommand{\arraystretch}{1.2}
\begin{tabular}{c|c|c|c|c}
  Operator & $\Delta$ & Spin & $\mathfrak{so}(6)_R$ irrep & Flat space \\
  \hline
  $S$    & 1   & 0   & $\bf 15$ & $\phi + \overline \phi$ \\ \hline
  $\chi$ & 3/2 & 1/2 & $\bf 6$  & $\chi^\pm$\\
  $F$    & 3/2 & 1/2 & $\bf 10$ & $F^\pm$ \\
  $\overline F$    & 3/2 & 1/2  & ${\overline{\bf 10}}$ & $F^\pm$ \\   \hline
  $P$ & 2 & 0 & $\bf 15$ & $i(\phi-\overline \phi)$ \\
  $J$ & 2 & 1 & $\bf 15$ & $g^\pm$ \\
  $j$ & 2 & 1 & $\bf 1$ & $a^\pm$ \\ \hline
  $\psi$ & 5/2 & 3/2 & $\bf 6$ & $\psi^\pm$ \\ \hline
  $T$ & 3 & 2 & ${\bf 1}$ & $h^\pm$ \\ \hline
\end{tabular}}
\caption{The conformal primary operators in the $\cN = 6$ stress tensor multiplet.  For each such operator, we list the scaling dimension, spin, $\mathfrak{so}(6)_R$ representation, and the particle whose scattering amplitudes it is related to in the flat space limit of the $AdS_4$ dual.}
\label{stressTable}
\end{center}
\end{table}

To impose superconformal invariance on a correlator, it is sufficient to impose conformal invariance, R-symmetry invariance, and invariance under the Poincar\'e supercharge $Q^{\alpha I}$. We have already seen how to impose the first two symmetries on the $\langle SSSS\rangle$, and it is straightforward to expand other correlators in the multiplet as a sum of conformal and R-symmetry invariants. Explicit expressions for these can be found in Appendix \ref{WARDAP}. The supersymmetric Ward identities then follow by imposing that the $Q$ variations vanish:
\begin{equation}
 \label{firstVars}
  \delta\langle SSS\chi\rangle = 0\,,\qquad 
   \delta \langle SSSF\rangle = 0 \,.
\end{equation}
From \eqref{firstVars} we can derive
\es{SSSSward}{
\partial_U \cS^6(U,V) &= \frac 1 {2U^2}\bigg[-(U^3\partial_U+U^2V\partial_V)\cS^1+(1-V+U(V-1)\partial_U+UV\partial_V)\cS^2\\
&\ \ \ +(1-U-V-U(1-2U+U^2-V)\partial_U+U(1-U)V\partial_V)\cS^3 \\
&\ \ \ +(2-U-2V+2U(U+V-1)\partial_U+2UV\partial_V)\cS^4\\
&\ \ \ -U(1+2U(U-1)\partial_U+2UV\partial_V)\cS^5 + U\cS^6\bigg] \,,\\
\partial_V \cS^6(U,V) &= \frac 1 {2U}\bigg[ U(U\partial_U+(V-1)\partial_V)\cS^1+(1-U\partial_U-U\partial_V)\cS^2\\
&\ \ \ +(1+U(U-1)\partial_U+UV\partial_V)\cS^3+(2-2U\partial_U)\cS^4\\
&\ \ \ +(2U^2\partial_U+2UV\partial_V)\cS^5\bigg] \,.
}

We can use \eqref{firstVars} as well as other similar SUSY Ward identities in order to determine the relations between $\langle SSSS \rangle$ and other four-point functions of operators in the stress tensor multiplet.  Note, however, that we will not be able to determine the four-point function of the stress tensor multiplet completely.  This should already be clear from the flat space limit, where we can ask the analogous question for the flat space scattering amplitudes:  given $A[\phi \phi \overline{\phi} \overline{\phi}]$, can we determine all the other component amplitudes?  The answer is no, because it is only the superamplitude $\cA[\Phi \Phi \Psi \Psi]$ that contributes to $A[\phi \phi \overline{\phi} \overline{\phi}]$.  Therefore, knowing $A[\phi \phi \overline{\phi} \overline{\phi}]$ allows us to determine the function $f_1(s, t)$ in \eqref{FampSol} via \eqref{ATof} and leaves the complex functions $f_2(s, t)$ and $f_3(s, t)$ undetermined.  In other words, $A[\phi \phi \overline{\phi} \overline{\phi}]$ determines only one out of five super-amplitudes.

The situation is better for ${\cal N} = 6$ SCFTs where from $\langle SSSS \rangle$ we can determine more than just one out of five superconformal invariants.  The reason for this improvement is that although some of the superconformal invariants do not contribute to $\langle SSSS \rangle$ in the flat space limit, they do contribute at subleading orders in $1/L$.  It can be argued that $\langle SSSS \rangle$ is related to two out of the five super-invariants as follows.  While the stress tensor multiplet forms a representation of the superconformal group $OSp(6|4)$, it also forms a representation of a larger group that includes two $\Z_2$ transformations:  a parity transformation ${\cal P}$ and discrete R-symmetry transformation ${\cal Z}$ whose precise definitions are given in Appendix~\ref{DISCRETESCFT}.  Moreover, the superconformal Ward identity relates four-point structures that have the same ${\cal P}$ and ${\cal Z}$ charges.  Because the $\langle SSSS \rangle$ correlator is ${\cal P}$-even and ${\cal Z}$-even, and only one other structure has this property, it follows that from $\langle SSSS \rangle$ we can determine at most two out of the five superconformal structures.  Explicit computations show that we can indeed determine two superconformal invariants.  In Table~\ref{SCFTDet}, we give examples of operators that contribute to each superconformal structure. The correlator $\langle SSSS \rangle$ allows us to determine the conformal structures in the second and fifth columns of this table.

\begin{table}[htp]
\begin{small}
\begin{center}
\begin{tabular}{c||c|c|c|c|c|}
 & ${\cal A}[\Phi \Phi \Psi \Psi]$ & \multicolumn{2}{|c|}{${\cal A}[\Phi \Phi \Phi \Psi]$, ${\cal A}[\Psi \Psi \Psi \Phi]$} & \multicolumn{2}{|c|}{${\cal A}[\Phi \Phi \Phi \Phi]$, ${\cal A}[\Psi \Psi \Psi \Psi]$}\\ \hline 
$\cal Z$         & $+$ & $-$ & $-$ & $+$ & $+$ \\
$\cal P$, $\cC \cT$         & $+$ & $+$ & $-$ & $+$ & $-$ \\
\hline \hline
$\langle SSSS \rangle$ & $A[\phi \phi \overline{\phi} \overline{\phi}]$ & None & None & Subleading & None \\
$\langle SSPP \rangle$ & & & & & \\
$\langle PPPP \rangle$ & & & & & \\
\hline 
$\langle SSSP\rangle$  & None & None & Subleading & None & Subleading \\
$\langle SPPP\rangle$  &  &  &  &  &  \\
\hline
$\langle SSFF \rangle$ & $A[\phi F^+ F^- \overline{\phi}]$ & Subleading & Subleading & Subleading & Subleading \\
$\langle SPFF \rangle$ & & & & & \\
$\langle PPFF \rangle$ & & & & & \\
\hline
$\langle SS \chi \chi \rangle$ & $A[\phi \chi^- \chi^+ \overline{\phi}]$ & None & None & Subleading & Subleading\\
$\langle SP\chi\chi \rangle$ & & & & & \\
$\langle PP\chi\chi \rangle$ & & & & & \\
\hline
$\langle SS S j \rangle$ & Subleading & $A[\phi \phi \phi a^+]$& $A[\phi \phi \phi a^+]$ & Subleading & Subleading \\ 
$\langle SS P j \rangle$ & & & & & \\
$\langle SP P j \rangle$ & & & & & \\
$\langle PP P j \rangle$ & & & & & \\
\hline
$\langle FFFF \rangle$ & $A[F^+ F^+ F^- F^-]$ & $A[F^+ F^+ F^+ F^-]$& $A[F^+ F^+ F^+ F^-]$ & $A[F^+ F^+ F^+ F^+]$ & $A[F^+ F^+ F^+ F^+]$ \\
etc. & & & & & 
\end{tabular}
\caption{Examples of CFT four-point correlators that contribute to the five superconformal invariants.  Each superconformal invariant can be labeled by its transformation properties under the discrete symmetries $\cal P$ and $\cal Z$.  For every CFT correlator in the first column, we list how it contributes to the superconformal invariants in Mellin space:  either at leading order, in which case we list the scattering amplitude it should match at this order;  either at subleading order, in which case we write ``Subleading'';  or it does not contribute, in which case we write ``None.''}
\label{SCFTDet}
\end{center}\end{small}
\end{table}%



In the next section, we will need to know the relation between $\langle SSPP\rangle$ and $\langle SSSS \rangle$.  To derive this relation, we need to consider one more variation, $\delta\langle SSP\chi\rangle.$ Using the results of \eqref{firstVars} and the variation $\delta\langle SSP\chi\rangle $, we can compute $\langle SSPP\rangle$, along with $\langle SP\chi \chi\rangle$ and $\langle SP\chi F\rangle$. 
More details can be found in Table \ref{wardCorr} and in Appendix \ref{WARDAP}.  Note that while, as discussed above, the superconformal Ward identities fall short of making it possible to determine the all five superconformal invariants (for instance, we cannot determine $\langle SSFF \rangle$ fully), we will be able to completely determine the correlators $\langle SSPP \rangle$ and, if we wish, $\langle PPPP \rangle$ in terms of $\langle SSSS \rangle$.

\begin{table}
\begin{center}
\begin{tabular}{|c|l l l|l l l l|}\hline
Variation & \multicolumn{3}{c|}{Correlators Used} & \multicolumn{4}{c|}{Correlators Obtained} \\ \hline
$\delta\langle SSS\chi\rangle$ & $\langle SSSS\rangle$       &                           &                       & $\langle SS\chi\chi\rangle$ & $\langle SS\chi F\rangle$       & $\langle SSSj\rangle$ & \\
$\delta\langle SSSF\rangle$    & $\langle SSSS\rangle$       & $\langle SSFF\rangle$     &                       & $\langle SS\chi F\rangle$   & $\langle SSF\overline F\rangle$ & $\langle SSSJ\rangle$ &  \\
$\delta\langle SSP\chi\rangle$ & $\langle SS\chi\chi\rangle$ & $\langle SS\chi F\rangle$ &                       & $\langle SP\chi\chi\rangle$ & $\langle SP\chi F\rangle$       & $\langle SSPP\rangle$ & $\langle SSPj\rangle$ \\ \hline
\end{tabular}
\end{center}
\caption{Taking supersymmetric variations to compute correlators. By setting the variation in the first column to zero, we can use the correlators in the second column to compute the correlators in the third column. For each correlator we only compute the $\cP$ and $\cZ$ invariant structures. In the table we have not included correlators involving $\overline F$ which are related to those with $F$ by Hermitian conjugation. }
\label{wardCorr}
\end{table}

\subsection{The local Mellin amplitudes $M_3^i$ and $M_4^i$}
\label{BULKLOC}

We will now use these Ward identities to find the degree $m$ polynomial Mellin amplitudes $M^i_m(s,t)$  with $m=3,4$.   

\subsubsection{$M_4^i$}
\label{secM4}

The amplitude $M_4^i$ can be obtained from existing results in the literature as follows.  A particular case of ${\cal N} = 6$ SCFTs are ${\cal N} = 8$ SCFTs.  In an ${\cal N} = 8$ SCFT, the stress tensor multiplet has as its bottom component $\Delta = 1$ scalar operators $\overline{S}_{AB}(\vec{x})$ transforming in the ${\bf 35}_c$ irrep of the $\mathfrak{so}(8)_R$ R-symmetry.\footnote{The fact that this representation is the ${\bf 35}_c$ as opposed to one of the other two 35-dimensional irreducible representations of $\mathfrak{so}(8)_R$ assumes a choice of the triality frame.}  (Here $\overline{S}_{AB}(\vec{x})$, with $A, B = 1, \ldots, 8$ being ${\bf 8}_c$ indices, is a traceless symmetric tensor.)  Like in the ${\cal N} = 6$ case, we can use an index-free notation by contracting $\overline{S}_{AB}(\vec{x})$ with a symmetric traceless $8 \times 8$ matrix $\overline{X}_{AB}$.  The four-point function of the ${\bf 35}_c$ scalar operator is restricted by conformal invariance and $\mathfrak{so}(8)_R$ to take the form 
 \es{SbarFour}{
  \langle \overline{S}(\vec{x}_1, \overline{X}_1) \cdots \overline{S}(\vec{x}_4, \overline{X}_4)
   \rangle &=  \frac 1{x_{12}^2 x_{34}^2}\Bigg[\overline{\cS}^1(U,V) \overline{A}_{12}\overline{A}_{34}+\overline{\cS}^2(U,V)\overline{A}_{13}\overline{A}_{24}+\overline{\cS}^3(U,V)\overline{A}_{14}\overline{A}_{23} \\
&{}+ \overline{\cS}^4(U,V) \overline{B}_{1423} + \overline{\cS}^5(U,V) \overline{B}_{1234}+\overline{\cS}^6(U,V) \overline{B}_{1342}\Bigg] \,,
 }
where we defined\footnote{Despite the use of matrix $\mathfrak{so}(8)$ polarizations here, the $\overline{\cS}^i(U, V)$ here are equal to the $\cS_i(U, V)$ in \cite{Chester:2018aca}.}
 \es{ABbarDef}{
  \overline{A}_{ij} \equiv \tr(\overline{X}_i\overline{X}_j) \,, \qquad \overline{B}_{ijkl} \equiv \tr(\overline{X}_i \overline{X}_j \overline{X}_k \overline{X}_l) \,.
 }

The Mellin transforms of $\overline{S}^i$ corresponding to contact interactions were found in \cite{Chester:2018aca}.  With our definition \eqref{melDef} (with ${\cal S}^i_\text{conn}$ replaced by $\overline{\cal S}^i_\text{conn}$ and $M^i$ replaced by $\overline{M}^i$), the result in \cite{Chester:2018aca} for the quartic amplitude is
\es{R4MellinBar}{
 \overline{M}^1_4&= \frac 1 {35} (t-2)(u-2)(35tu+100s-112)\,,\\
 \overline{M}^4_4&= \frac 2 {35}(s-2)(35 s t u-90(t^2+u^2)-280 tu -324s+1072)\,.
}

To relate \eqref{R4MellinBar} to $M_4^i(s, t)$ we should relate the $\mathfrak{so}(8)_R$ structures \eqref{ABbarDef} to the $\mathfrak{su}(4)_R$ ones defined in \eqref{ABRDef}.  Under the decomposition $\mathfrak{so}(8) \to \mathfrak{su}(4)$, we have ${\bf 8}_c \to {\bf 4} + \overline{\bf 4}$, which implies ${\bf 35}_c \to {\bf 10} + \overline{\bf 10} + {\bf 15}$.  To select the ${\bf 15}$, we should restrict the $8 \times 8$ matrices $\overline{X}$ to take the form
 \es{XbToX}{
  \overline{X} = \frac{1}{\sqrt{2}} \biggl[ (\Re X) \otimes I_2 + (\Im X) \otimes (i \sigma_2) \biggr] \,,
 }
where $X$ is a $4 \times 4$ traceless hermitian matrix, $I_2$ is the $2\times 2$ identity matrix, and $\sigma_2$ is the second Pauli matrix.  (See Eq.~(3.16) of \cite{Chester:2018aca}.\footnote{The factor of $1/\sqrt{2}$ is just a choice of normalization.})  Then it is straightforward to check that 
 \es{AbBbToAB}{
  \overline{A}_{ij} = A_{ij} \,, \qquad 
   \overline{B}_{ijkl} = \frac{1}{4} B_{ijkl} \,.
 }
This implies that ${\cal S}^i = \overline{\cal S}^i$ for $i = 1, 2, 3$ and ${\cal S}^i = \frac 14 \overline{\cal S}^i$ for $i = 4, 5, 6$ and analogously for the Mellin amplitudes.  Thus, 
 \es{R4Mellin}{
M_4: \qquad M^1_4&= \frac 1 {35} (t-2)(u-2)(35tu+100s-112)\,,\\
M^4_4&= \frac 1 {70}(s-2)(35 s t u-90(t^2+u^2)-280 tu -324s+1072)\,,
}
where the other $M_4^i$ are given by crossing \eqref{CrossingM}.   The Melin amplitudes $M_4^i$  are normalized so that at large $s, t$ they obey \eqref{largeA4}.

\subsubsection{$M_3^i$}
The degree 3 polynomial Mellin amplitude $M^i_3$ is not allowed by $\mathcal{N}=8$ supersymmetry, and so we must compute it using the $\mathcal{N}=6$ Ward identities derived in the previous section. In particular, we impose the following constraints to find $M_3$:
\begin{enumerate}
	\item $M_3^i$ must satisfy crossing symmetry \eqref{CrossingM}.
	\item $M_3^i$ must be a degree 2 polynomial solution of the $\langle SSSS\rangle$ Ward identities given in position space \eqref{SSSSward}, which can be translated into Mellin space using the rules \eqref{3DMellin}. The ansatz for $M_3$ is only degree 2, since in the previous section we showed that $\mathcal{A}_3$ does not appear in the scattering of four scalars, so $M_3$ must vanish in the flat space limit.
	\item $M_3$ must remain a polynomial when expressed as correlator of other operators in the stress tensor multiplet using the Ward identities in the previous section.\footnote{Instead of imposing this requirement, we could alternatively impose the condition that certain operators in the $S \times S$ OPE do not acquire anomalous dimensions.  For instance, we can uniquely determine $M_3$ if we impose this requirement for the spin $0$ operators of dimension $2$ in the ${\bf 84}$, ${\bf 20}'$, and ${\bf 15}_s$ irreps of $SO(6)_R$, as well as for the spin $1$ operator of dimension $3$ in the ${\bf 45} \oplus \overline{\bf 45}$, all of which belong to protected multiplets and do not mix with unprotected operators.} The degree of these polynomials is at most 2 if the corresponding flat space amplitude vanishes, and 3 otherwise.
\end{enumerate}
   
Condition 3 was trivially satisfied in maximally supersymmetric cases considered before in various dimensions \cite{Binder:2018yvd,Binder:2019jwn}, where polynomial Mellin amplitudes for $\langle SSSS\rangle$ remained polynomials for all other stress tensor multiplets correlators. In our case though, we find that just imposing conditions 1 and 2 leads to five linearly independent solutions: a degree 0, a degree 1, and three degree 2:
 \es{Putative}{
   \text{degree $0$:} \qquad M^1 &= 1 \,, \qquad M^4 = 1 \,, \\
   \text{degree $1$:} \qquad M^1 &= s \,, \qquad M^4 = \frac{s-4}{2} \,, \\
   \text{1st degree $2$:} \qquad M^1 &= (t-2)(u-2)\,,\qquad M^4=\left(s-\frac43\right)(s-2) \,, \\
    \text{2nd degree $2$:} \qquad M^1 &= t u \,, \qquad M^4 = \frac{s (s - 4)}{2} \,, \\
   \text{3rd degree $2$:} \qquad  M^1 &= s^2 \,, \qquad
    M^4 = s^2 + t u  - 3s \,.
 }

To reduce these to a unique amplitude, we must consider the other Ward identities $\langle SS\chi\chi\rangle$, $\langle SS\chi F\rangle$, $\langle SSFF\rangle$, and $\langle SSF\overline F\rangle$ given in Appendix \ref{WARDAP}, which we can transform into Mellin space as in \ref{scalscalfermferm}.   
We can write $\langle SS\chi\chi\rangle$ in terms of the structures $\cC^{a,I}(U,V)$ defined in \eqref{sscc4}, where the indices $a=1,2,3$ and $I=1,2$ refer to the various R-symmetry and conformal structures, respectively. The Mellin transform $M^{SS\chi\chi}_{a,I}(s,t)$ of these $\cC^{a,I}(U,V)$ is then defined by \eqref{MellinDefFerm}. We can relate $M^{SS\chi\chi}_{a,I}(s,t)$ to $M^i(s,t)$ as
\es{MellinRelation}{
M^{SS\chi\chi}_{a,1}=&\left(1-\frac s2\right)^{-1}\widehat {\mathcal{D}_{ai,1}^C}{(U,V,\partial_U,\partial_V)} M^i(s,t)\,,\\
M^{SS\chi\chi}_{a,2}=&\left(1-\frac s2\right)^{-2}\widehat {\mathcal{D}_{ai,2}^C}{(U,V,\partial_U,\partial_V)} M^i(s,t)\,,\\
}
where the $\langle SS\chi\chi\rangle$ Ward identity ${\mathcal{D}_{ai,1}^C}$ is given in position space in \eqref{Cward},  we express derivatives and powers of $U$ and $V$ in Mellin space using the rules \eqref{3DMellin}, and $s$-dependent  prefactors come from the difference in the definition of the scalar and fermion Mellin amplitudes in \eqref{melDef} and \eqref{MellinDefFerm}.  We find that degree $0$ amplitude in \eqref{Putative} gives 
\es{fake0C}{
\text{degree $0$:}  
&   M^{SS\chi\chi}_{1,1}(s,t)=0\,,\quad   M^{SS\chi\chi}_{2,1}(s,t)=\frac{1}{16}\,, \quad M^{SS\chi\chi}_{3,1}(s,t)=\frac{2-t}{16u}\,,\\ 
&   M^{SS\chi\chi}_{1,2}(s,t)=0\,,\quad  M^{SS\chi\chi}_{2,2}(s,t)=\frac{1}{8t}\,,\quad M^{SS\chi\chi}_{3,2}(s,t)=\frac{1}{8u}\,,
}
which contain poles, and so must be discarded. 

When we apply this method to the Ward identities for $\langle SSFF\rangle$ and $\langle SSF\overline F\rangle$, a new subtlety is that these Ward identities \eqref{SSFFward}, \eqref{SSFFward2}, and \eqref{SSFGward} depend on both $\langle SSSS\rangle$ and $\langle SSFF\rangle$, and in particular can be written in terms of $\cS^1(U,V)$ and $\cS^4(U,V)$, as well as the first conformal structure $ \cF^{a,1}(U,V)$ for $\langle SSFF\rangle$ as defined in \eqref{sscc4}, where here $a=1,2$ for the two R-symmetry structures. So to get the constraints from these Ward identities up to degree 2, we must also consider a degree 2 polynomial ansatz for the Mellin transform $M^{SSFF}_{a,1}(s,t)$ of $ \cF^{a,1}(U,V)$, which satisfies the crossing relations
\es{SSFFcross}{
M^{SSFF}_{1,1}(s,t)&=M^{SSFF}_{2,1}(s,u)+\left(1-\frac s2\right)M^{SSFF}_{2,2}(s,u)\,,\\
M^{SSFF}_{2,1}&=M^{SSFF}_{1,1}(s,u)+\left(1-\frac s2\right) M^{SSFF}_{1,2}(s,u)\,,\\
M^{SSFF}_{1,2}(s,t)&=-\left(1-\frac s2\right)M^{SSFF}_{2,2}(s,u)\,,\\
M^{SSFF}_{2,2}(s,t)&=-\left(1-\frac s2\right)M^{SSFF}_{1,2}(s,u)  \,, 
}
where the $s$-dependent prefactors come from the difference in the definition of the fermion Mellin amplitudes in \eqref{MellinDefFerm} for the two different conformal structures. After imposing the $\langle SS\chi F\rangle$, $\langle SSF F\rangle$, and $\langle SSF\overline F\rangle$ Ward identities, just as we did for $\langle SS\chi\chi\rangle$ above, and demanding that all poles vanish, we find that $M^{SSFF}_{a,1}(s,t)$ is completely fixed in terms of $M^i(s,t)$ up to degree 2, and that only a single degree 2 solution for $M^i(s,t)$ survives:
\es{degree2}{
M_3:&\qquad M_3^1=(t-2)(u-2)\,,\qquad M_3^4=\left(s-\frac43\right)(s-2)\,,\\
}
which in fact corresponds to the degree 3 Mellin amplitude $M_3(s,t)$ as discussed before.

\subsection{Supergravity exchange Mellin amplitude}
\label{superSec}
We will also use the supergravity amplitude $M^i_\text{SG}(s,t)$, which contains an infinite series of poles that correspond to the stress tensor multiplet operators (or the exchange of the graviton multiplet in the bulk) and their descendants.  This amplitude is unique and can be derived using the method we used above for determining $M_4^i$ by translating the ${\cal N} = 8$ SCFT results into ${\cal N} = 6$ language.  For the case of 3d $\cN = 8$ CFTs, $M^i_\text{SG}$ 
was derived in \cite{Zhou:2017zaw}.
From Eqs. (E.1) and (4.8) of \cite{Binder:2018yvd} , and converting to $\mathcal{N}=6$ notation as we did before in subsection \ref{secM4}, we find that
\es{SugMellin}{
M^1_\text{SG}&=-\frac{(t-2)(u-2)}{s(s+2)}\left(\frac{4\Gamma\left(\frac{1-s}2\right)}{\sqrt{\pi}\Gamma\left(1-\frac s2\right)}-(4+s)\right)
\,,\\
   M^4_\text{SG}&= -\frac{s-2}{2t u}\left(\frac{2u\Gamma\left(\frac{1-t}2\right)}{\sqrt\pi \Gamma\left(1-\frac t2\right)}+\frac{2t\Gamma\left(\frac{1-u}2\right)}{\sqrt\pi \Gamma\left(1-\frac u2\right)}+2s-tu-8\right)  \,,\\
}
where the other $M_\text{SG}^i$ are given by crossing \eqref{CrossingM}. We normalize $M_\text{SG}^i$ so that at large $s, t$ they obey \eqref{SGLimit} with $g_1=1$.

\section{Constraints from supersymmetric localization}
\label{locloc}

In order to determine the coefficients of the Mellin amplitudes $M_3$ and $M_4$ derived in the previous section in the case of ABJM theory, we will use information from supersymmetric localization.  Similarly to \cite{Chester:2018aca,Binder:2018yvd,Binder:2019jwn}, we will focus on the mass-deformed partition function of ABJM theory on a round $S^3$.  While it would be interesting to also obtain constraints coming from the partition function on a squashed $S^3$ \cite{Hama:2011ea}, in this work we will use the round sphere simply because the mass-deformed partition function can be computed \cite{Nosaka:2015iiw} using the Fermi gas formalism developed in \cite{Marino:2011eh} to all orders in the $1/N$ expansion.  A similar result for the squashed sphere partition function is not currently available.

\subsection{Integrated correlators on $S^3$}
\label{INTCOR}

To set the stage, let us begin with the result for the $S^3$ partition function in the presence of a mass deformation.  On $S^3$, there are two classes of mass deformations of ABJM theory that one can consider:  in $\cN = 2$ notation, there are superpotential mass deformations and real mass deformations.  The $S^3$ partition function has no dependence on the superpotential mass parameters, so we will focus on the real mass parameters.  These real masses are associated with global symmetries, because they can be constructed by coupling the conserved currents of the $\cN = 2$ theory to background vector multiplets and giving expectation values proportional to the mass parameters to the scalars in the vector multiplets.  Since ABJM theory has $\cN = 6$ SUSY for arbitrary $k$, it has an $SO(6)_R$ R-symmetry as well as an $U(1)_F$ global symmetry, with both the $SO(6)_R$ and $U(1)_F$ conserved currents belonging to the same multiplet as the stress-energy tensor, as discussed in the previous section.  When passing to $\cN = 2$ notation, a $U(1)_R$ subgroup of $SO(6)_R$ can be viewed as the $\cN = 2$ R-symmetry, and in $SO(6)_R \times U(1)_F$ there are three other $U(1)$'s that commute with one another and with $U(1)_R$.  (They are the Cartans of an $SO(4) \times U(1)$ flavor symmetry from the $\cN = 2$ point of view.)  Each of these $U(1)$'s can be coupled to an Abelian background vector multiplet, so for each of them one may consider introducing a real mass parameter.  There are thus three distinct real mass parameters.

For simplicity, in this work we will focus on only two of the three real mass parameters of ABJM theory.\footnote{In terms of symmetries, the two mass parameters that we consider correspond to linear combinations of $U(1)_F$ and one of the Cartans of an $SU(2)$ factor inside $SO(4) \cong SU(2) \times SU(2)$.}   Recall that ABJM theory in $\cN = 2$ notation is a theory of two $U(N)$ vector multiplets coupled to bifundamental chiral multiplets $\cW_i$, $i = 1, 2$ in $({\bf N}, \overline {\bf N})$ and $\cZ_i$ in $(\overline {\bf N}, {\bf N})$ of $U(N) \times U(N)$.  The two mass parameters we consider, denoted $m_+$ and $m_-$, correspond to giving masses $(m_+/2, m_-/2, -m_+/2, -m_-/2)$ to $\cW_1, \cW_2, \cZ_1, \cZ_2$, respectively.  The partition function can be written as \cite{Kapustin:2009kz,Jafferis:2010un}:
 \es{ZS3}{
  Z = \int d^N \lambda \, d^N \mu \, 
   \frac{e^{i \pi k \sum_i (\lambda_i^2 - \mu_i^2)} \prod_{i< j} 16 \sinh^2 \left[ \pi( \lambda_i - \lambda_j ) \right]  \sinh^2 \left[ \pi( \mu_i - \mu_j ) \right] }
   {\prod_{i, j} 4 \cosh \left[ \pi (\lambda_i - \mu_j) + \frac{\pi m_+}{2} \right]\cosh \left[ \pi (\lambda_i - \mu_j) + \frac{\pi m_-}{2} \right] } \,.
 }
The purpose of this section is to relate the mixed derivatives
 \es{derivs}{
  \frac{\partial^4 \log Z}{\partial m_+^4} \,, \qquad
   \frac{\partial^4 \log Z}{\partial m_-^4} \,, \qquad
    \frac{\partial^4 \log Z}{\partial m_+^2 \partial m_-^2} \,,
 }
all evaluated at $m_+ = m_- = 0$, to the correlation functions of the $S_a{}^b$ operators introduced in the previous section.

In the ABJM Lagrangian on a unit radius $S^3$, the parameters $m_+$ and $m_-$ appear at linear order as
 \es{ABJMLagMass}{
  m_+ \int \left( i J_+ + K_+ \right)  + m_- \int \left( i J_- + K_- \right) + O(m_\pm^2)  \,,
 }
where $J_\pm$ are linear combinations of the $S$'s and $K_\pm$ are linear combinations of the $P$'s.  In terms of the Lagrangian fields, the $J_\pm$ are scalar bilinears which are quadratic in the bottom components of the chiral multiplets $\cW_i = (W_i, \chi_i)$ and $\cZ_i = (Z_i, \psi_i)$, while the $K_\pm$ are fermion mass terms quadratic in the fermions in the same chiral multiplets:
 \es{JK}{
  J_+ &=  \frac 12 \tr \left( \abs{W_1}^2 -  \abs{Z_1}^2 \right) \,,  \qquad
   J_- =  \frac 12  \tr \left( \abs{W_2}^2 -  \abs{Z_2}^2 \right) \,, \\
  K_+ &= \frac 12   \tr \left( \chi_1^\dagger \chi_1 -  \psi_1^\dagger \psi_1 \right)\,, 
   \qquad K_- = \frac 12   \tr \left( \chi_2^\dagger \chi_1 -  \psi_2^\dagger \psi_2 \right) \,.
 }
The mixed derivatives \eqref{derivs} are given in terms of connected correlation functions as 
 \es{derivsCorrels}{
  \frac{\partial^4 \log Z}{\partial m_+^4} &= \left \langle \left( \int \left( i J_+ + K_+ \right) \right)^4 \right \rangle_\text{conn} 
   + \text{($2$- and $3$-pt functions)} \,, \\
  \frac{\partial^4 \log Z}{\partial m_-^4} &= \left \langle \left( \int \left( i J_- + K_- \right) \right)^4 \right \rangle_\text{conn} + \text{($2$- and $3$-pt functions)} \,, \\
  \frac{\partial^4 \log Z}{\partial m_+^2 \partial m_-^2} &= \left \langle  \left( \int \left( i J_+ + K_+ \right) \right)^2 
    \left( \int \left( i J_- + K_- \right) \right)^2 \right \rangle_\text{conn} + \text{($2$- and $3$-pt functions)} \,.
 }
where the $2$- and $3$-point function terms not written in \eqref{derivsCorrels} come from the $O(m^2)$ terms not written in \eqref{ABJMLagMass}.  We will not write down these $2$- and $3$-point function contributions because they will be automatically taken into account in the final formulas, by analogy with the similar situation encountered in \cite{Binder:2019jwn}.

To determine how $J_\pm$ and $K_\pm$ are related to $S$ and $P$, let us first note that $C^a = (W_1, Z_1^\dagger, W_2, Z_2^\dagger)$ and $\Psi^a = (\psi_2^\dagger, \chi_2, \psi_1^\dagger, \chi_1)$ transform as fundamentals of $SU(4)_R$,\footnote{The reason why the components of the chiral multiplets do not appear in the same order in this expression is that we require the $U(1)_R$ symmetry to be generated by the $\mathfrak{su}(4)_R$ matrix $\diag\{ 1/2, -1/2, 1/2, -1/2 \}$.} so $J_\pm$ and $K_\pm$ can be written as
 \es{JpmKpmAgain}{
  J_\pm =  \frac 12  (X_\pm)_a{}^b \tr (C_b^\dagger C^a) \,, \qquad
   K_\pm =  -\frac 12  (X_\mp)_a{}^b \tr (\Psi_b^\dagger \Psi^a)
 }
where we defined
 \es{GotMpm}{
  X_+ \equiv \diag\{1, -1, 0, 0\} \,, \qquad X_- \equiv \diag\{0, 0, 1, -1\}  \,.
 } 
Because $\tr C^\dagger_a C^b$ and $\tr \Psi^\dagger_a \Psi^b$ transform in the ${\bf 15}$ of $SU(4)_R$, they must be proportional to $S_a{}^b$ and $P_a{}^b$, respectively.  Eq.~\eqref{JpmKpmAgain} then implies that
 \es{JKpm}{
  J_\pm(\vec{x})  = N_J S(\vec{x}, X_\pm) \,, \qquad  K_\pm(\vec{x})  = N_K P(\vec{x}, X_\mp) \,, 
 }
where $N_J$ and $N_K$ are normalization constants.  

On general grounds, the two-point functions of $J_\pm$ and $K_\pm$ must be proportional to the coefficient $c_T$ appearing in the two-point function of the canonically normalized stress-energy tensor.  Because the two-point functions of $S$ and $P$ are both normalized as in \eqref{SNorm}, knowing that $N_J^2$ and $N_K^2$ are proportional to $c_T$ allows us to determine them in a free theory, such as the $k\to \infty$ limit of the $U(1)_k \times U(1)_{-k}$ ABJM theory.  In this limit, the $\cW_i$ and $\cZ_i$ chiral multiplets are free, and $\langle C^a(\vec{x}_1) C_b^\dagger(\vec{x}_2)  \rangle = \delta_b^a / (4 \pi \abs{\vec{x}_{12}})$ and $\langle \Psi^a(\vec{x}_1) \Psi_b^\dagger(\vec{x}_2) \rangle = \delta_b^a \gamma_\mu x_{12}^\mu / (4 \pi \abs{\vec{x}_{12}})$.  From the definition \eqref{JpmKpmAgain}, we then have 
 \es{FreeTwoPt}{
  \text{free theory:}\qquad  \langle J_\pm(\vec{x}_1) J_\pm(\vec{x}_2) \rangle = \frac{1}{32 \pi^2 \abs{\vec{x}_{12}}^2} \,, \qquad
   \langle K_\pm(\vec{x}_1) K_\pm(\vec{x}_2) \rangle = \frac{1}{16 \pi^2 \abs{\vec{x}_{12}}^4} \,.
 }
These expressions should be compared with what we obtain from \eqref{JKpm} and \eqref{SNorm}, which is
 \es{FreeTwoPtAgain}{
  \text{free theory:}\qquad  \langle J_\pm(\vec{x}_1) J_\pm(\vec{x}_2) \rangle = \frac{2 N_J^2}{\abs{\vec{x}_{12}}^2} \,, \qquad
   \langle K_\pm(\vec{x}_1) K_\pm(\vec{x}_2) \rangle = \frac{2 N_K^2}{\abs{\vec{x}_{12}}^4} \,.
 }
Thus, for a free theory, we have $N_J^2 = 1/(64 \pi^2)$ and $N_K^2 = 2 N_J^2$.  In conventions in which a free massless real scalar or a free real Majorana fermion has $c_T = 1$, as in \eqref{CanStress}, the free theory has $c_T = 16$.  From this, and the fact that $N_J^2$ and $N_K^2$ should be proportional to $c_T$, we conclude that we must have
 \es{GotNJNK}{
  N_J^2 = \frac{c_T}{2^{10}\pi^2} \,, \qquad  N_K^2 = 2N_J^2 \,.
 }
Note that the second derivatives of $Z$ are $\frac{\partial^2 \log Z}{\partial m_\pm^2} \big|_{m_\pm=0} 
   =  \langle \left( \int \left( i J_+ + K_+ \right) \right)^2  \rangle$.  
Using \eqref{FreeTwoPtAgain} and \eqref{GotNJNK} and explicitly evaluating the integrals gives \cite{Closset:2012vg}
 \es{Gottau}{
  c_T = -\frac{64}{\pi^2}\frac{\partial^2 \log Z}{\partial m_\pm^2} \bigg|_{m_\pm=0}  \,.
 }

Having determined the normalization factors in \eqref{JKpm}, we can then evaluate \eqref{derivsCorrels} .  The result will be given in terms of the functions ${\cal S}^i$ that appear in the $\langle SSSS \rangle$ correlator in Eq.~\eqref{SSSScor} as well as analogous functions that appear in $\langle SSPP \rangle$ and $\langle PPPP \rangle$.   While this is certainly a valid procedure,\footnote{The result is  
\es{derivsCorrelsAgain}{
   \frac{\partial^4 \log Z}{\partial m_+^4}
    &= \frac{\partial^4 \log Z}{\partial m_-^4}  = 
   4  \sum_{i=1}^6  \left( N_J^4  I_{1, 1} [{\cal S}^i]  + N_K^4  I_{2, 2} [{\cal P}^i]   \right)
    - 24 N_J^2 N_K^2  I_{2, 1} [{\cal R}^1]  + \text{($2$- and $3$-pt functions)} \,, \\
   \frac{\partial^4 \log Z}{\partial m_+^2 \partial m_-^2}
    &= 
   \frac{4}{3}  \sum_{i=1}^3  \left( N_J^4  I_{1, 1} [{\cal S}^i]  + N_K^4  I_{2, 2} [{\cal P}^i]  \right) 
    -8  N_J^2 N_K^2 \left( I_{2, 1} [{\cal R}^2 + {\cal R}^3] + \sum_{i=1}^6    I_{2, 1} [{\cal R}^i] \right) \\
    &{}+ \text{($2$- and $3$-pt functions)} \,,
 }
where ${\cal S}^i$ are the functions appearing in  \eqref{SSSScor}, ${\cal R}^i$ are the functions appearing in the $\langle SSPP \rangle$ correlator given in \eqref{SSPP}, ${\cal P}^i$ are the six functions appearing in the $\langle PPPP \rangle$ correlator defined as in \eqref{SSSScor} but with $S \to P$ and ${\cal S}^i \to {\cal P}^i$, and 
 \es{IDef}{
  I_{\Delta_A, \Delta_B} [{\cal G} ] 
   = \int \left(\prod_{i=1}^4 d^3 \vec{x}_i \right)
    \frac{ \left[ \Omega(\vec{x}_1) \Omega(\vec{x}_2) \right]^{3 - \Delta_A} \left[ \Omega(\vec{x}_3) \Omega(\vec{x}_4) \right]^{3 - \Delta_B} }
     {\vec{x}_{12}^{2 \Delta_A} \vec{x}_{34}^{2 \Delta_B} } {\cal G}(U, V) \,, \qquad
      \Omega(\vec{x}) \equiv \frac{1}{1 + \frac{x^2}{4} } \,.
 } 
The powers of $\Omega$ in \eqref{IDef} appear because the operators are integrated over $S^3$ as opposed to $\R^3$. 
} it is possible to obtain simpler formulas by making use of the fact that all ${\cal N} \geq 4$ SCFTs in 3d have a 1d topological sector \cite{Beem:2013sza,Chester:2014mea,Beem:2016cbd,Dedushenko:2016jxl,Dedushenko:2017avn,Dedushenko:2018icp}.  

In general, a 3d ${\cal N} = 4$ SCFT has $SU(2)_H \times SU(2)_C$ R-symmetry, and one can consider $1/2$-BPS operators that have scaling dimension $\Delta = j_H$, where $j_H$ is the $SU(2)_H$ spin, and are invariant under $SU(2)_C$. Such operators can be written as rank-$2j_H$ symmetric tensors ${\cal O}_{a_1 a_2 \ldots a_{2j_H}}(\vec{x})$ where $a_i = 1, 2$ are $SU(2)_H$ spinor indices.  From these operators, one can construct 1d topological operators by inserting them on a line, say the line $(0, 0, x)$, and contracting the $SU(2)_H$ indices with position-dependent polarizations:
 \es{TopOp}{
  \tilde {\cal O}_{\R^3}(x) = {\cal O}_{a_1 a_2 \ldots a_{2j_H}}(0, 0, x) u^{a_1}(x) \cdots u^{a_{2 j_H}}(x) \,,
 }
where we can take\footnote{In the notation of \cite{Dedushenko:2016jxl} this choice corresponds to $h_a{}^b = (\sigma_3)_a{}^b$.}
 \es{uDef}{
  u^a(x) = \begin{pmatrix}
   1 + \frac{i x}{2} \\
   1 - \frac{i x}{2} 
  \end{pmatrix} \,.
 }
If we want to express the topological operator in terms of the operator ${\cal O}_{a_1 a_2 \ldots a_{2j_H}}$ when the theory is placed on $S^3$, we have 
 \es{S3Op}{
   \tilde {\cal O}(x) = \frac{1}{\left(1 + \frac{x^2}{4}\right)^{j_H}}  {\cal O}_{a_1 a_2 \ldots a_{2j_H}}(0, 0, x) u^{a_1}(x) \cdots u^{a_{2 j_H}}(x) \,,
 }
where the extra factor accounts for the fact that the operators on $\R^3$ and those on $S^3$ differ by a Weyl factor.  In this case, the 1d topological theory lives on a circle parameterized by $x$, with the point at $x = + \infty$ being identified with the point at $x = -\infty$.

To connect this discussion to the ${\cal N} = 6$ ABJM theory, let us embed the ${\cal N} = 4$ $SU(2)_H \times SU(2)_C$ R-symmetry into $SU(4)_R$ such that $SU(2)_H$ corresponds to the top left $2\times 2$ block of an $SU(4)_R$ matrix written in the fundamental representation and $SU(2)_C$ corresponds to the bottom right $2\times 2$ block.  Raising and lowering indices with the epsilon symbol, Eqs.~\eqref{TopOp} and \eqref{S3Op} applied to $S$ give
 \es{STilde}{
  \tilde S(x) = \frac{\left( 1 + \frac{i x}{2} \right)^2}{1 + \frac{x^2}{4}} S_1{}^2(0, 0, x) - \frac{\left( 1 - \frac{i x}{2} \right)^2}{1 + \frac{x^2}{4}} S_2{}^1(0, 0, x) +  S_1{}^1(0, 0, x) - S_2{}^2(0, 0, x) 
 }
on $S^3$ and $\tilde S_{\R^3} (x) = \left(1 + \frac{x^2}{4} \right) S(x)$ on $\R^3$.  It is straightforward to check that the superconformal Ward identities \eqref{SSSSward} imply that the four-point function of $\tilde S_{\R^3}$, namely
 \es{SSSSTop}{
  \langle \tilde S_{\R^3}(x_1) \tilde S_{\R^3}(x_2) \tilde S_{\R^3}(x_3) \tilde S_{\R^3}(x_4) \rangle
   &= {\cal S}^1 + \frac{{\cal S}^2}{z^2} + \frac{(1-z)^2 {\cal S}^3}{z^2} \\
    &{}+ \frac{2 (1-z) {\cal S}^4}{z^2}
     - \frac{2 (1-z) {\cal S}^5}{z} 
      + \frac{2 {\cal S}^6}{z} \bigg|_{\substack{U = z^2 \\
      V = (1-z)^2 }} \,,
 }
where $z \equiv \frac{(x_1 - x_2) (x_3 - x_4) }{(x_1 - x_3) (x_2 - x_4)}$, is piece-wise constant.

The advantage of the topological sector is that we can replace the integrated operator $\int_{S^3} d^3 \vec{x} \, \sqrt{g}  (i J_+ + K_+)$ by a different operator that is integrated only along the circle.  Such a  replacement can be rigorously justified in the class of ${\cal N} = 4$ theories studied in \cite{Dedushenko:2016jxl,Dedushenko:2017avn,Dedushenko:2018icp} where it was shown how one can obtain a 1d action for the topological sector by using supersymmetric localization in the 3d ${\cal N} = 4$ theory.  Unfortunately, ABJM theory with $k>1$ falls outside the range of theories studied in \cite{Dedushenko:2016jxl,Dedushenko:2017avn,Dedushenko:2018icp}.  Nevertheless, as explained in Section 3.1 of \cite{Agmon:2017xes}, one expects that such a replacement should be possible in ABJM theory as well.  In particular, one expects \cite{Agmon:2017xes}
\es{identity}{
4 \pi \int \frac{dx}{1 + \frac{x^2}{4}} i \tilde J(x) = \int d^3 \vec{x}\, \sqrt{g} (i J_+ + K_+) + (\text{$Q$-exact terms} ) \,, 
}
where\footnote{A quick check of the normalization is as follows.  The two-point function of the RHS of \eqref{identity} equals $4 \pi^2 N_J^2 \int d^3 \vec{x}\left( - \frac{\Omega(\vec{x})^2}{\abs{\vec{x}}^2} + 2  \frac{\Omega(\vec{x})^1}{\abs{\vec{x}}^4} \right)
 =  -16 \pi^4 N_J^2 $.  The two-point function of the LHS gives $- 16 \pi^2 (N_J^2 / 4) 4 \pi^2  = - 16 \pi^4 N_J^2$.}
 \es{tildeJ}{
  \tilde J(x) = \frac{N_J}{2} \tilde S(x) \,.
 }

Thus, instead of \eqref{derivsCorrels}, we may write 
 \es{DerSimp}{
  \frac{\partial \log Z}{\partial m_+^4} &= (4 \pi)^4 \left \langle 
    \left( \int \frac{dx}{1 + \frac{x^2}{4}} i \tilde J(x) \right)^4 
   \right \rangle_\text{conn} \,, \\
   \frac{\partial \log Z}{\partial m_+^2 \partial m_-^2} &= (4 \pi)^2 \left \langle 
    \left( \int d^3\vec{x} \, \sqrt{g} \left( i J_-(\vec{x}) + K_-(\vec{x}) \right) \right)^2
    \left( \int \frac{dx}{1 + \frac{x^2}{4}} i \tilde J(x) \right)^2 
   \right \rangle_\text{conn} \,.
 } 

Because the correlation function $\langle \tilde J \tilde J \tilde J \tilde J \rangle$ is topological, we can place the four operators at any four locations of our choosing and multiply the answer by $(2 \pi)^4$.  Using \eqref{STilde}, we have  
 \es{DerSimpOneMass}{
  \frac{\partial \log Z}{\partial m_+^4} &= 128\, \pi^8 N_J^4 I_{++}[\cS^i] \,,
}
where 
 \es{Ipp}{
   I_{++}[\cS^i] &= 2 \biggl[
    {\cal S}^1 + \frac{{\cal S}^2}{z^2} + \frac{(1-z)^2 {\cal S}^3}{z^2}
    + \frac{2 (1-z) {\cal S}^4}{z^2}
     - \frac{2 (1-z) {\cal S}^5}{z} 
      + \frac{2 {\cal S}^6}{z} \bigg|_{\substack{U = z^2 \\
      V = (1-z)^2 }} 
   \biggr] -6 \,,
 }
where the $-6$ comes from subtracting the disconnected part.  After relating $N_J$ to $c_T$ using \eqref{GotNJNK}, we obtain
 \es{DerSimpOneMassFinal}{
 \frac{\partial \log Z}{\partial m_+^4} = \frac{\pi^4 c_T^2}{2^{13}}I_{++}[\cS^i] \,.
 }
The quantity $I_{++}[\cS^i]$ is independent of $z$.  It can be simplified significantly using the conformal block expansion introduced in Eq.~\eqref{blockExp}.  Indeed, \eqref{Ipp} can be written as
 \es{IppRewrite}{
  I_{++}[\cS^i] &= 2 \biggl[ \cS_{\bf 1} + \cS_{{\bf 15}_a} \frac{2 (z - 2)}{z} + \cS_{{\bf 15}_s} + 2 \cS_{{\bf 20}'}
  + \cS_{{\bf 45} \oplus \overline{\bf 45}} \frac{4 - 2z}{z} \\
  &{}+ \cS_{\bf 84} \left( \frac{16}{z^2} - \frac{16}{z} + \frac{44}{15} \right) \biggr] \bigg|_{\substack{U = z^2 \\
      V = (1-z)^2 }} - 6  \,.
 }
Each $\cS_R$ must be expanded in conformal blocks, which as $z \to 0$ behave as $(z/4)^\Delta$ where $\Delta$ is the scaling dimension of the corresponding conformal primary.  Since $I_{++}$ is independent of $z$, it follows that the only conformal primaries that can contribute must have either $\Delta = 0$ in the ${\bf 1}$, ${\bf 15}_s$, ${\bf 20}'$ channels, $\Delta = 1$ in the ${\bf 15}_a$ and ${\bf 45} \oplus \overline{\bf 45}$ channels, or $\Delta = 2$ in the ${\bf 84}$ channel.  The only $\Delta = 0$ operator is the identity operator and it appears in the ${\bf 1}$ channel with squared OPE coefficient $\lambda_{0, 0, {\bf 1}}^2 = 1$ by convention.   The ${\bf 15}_a$ and ${\bf 45} \oplus \overline{\bf 45}$ channels contain only odd spin operators, and for them $\Delta = 1$ would violate the unitarity bound.  Thus, there are no $\Delta = 1$ operators contributing to \eqref{IppRewrite}.   Consequently, the only operators that can contribute to \eqref{IppRewrite} are the identity operator and any $\Delta = 2$ operators in the ${\bf 84}$.  Such operators must be scalars because these are the operators that are non-trivial in the 1d theory \cite{Chester:2014mea}.  Using $G_{2, 0}(U, V) \approx U/16$ at small $U$, we have
 \es{IppRewrite2}{
  I_{++}[\cS^i] = -4 + 2 \lambda_{2, 0, {\bf 84}}^2  \,.
 }
As explained in more detail in Appendix~\ref{OPEEXTRACTION}, the OPE coefficient $\lambda_{2, 0, {\bf 84}}^2$ can be written in terms of the Mellin amplitude corresponding to the ${\bf 84}$ channel, which is defined as $M_{\bf 84}(s, t) = (M^2 + M^3 + 2 M^4 )/ 16$.  The final expression for $I_{++}[\cS^i]$ is
 \es{IppRewrite3}{
  I_{++}[\cS^i]  =  32 \pi^2 \lim_{s \to 2} \lim_{t \to 3-s} \frac{(u-1)M_{{{\bf 84}}}(s, t) }{s-2}  + \frac{16 i}{\pi}  \int_{-i \infty}^{i\infty} dt\,  \Gamma^2\left( 1 - \frac{t}{2} \right) \Gamma^2 \left( \frac t2 \right)  \lim_{s \to 2} \frac{M_{\bf 84}(s, t)}{s - 2} 
 }
with the contour in the $t$ integral obeying $0 < \Re t < 1$.  For a derivation, see Appendix~\ref{OPEEXTRACTION}.


For the mixed derivative, let us take the first $\tilde J$ to be at $x_3 = 0$ and the second at $x_4 = \infty$ and multiply by $(2\pi)^2$.  Then, relating all the operators in the second line of \eqref{DerSimp} to ${\cal S}^i$ and ${\cal R}^i$ (where ${\cal R}^i$ are defined in \eqref{SSPP}), and computing the required traces of $M$ matrices, we obtain
  \es{DerSimpTwoMasses}{
  \frac{\partial \log Z}{\partial m_+^2 \partial m_-^2} 
   = 16 \pi^4 N_J^2 \left[N_J^2  \tilde I_1[2 \cS^1]
    -  N_K^2  \tilde I_2[2 \cR^1 +  \cR^2 +  \cR^3 + 2 \cR^5 + 2 \cR^6] \right]
   \,,
 }
where
 \es{ItildeDef}{
  \tilde I_\Delta[\cG] \equiv
   \int  d^3 \vec{x}_1 \, d^3 \vec{x}_2\,  
    \frac{ \left[ \Omega(\vec{x}_1) \Omega(\vec{x}_2) \right]^{3 - \Delta}  }
     {\vec{x}_{12}^{2 \Delta} } {\cal G} \left( \frac{\vec{x}_{12}^2 }{\vec{x}_{1}^2 },  \frac{\vec{x}_2^2}{\vec{x}_{1}^2} \right)  \,, \qquad
      \Omega(x) = \frac{1}{1 + \frac{x^2}{4}} \,.
 }

We can evaluate \eqref{ItildeDef} as follows.  Using rotational symmetry, we can set $\vec{x}_1 = (r_1, 0, 0)$ and $\vec{x}_2 = (r_2 \cos \theta, r_2 \sin \theta, 0)$ and perform the angular integrals which give $4 \pi \times 2 \pi = 8 \pi^2$.  Thus
 \es{ItildeDef2}{
  \tilde I_\Delta[\cG] \equiv
    8 \pi^2 \int  dr_1\, dr_2\, d\theta\, r_1^2 r_2^2 \sin \theta \,  
    \frac{ \left[ \left( 1 + \frac{r_1^2}{4} \right)  \left( 1 + \frac{r_2^2}{4} \right)  \right]^{\Delta  - 3}  }
     { (r_1^2 + r_2^2 - 2 r_1 r_2 \cos \theta)^{ \Delta} } {\cal G} \left( \frac{r_1^2 + r_2^2 - 2 r_1 r_2 \cos \theta }{r_{1}^2 },  \frac{r_2^2}{r_{1}^2} \right)  \,.
 }
Let us now change variables by setting $r_1 = 2 \rho$ and $r_2 = 2 r \rho$.  Then \eqref{ItildeDef2} becomes
 \es{ItildeDef3}{
    \tilde I_\Delta[\cG] \equiv
      2^{9 - 2 \Delta} \pi^2 \int  d\rho \, dr \, d\theta\,  \rho^{5 - 2 \Delta} r^2 \sin \theta \,  
    \left[ \left( 1 + \rho^2 \right)  \left( 1 +r^2 \rho^2 \right)  \right]^{\Delta  - 3} 
    \frac{{\cal G} \left( 1 + r^2 - 2 r \cos \theta ,  r^2 \right) }{(1 + r^2 - 2 r \cos \theta)^{ \Delta} } \,.
 }
The $\rho$ integral can be done analytically.  For the cases of interest, namely $\Delta = 1$ and $2$, the result is 
 \es{ItildeResult}{
  \tilde I_1[\cG] &= 2^{7} \pi^2 \int  dr \, d\theta\,  r^2 \sin \theta \,  
   \frac{1 - r^2 + (1 + r^2) \log r}{(r^2 - 1)^3}
    \frac{{\cal G} \left( 1 + r^2 - 2 r \cos \theta ,  r^2 \right) }{1 + r^2 - 2 r \cos \theta }  \,, \\
   \tilde I_2[\cG] &= 2^{5} \pi^2 \int  dr \, d\theta\,  r^2 \sin \theta \,  
   \frac{\log r}{r^2 - 1}
    \frac{{\cal G} \left( 1 + r^2 - 2 r \cos \theta ,  r^2 \right) }{(1 + r^2 - 2 r \cos \theta)^2 }   \,.
 }

The expression \eqref{DerSimpTwoMasses} can be simplified further after using the Ward identity relating ${\cal R}^i$ to ${\cal S}^i$ in Eqs.~\eqref{RFirst}--\eqref{RLast}, and integrating by parts.  We find
 \es{tildeI2R}{
  \tilde I_2[2\cR^1 + \cR^2 + \cR^3 + 2 \cR^5 + 2 \cR^6] &=  \int  dr \, d\theta\, 
    {\cal S}^1 \left( 1 + r^2 - 2 r \cos \theta ,  r^2 \right)  \\
    &{}\times \left( -16 \pi^2 \sin \theta \frac{-1 - 5 r^2 + 5r^4 + r^6 - 8 (r^2 + r^4) \log r}{(r^2 - 1)^3 (1 + r^2 - 2 r \cos \theta)} \right) \,.
 }
Combining with \eqref{DerSimpTwoMasses}, we obtain
\es{DerSimpTwoMassesAgain}{
\frac{\partial \log Z}{\partial m_+^2 \partial m_-^2} 
= 2^9 \pi^6 N_J^4  \int dr\, d\theta\, \sin \theta \, \frac{{\cal S}^1 \left( 1 + r^2 - 2 r \cos \theta ,  r^2 \right) }{1 + r^2 - 2 r \cos \theta} \,.
 }
Once again we can view the right-hand side as a linear functional defined on $\cS$, defining
\begin{equation} \label{Ipm}
I_{+-}[\cS^i] = \int dr\, d\theta\, \sin \theta \, \frac{{\cal S}^1 \left( 1 + r^2 - 2 r \cos \theta ,  r^2 \right) }{1 + r^2 - 2 r \cos \theta}\end{equation}
so that
\begin{equation}\label{DerSimpTwoMassesFinal}
\frac{\partial \log Z}{\partial m_+^2 \partial m_-^2} = \frac{\pi^2 c_T^2}{2^{11}}I_{+-}[\cS^i].
\end{equation}
See Appendix~\ref{INTEGRALS} for an expression for $I_{+-}[\cS^i]$ in terms of the Mellin amplitude corresponding to ${\cal S}^i$.


\subsection{Large $c_T$ expansion}
\label{LARGECT}


We will now show how integrated correlators can be expanded to all orders in $1/c_T$. Using the Fermi gas method \cite{Marino:2011eh}, the localization formula \eqref{ZS3} for the mass deformed partition function was computed to all orders in $1/N$ \cite{Nosaka:2015iiw}:
 \es{GotZABJM}{
 & Z \approx e^A C^{-\frac 13} \text{Ai}\left[C^{-\frac 13} (N-B) \right] \,,\\
  C &= \frac{2}{\pi^2 k (1 + m_+^2) (1 + m_-^2)} \,, \qquad
   B = \frac{\pi^2 C}{3} - \frac{1}{6k} \left[ \frac{1}{1 + m_+^2} + \frac{1}{1 + m_-^2} \right] + \frac{k}{24} \,, \\
  A&= \frac{{\cal A}[k(1 + i m_+)] + {\cal A}[k(1 - i m_+)] +  {\cal A}[k(1 + i m_-)] + {\cal A}[k(1 - i m_-)] }{4}  \,,
 } 
 where the constant map function ${\cal A}$ is given by
\es{constantMap}{
{\cal A}(k)&=\frac{2\zeta(3)}{\pi^2k}\left(1-\frac{k^3}{16}\right)+\frac{k^2}{\pi^2}\int_0^\infty dx\frac{x}{e^{kx}-1}\log\left(1-e^{-2x}\right)\\
&=-\frac{\zeta(3)}{8\pi^2}k^2+2\zeta'(-1)+\frac{\log\left[\frac{4\pi}{k}\right]}{6}+\sum_{g=0}^\infty\left(\frac{2\pi i}{k}\right)^{2g-2}\frac{4^gB_{2g}B_{2g-2}}{(4g)(2g-2)(2g-2)!}\,,
}
 and in the second line we wrote $\cA$ in the large $k$ expansion \cite{Hanada:2012si}. We will be interested in derivatives of $Z(m_\pm)$ at $m_\pm=0$, in which case we expect the non-perturbative corrections to take the form $e^{-\sqrt{N k}}$ and $e^{-\sqrt{N/k}}$, which is known for $Z(0)$ that has been computed exactly for all $N$ and $k$ in \cite{Marino:2016new,Hatsuda:2013gj,Hatsuda:2012dt,Calvo:2012du,Marino:2011eh,Drukker:2010nc,Marino:2009jd,Hatsuda:2013oxa}. The large $N$ expansion is then expected to apply to the finite $k$, the strong coupling 't Hooft limit with 't Hooft coupling
  \es{thooft}{
 \lambda=\frac{N}{k}-\frac{1}{3k^2}-\frac{1}{24}\,,
 }
 and the finite $\mu\equiv N/k^5$ limit discussed in the Introduction, which interpolates between finite $k$ as $\mu\to\infty$ and the strong coupling 't Hooft limit as $\mu\to0$. In particular, the non-perturbative corrections $e^{-\sqrt{N k}}$ and $e^{-\sqrt{N/k}}$ do not allow for any non-perturbative corrections in $\mu$.
 
 For each of these limits, we can use  \eqref{GotZABJM} and \eqref{Gottau} to expand $c_T$, $\frac{\partial^4 \log Z}{\partial m_\pm^4}$, and $\frac{\partial^4 \log Z}{\partial m_+^2 \partial m_-^2}$ to all orders in $1/N$, and then rewrite the latter two quantities as expansions to all orders in $1/c_T$. For the finite $k$ limit, we find
\es{finitekupppp}{
\text{finite $k$:}\quad\frac{1}{c_T^2}\frac{\partial^4 \log Z}{\partial m_\pm^4} &=\frac{3 \pi^2}{64}\frac{1}{c_T}+\frac{3^{\frac43}\pi ^{4/3}}{2^{\frac83}
   k^{2/3}}\frac{1}{c_T^{\frac53}}+\frac{k^4 \cA^{(4)}(k)-3 k^2 \cA''(k)-3}{2 {c_T}^2}+O(c_T^{-\frac73})\,,\\
\frac{1}{c_T^2}\frac{\partial^4 \log Z}{\partial m_+^2 \partial m_-^2} &=-\frac{\pi^2}{64}\frac{1}{c_T}+\frac{5 \pi ^{4/3}}{4\ 6^{2/3}  k^{2/3}}\frac{1}{c_T^{\frac53}}+\frac{k^2 \cA''(k)-1}{2 {c_T}^2}+O(c_T^{-\frac73})\,,
}
where we have only shown the lowest couple terms in $1/c_T$ for simplicity. We can evaluate $\cA^{(4)}(k)$ and $\cA''(k)$ using the definition in the first line of \eqref{constantMap}, which holds for finite $k$, in which case the $\zeta(3)$ term is cancelled by the integral term.\footnote{For instance, for $k=1,2$ these values are \cite{Agmon:2017xes}
\es{Aprime}{
{\cal A}''(1)&=\frac16+\frac{\pi^2}{32}\,,\qquad\qquad \,\, \,\,{\cal A}''(2)=\frac{1}{24}\,,\\
{\cal A}''''(1)&=1+\frac{4\pi^2}{5}-\frac{\pi^4}{32}\,,\qquad {\cal A}''''(2)=\frac{1}{16}+\frac{\pi^2}{80}\,.
} 
} 

For the strong coupling 't Hooft limit, we find the all orders in $1/\lambda$ and $1/c_T$ result
 \es{largeLam}{
 \text{'t Hooft:} \quad& \frac{1}{c_T^2}\frac{\partial^4\log Z}{\partial m_\pm^4}=\left[\frac{3\pi^2}{64}+\frac{9\zeta(3)}{512\sqrt{2}\pi}\frac{1}{\lambda^{\frac32}}+\frac{27 \zeta(3)^2}{8192 \pi^4 }\frac{1}{\lambda^3}+O(\lambda^{-\frac92})\right]\frac{1}{c_T}\\
   &\qquad+\left[\frac32\pi\sqrt{2\lambda}-\frac54-\frac{9 \zeta (3)}{16 \pi ^2
    }\frac{1}{\lambda}+\frac{15 \zeta
   (3)}{32 \sqrt{2} \pi ^3} \frac{1}{\lambda^{\frac32} }+O(\lambda^{-\frac52})\right]\frac{1}{c_T^2}+O(c_T^{-3})\,,\\
&  \frac{1}{c_T^2}\frac{\partial^4\log Z}{\partial m_+^2\partial m_-^2}=\left[-\frac{\pi^2}{64}-\frac{3\zeta(3)}{512\sqrt{2}\pi}\frac{1}{\lambda^{\frac32}}-\frac{9 \zeta (3)^2}{8192 \pi ^4 }\frac{1}{\lambda^3}+O(\lambda^{-\frac92})\right]\frac{1}{c_T}\\
  &\qquad+\left[\frac56\pi\sqrt{2\lambda}-\frac{5}{12}+\frac{3 \zeta (3)}{16 \pi ^2
   \lambda }-\frac{5 \zeta
   (3)}{32 \sqrt{2} \pi ^3} \frac{1}{\lambda^{\frac32} }+O(\lambda^{-\frac52})\right]\frac{1}{c_T^2}+O(c_T^{-3})\,,\\
}
where we used the large $k$ formula for $\cA(k)$ in the second line of \eqref{constantMap}, so $\zeta(3)$ terms appear. In fact, $\zeta(3)$ and $\pi$ are the only transcendental numbers that appear to any order in $1/\lambda$ and $1/c_T$.

  Finally, for the finite $\mu$ limit we find
\es{finiteMupppp}{
\text{finite $\mu$:}\qquad\frac{1}{c_T^2}\frac{\partial^4 \log Z}{\partial m_\pm^4} &=\frac{3 \pi^2}{64}\frac{1}{c_T}+\frac{3^{\frac54} \left(4 \sqrt{2} \pi ^3 \sqrt{\mu
   }+\zeta (3)\right)}{16\ 2^{5/8} \pi ^{7/4} \mu ^{3/8}}\frac{1}{c_T^{\frac74}}-\frac54\frac{1}{c_T^2}+O(c_T^{-\frac94})\,,\\
\frac{1}{c_T^2}\frac{\partial^4 \log Z}{\partial m_+^2 \partial m_-^2} &=-\frac{\pi^2}{64}\frac{1}{c_T}+\frac{20 \sqrt{2} \pi ^3 \sqrt{\mu }-3 \zeta (3)}{16\
   2^{5/8} 3^{3/4} \pi ^{7/4} \mu ^{3/8}}\frac{1}{c_T^{\frac74}}-\frac{5}{12}\frac{1}{c_T^2}+O(c_T^{-\frac94})\,,
}
where we again used the large $k$ formula for $\cA(k)$. 

From the finite $\mu$ limit we can derive both the 't Hooft limit and the finite $k$ limit by taking $\mu\rightarrow0$ and $\mu\rightarrow\infty$ respectively. To reproduce the 't Hooft limit \eqref{largeLam} we first solve for $\mu$ in terms of $\lambda$ and $c_T$ using \eqref{Gottau} and \eqref{thooft}, which at leading order in $1/c_T$ gives
\begin{equation}\label{muTotHooft}
\mu = \frac{8192\lambda^4}{9c_T^2\pi^2} + \dots\,.
\end{equation}
We then take the large $c_T$ limit followed by the large $\lambda$ limit. The $\zeta(3)\mu^{-\frac38}c_T^{-\frac74}$ and $\mu^{\frac18}c_T^{-\frac74}$ terms give rise to the $\zeta(3)\lambda^{-\frac32}c_T^{-1}$ and $\sqrt{\lambda}c_T^{-2}$ terms in \eqref{largeLam}, respectively.

To extract the finite $k$ limit \eqref{finitekupppp} from \eqref{finiteMupppp} we solve for $\mu$ in terms of $c_T$ and $k$ using \eqref{Gottau}, which at leading order in $1/c_T$ gives
\begin{equation}\label{muTok}
\mu =  \frac{(3 \pi )^{2/3} {c_T}^{2/3}}{2^{\frac{13}{3}}k^{\frac{16}{3}}}+\dots\,.
\end{equation}
We then take the large $c_T$ limit. In this limit, the ratio $c_T^2 \mu^{-3}$ is finite, so we must sum infinitely many terms in the finite $\mu$ limit to recover the finite $k$ limit. This infinite sum cancels all the $\zeta(3)$ terms which appear at finite $\mu$. The $\mu^{\frac18}c_T^{-\frac74}$ term becomes a $c_T^{-\frac53}$ term at finite $k$.

\section{ $\cN = 6$ ABJM correlators at large $c_T$}
\label{CORRELATORS}

We will now combine the results of the previous to sections and determine the first few terms in the large $N$ expansion of the $\langle SSSS\rangle$ correlator in ABJM theory.  We will do this for the finite $k$, finite $\mu$, and strong coupling 't Hooft limits, which correspond to M-theory on $AdS_4 \times S^7 / \Z_k$ for the first limit, or to type IIA string theory on $AdS_4 \times \CP^3$ in the second and third limits.  

In each of these limits, we can use the Penedones formula \eqref{FlatLimit} to relate the $\langle SSSS \rangle$ Mellin amplitude to the four-point scattering amplitudes of gravitons and their superpartners in 11d (in the M-theory case) or 10d (in the type IIA case) flat space, with momenta restricted to lie within a four-dimensional subspace.  Of course, the flat space limit of the $\langle SSSS \rangle$ correlator in ABJM theory cannot give the four-point scattering amplitude of {\em all} massless particles in 11d or 10d.  Indeed, in either 11d M-theory or in 10d type IIA string theory, the massless particle spectrum consists of $128$ bosons and $128$ fermions that are related by maximal SUSY\@.  The flat space limit of the $\langle SSSS \rangle$ correlator must match the four-point scattering amplitude of only $15$ of the $128$ bosons, which all have the property that after restricting their momenta to lie within 4d, they can be thought of as scalars from the 4d point of view.\footnote{More generally, from all the 4-point CFT correlators of the ${\cal N} = 6$ stress tensor multiplet, we would be able to determine the 4-point scattering amplitudes of precisely half ($64$ bosons $+$ $64$ fermions) of the massless particles of both 11d M-theory and 10d type IIA string theory.}   Note that when using Eq.~\eqref{FlatLimit}, we should keep either the 11d Planck length $\ell_{11}$ or the 10d string length $\ell_s$ fixed as we send $L \to \infty$. In other words, we should more precisely send $L / \ell_{11}$ or $L / \ell_s$ to infinity.

As explained in Section~\ref{4POINT}, the ingredients we will use to construct the first few terms in the large $N$ expansion of the $\langle SSSS \rangle$ correlator are the Mellin amplitudes
 \es{MellinIngredients}{
  M_\text{SG}^i\,, \quad M_3^i \,, \quad M_4^i 
 } 
given in \eqref{SugMellin}, \eqref{degree2}, and \eqref{R4Mellin}, respectively.  $M_\text{SG}^i$ is the Mellin amplitude corresponding to an exchange Witten diagram with supergravity vertices.  $M_3^i$ is a polynomial Mellin amplitude that represents the $\langle SSSS \rangle$ component of a degree $3$ super-Mellin amplitude corresponding to a contact Witten diagram with an $F^2 R^2$ contact interaction vertex.  Likewise, $M_4^i$ is part of a degree $4$ super-Mellin amplitude corresponding to a contact Witten diagram with an $R^4$ super-vertex.  As explained in Section~\ref{4POINT}, if we apply the Penedones formula \eqref{FlatLimit} to each of the Mellin amplitudes \eqref{MellinIngredients}, we find that
\begin{equation}\begin{aligned}\label{flat34}
\frac 1 {L^2\cN(L)} M_\text{SG}^i(s,t)  &\underset{\text{flat space}}\longrightarrow &\cA^i_\text{SG}(s,t) &= \begin{pmatrix} 
\frac{tu}{s} &  \frac{su}{t}  &\frac{st}{u} & \frac{s}{2} & \frac{t}{2} & \frac{u}{2}
\end{pmatrix} \,,\\
\frac 1 {L^{6}\cN(L)} M_3^i(s,t) &\underset{\text{flat space}}\longrightarrow &\cA^i_3(s,t) &= 0 \,,\\
\frac 1 {L^{8}\cN(L)} M_4^i(s,t) &\underset{\text{flat space}}\longrightarrow &\cA^i_4(s,t) &= \frac{stu}{105} \cA^i_\text{SG}(s,t)  \,.
\end{aligned}\end{equation}
Here, the normalization constant $\cN(L)$ appearing in \eqref{FlatLimit} depends on our precise choice of normalization for the $\langle SSSS \rangle$ correlator. If we normalize this correlator such that the disconnected piece scales as $c_T^0$, then we should take $\cN (L) = \cN_0 L^D$, where $D = 7$ for the case of an 11d dual and $D = 6$ for the case of a 10d dual.  

In addition to \eqref{MellinIngredients}, we will also consider the contact Mellin amplitudes
 \es{MellinIngredients2}{
  M_{5, 1}^i \,, \qquad M_{5, 2}^i\,,
 }
which are part of degree-5 super-Mellin amplitudes corresponding to $D^2 R^4$ and $D^4 F^2 R^2$ interaction vertices, respectively.  While in Section~\ref{4POINT} we did not determine the forms of $M_{5, 1}^i$ and $M_{5, 2}^i$, we know that such Mellin amplitudes must exist because they must reproduce the scattering amplitudes in the 3rd line of Table~\ref{lowOrderTab} in the flat space limit.  Upon a convenient choice of normalization, the flat space limits of the Mellin amplitudes can be taken to be
\begin{equation}\begin{aligned}\label{flat5}
\frac 1 {L^{10}\cN(L)} M_{5,1}^i(s,t) &\underset{\text{flat space}}\longrightarrow &\cA^i_{5,1}(s,t) &= \frac{1}{945}stu \begin{pmatrix} s^2+3t^2+3u^2 & \cdots \end{pmatrix}  \,,\\
\frac 1 {L^{10}\cN(L)} M_{5,2}^i(s,t) &\underset{\text{flat space}}\longrightarrow &\cA^i_{5,2}(s,t) &= 0 \,.
\end{aligned}\end{equation}

It is important to note that the Mellin amplitudes $M_\text{SG}^i$, $M_3^i$, $M_4^i$, $M_{5, 1}^i$, and $M_{5, 2}^i$ are the only crossing-invariant Mellin amplitudes that obey the SUSY Ward identities and that grow at most as the fifth power of $s, t$ at large $s, t$.

\subsection{Strong coupling expansions}

Let us now analyze the (Mellin transform of) the $\langle SSSS \rangle$ correlator in each of the three large $N$ limits we consider.

\subsubsection{Large $c_T$, finite $k$}

At large $c_T$ limit with $k$ fixed, ABJM theory is dual to M-theory on $AdS_4\times S^7/\mathbb{Z}_k$. At leading order in $1/c_T$, we have the AdS/CFT relation \cite{Aharony:2008ug,Chester:2018aca}
\es{cTAdS}{
\frac{L^9}{\ell_{11}^9}=  \frac{3\pi k}{2^{11}}c_T+\dots \,,
}
with corrections suppressed in $1/c_T$.  From this relation, the flat space limits \eqref{flat34} and \eqref{flat5}, as well as the requirement that in the flat space limit the scattering amplitude should have an expansion in $\ell_{11}$ times momentum, we can infer that $M^i(s,t)$ can be expanded at large $c_T$ in terms of $M_n^i(s,t)$ as 
\es{finitekM}{
M^i(s,t)=&\frac{1}{c_T}A_\text{SG}^1 M^i_\text{SG}+\frac{1}{c_T^{\frac{13}{9}}}\left[ A_\text{SG}^3M^i_\text{SG}+A_3^3M^i_3 \right]+\frac{1}{c_T^{\frac53}}\left[ A_\text{SG}^4M^i_\text{SG}+A_3^4M^i_3+A_4^4M^i_4 \right]\\
&+\frac{1}{c_T^{\frac{17}{9}}}\left[ A_\text{SG}^5M^i_\text{SG}+A_3^5M^i_3+A_4^5M^i_4+A_{5,1}^5M^i_{5,1}+A_{5,2}^5M^i_{5,2} \right]+O(c_T^{-2})\,,
}
where $A_{i,j}^l$ are $k$-dependent numerical coefficients. In the flat space limit only the maximal degree Mellin amplitudes contribute at each order in $1/c_T$, and so from \eqref{flat34} and \eqref{flat5} we find that
\begin{equation}\begin{split}\label{msflat}
\cA^i(s,t) &= \ell_{11}^9 \left(A_\text{SG}^1\cA^i_\text{SG}+\left(\frac{3k\pi}{2^{11}}\right)^{2/3}\ell_{11}^6A_4^4\cA^i_4+\left(\frac{3k\pi}{2^{11}}\right)^{8/9}\ell_{11}^{8}A_{5,1}^5\cA^i_{5,1}+ \cdots \right) \,.
\end{split}\end{equation}
Note that neither $\cA_3^i$ nor $\cA_{5,2}^i$ give rise to scalar scattering amplitudes in flat space, which is why they do not appear in \eqref{msflat}. Comparing \eqref{msflat} to the known M-theory four-point scattering amplitude \cite{Green:1998by} 
\begin{equation}\label{Mamps}
\cA^{11} = \cA_\text{SG}^{11}\left[1+\ell_{11}^6\frac 1{3\cdot 2^7}stu+O(\ell_{11}^9)\right] \,, 
\end{equation}
where $\cA_\text{SG}^{11}$ is the 11d supergravity scattering amplitude, we can immediately deduce that
\begin{equation}\label{mRes1}
\frac{A_4^4}{A_{\text{SG}}^1} = 35\left(\frac{2}{9\pi^2k^2}\right)^{1/3}\,,\quad A_{5,1}^5 = 0 \,.
\end{equation}

Although $M_3^i$ and $M_{5,2}^i$ do not give rise to scattering amplitudes for the 11d super-gravitons that are scalars from the 4d point of view, they do contribute to the scattering of other particles in the same multiplet. The M-theory amplitude \eqref{Mamps} however encodes the scattering amplitudes for all such particles, and it does not contain any terms of order $\ell_{11}^{13}$ or $\ell_{11}^{17}$. From this we conclude that
\begin{equation}\label{mRes2}
A_3^3 = A_{5,2}^5 = 0 \,.
\end{equation}

As a final aside, note that the $O(c_T^{-2})$ term \eqref{finitekM} is not a local Mellin amplitude. It instead corresponds to the one-loop supergravity term, which is not analytic in $s$ and $t$.   We will not study this term further.

\subsubsection{'t Hooft strong coupling limit} 

We next consider the strong coupling 't Hooft limit of ABJM theory, whereby we first take $N\to\infty$ with fixed $\lambda$ (see \eqref{thooft} for the definition of $\lambda$), and then take $\lambda\to\infty$.  In this double limit, ABJM theory is dual to weakly coupled type IIA string theory on $AdS_4\times \mathbb{CP}^3$ \cite{Aharony:2008ug}. The leading order AdS/CFT relations are \cite{Aharony:2008ug,Chester:2018aca}
\es{class}{
{L^8\over \ell_s^8} = 4\pi^4{\lambda}^2 +\dots\,,\qquad g_s^2 =\frac{512\lambda^2}{3c_T}+\dots\,,
}
where both $\ell_{s}/L$ and the string coupling $g_s$ are small in this double expansion.  The ellipses in \eqref{class} stand for terms that are suppressed at large $c_T$ in both expressions.   Similarly to the M-theory limit discussed above, we can expand $M^i(s,t)$ in powers of $\ell_s/L$, with the appropriate powers of $\ell_s/L$ being such that after taking the flat space limit, the string theory scattering amptliude has an expansion in $\ell_s$ times momentum.  Unlike M-theory however, type IIA string theory has an additional dimensionless parameter, the string coupling constant $g_s$, that governs the strength of string interactions. Simultaneously expanding in both, we find that
\es{thooftM}{
M(s,t)=&\,\frac{1}{c_T}\left[B_\text{SG}^1 M_\text{SG}+\frac{1}{\lambda}\left( B_\text{SG}^3M_\text{SG}+B_3^3  M_3\right)+\frac{1}{\lambda^{\frac32}}\left( B_\text{SG}^4M_\text{SG}+B_3^4  M_3+B_4^4 M_4\right)\right.\\
&\left.+\frac{1}{\lambda^{2}}\left( B_\text{SG}^5M_\text{SG}+B_3^5  M_3+B_4^5 M_4+B_{5,1}^5 M_{5,1}+B_{5,2}^5 M_{5,2}\right)+O(\lambda^{-\frac52})\right]\\
&+\frac{1}{c_T^{2}}\left[{\lambda}^2 \widetilde B_\text{SG}^1 M_\text{SG}+{\lambda}\left( \widetilde B_\text{SG}^3M_\text{SG}+\widetilde B_3^3  M_3 \right)\right.\\
&\left.+ \sqrt{\lambda}\left( \widetilde B_\text{SG}^4M_\text{SG}+\widetilde B_3^4  M_3+\widetilde B_4^4 M_4 \right)+O(\lambda^0) \right]+O(c_T^{-3})\,,
}
where $B_{i,j}^l$ and $\widetilde B_{i,j}^l$ are numerical coefficients. The leading order $1/c_T$ behavior corresponds to tree-level string theory, and the higher order terms are loop corrections. At fixed order in $1/c_T$ and $1/\lambda$ only the maximal degree Mellin amplitudes contribute in the flat space limit, and so we find that
\begin{equation}\begin{split}\label{IIAWflat}
\cA^i(s,t) &= \frac{3\pi^4}{128}g_s^2\ell_{s}^8 \left(B_\text{SG}^1\cA^i_\text{SG}+2\sqrt{2}\pi^3\ell_s^6 B_4^4\cA^i_4+4\pi^4\ell_s^8B_{5,1}^5\cA^i_{5,1}+ \cdots\right) \\
&+\frac{9\pi^4}{2^{16}}g_s^4\ell_s^8\left(\tilde B_\text{SG}^1\cA^i_\text{SG}+2\sqrt{2}\pi^3\ell_s^6\tilde B_4^4\cA^i_4+ \cdots \right) \,.
\end{split}\end{equation}
Although the $1/c_T^2$ terms are one-loop corrections, non-analytic Mellin amplitudes will occur first at $\lambda^0/c_T^2$ corresponding to the one-loop correction in supergravity. Comparing this to the IIA $S$-matrix at weak coupling \cite{GREEN1982444}
\begin{equation}\label{IIAAMP}
\cA_{\text{IIA}}^{10} = \cA^{10}_\text{SG}\left[\left( 1 + \ell_{s}^6  \frac{\zeta(3)}{32}stu +O(\ell_s^{10}) \right) +g_s^2 \left(\ell_s^6  \frac{\pi^2}{96}stu +O(\ell_s^8)\right)+O(g_s^4)\right]\,,
\end{equation}
we find that
\begin{equation}\label{IIARes1}
\frac{B^4_4}{B^1_\text{SG}} = \frac{105\zeta(3)}{64\sqrt{2}\pi^3}\,,\qquad \frac{\widetilde B^4_4}{B^1_\text{SG}} = \frac{140\sqrt 2}{3\pi}\,,\qquad B^5_{5,1} = \widetilde B^1_\text{SG} = 0\,.
\end{equation}
Like the M-theory amplitude, the type IIA super-amplitude does not contain any terms which could correspond to $M_3^i$ or $M_{5,2}^i$, which in 10d contribute at $\ell_s^{12}$ and $\ell_s^{16}$. We hence conclude that these terms do not contribute at leading order:
\begin{equation}\label{IIARes2}
B^3_{3} = \widetilde B^3_3 = B_{5,2}^5 = 0 \,.
\end{equation}


\subsubsection{Large $c_T$, finite $\mu$} 

Finally, we consider the large $c_T$ expansion of ABJM at finite $\mu \equiv N/k^5$. Like the 't Hooft strong coupling limit, ABJM theory in this limit is dual to type IIA string theory on $AdS_4\times \mathbb{CP}^3$, except now the string coupling $g_s$ is finite. The AdS/CFT relations are \cite{Aharony:2008ug,Chester:2018aca} 
\es{cTAdSmu}{
{L^8\over \ell_s^8} = \frac{3 c_T \pi^5 \sqrt{\mu}}{16 \sqrt{2}}+\cdots \,, \qquad
 g_s^4 = 32 \pi^2 \mu + \cdots \,,
 }
with corrections suppressed at large $c_T$.    The relation \eqref{cTAdSmu} implies that $M^i(s,t)$ can be expanded at large $c_T$ in terms of $M_n^i(s,t)$ as
\es{finitemuM}{
M^i(s,t)=&\frac{1}{c_T}C_\text{SG}^1M^i_\text{SG}+\frac{1}{c_T^{\frac{3}{2}}}\left[ C_\text{SG}^3M^i_\text{SG}+C_3^3M^i_3 \right]+\frac{1}{c_T^{\frac74}}\left[ C_\text{SG}^4M^i_\text{SG}+C_3^4M^i_3+C_4^4M^i_4 \right]+O(c_T^{-2})\,,
}
where now $C_{i,j}^l$ are $\mu$-dependent numerical coefficients.  (This expansion is nothing but a reorganized version of the double expansion \eqref{thooftM}.)  Unlike in the previous limits, we do not include the two amplitudes $M_{5,1}^i$ and $M_{5,2}^i$ because in this case they contribute at the same order in $1/c_T$ as the one-loop supergravity Mellin amplitude.   Taking the flat space limit of \eqref{finitemuM} we find that
\begin{equation}\label{muToAmp}
\cA^i(s,t) = \frac{3\pi^4}{128 } g_s^2 \ell_s^8\left(C^1_\text{SG}\cA^i_\text{SG} + \ell_s^6\left(\frac{9\pi^{8}g_s^4}{2^{14}}\right)^{3/8} C^4_4\cA^i_4 + O(\ell_s^{8})\right)
\end{equation}
This expression can be compared with the type IIA scattering amplitude at fixed $g_s$, which is given by \cite{Green:2008uj}
\begin{equation}\label{IIAAMPAgain}
\cA_{\text{IIA}}^{10} = \cA^{10}_\text{SG}\left[1 + \ell_{s}^6  stu \left( \frac{\zeta(3)}{32}+g_s^2 \frac{\pi^2}{96} \right) +O(\ell_s^8) \right]\,.
\end{equation}
Note that the $\ell_s^6$ term only receives contributions from tree-level and one-loop, and it does not have any other perturbative or non-perturbative corrections.

From comparing \eqref{IIAAMPAgain} and \eqref{muToAmp}, we conclude that
 \begin{equation}\label{IIAResAgain}
\frac{C^4_4}{C^1_\text{SG}} =  \frac{35}{2 \pi^4} \left( \frac{9 \pi^2}{32 \mu^3} \right)^{1/8} \left( \zeta(3) + \frac{4}{3} \sqrt{2 \mu} \pi^3 \right) \,.
\end{equation}

We can recover both the finite $k$ and strong coupling 't Hooft limit expansions from \eqref{finitemuM} by taking the $\mu\to\infty$ and $\mu\to0$ limits respectively, as we explain at the end of Section~\ref{LARGECT}. Using the relations \eqref{muTotHooft} and \eqref{muTok}, we find that the $c_T^{-\frac74}$ term becomes the $c_T^{-\frac53}$ term at finite $k$, and gives rise to both the $c_T^{-1}\lambda^{-\frac32}$ and $c_T^{-2}\lambda^{\frac12}$ terms in the strong coupling 't Hooft limit.

\subsection{Fixing the SUGRA coefficients}

Our goal is now to fix the coefficients $A_{i,j}^l$, $B_{i,j}^l$, $\widetilde B_{i,j}^l$, and $C_{i,j}^l$ in each expansion considered above, purely using CFT data. We will begin with the supergravity coefficients $A_\text{SG}^l$, $B_\text{SG}^l,$ $\widetilde B_\text{SG}^l$, and $C_\text{SG}^l$, which we fix by determining how the various Mellin amplitudes contribute to the squared OPE coefficient $\lambda_{1, 0, {\bf 15}_s}^2$ with which the $S$ operator appears in the $S \times S$ OPE\@.  As we will explain, this OPE coefficient is proportional to $1/c_T$, and this fact will allow us to determine all $A_\text{SG}^l$, $B_\text{SG}^l,$ $\widetilde B_\text{SG}^l$, and $C_\text{SG}^l$ exactly.

Our starting point is the expression \eqref{Gotlam15} for $\lambda_{1, 0, {\bf 15}_s}^2$ in terms of the Mellin amplitude $M_{{\bf 15}_s} = \frac 16 \left( M^2 + M^3 - M^4 \right) + \frac 12 (M^5 + M^6)$ corresponding to the ${\bf 15}_s$ channel in the $S\times S$ OPE\@.  For the reader's convenience, we reproduce it here 
 \es{Gotlam15Main}{
  \lambda_{1, 0, {\bf 15}_s}^2 = -\frac{1}{2i }  \int_{ -i \infty}^{ i \infty} dt\,  
   \Gamma^2 \left( 1 - \frac{t}{2} \right) \Gamma^2 \left( \frac{t-1}{2} \right)
    \lim_{s\to 1} \biggl[ (s-1) M_{{{\bf 15}_s}}(s, t) \biggr] \,,
 }
and refer the reader to Appendix~\ref{OPEEXTRACTION} for a derivation.  As can be seen from \eqref{Gotlam15Main}, it is only the pole as $s \to 1$ in $M_{{{\bf 15}_s}}$ that contributes to $ \lambda_{1, 0, {\bf 15}_s}^2$.  
Therefore local Mellin amplitudes cannot contribute to $\lambda_{1,0,{\bf 15}_s}^2$, so the only contribution will come from the supergravity exchange Mellin amplitude.  Indeed, the supergravity exchange amplitude $M_\text{SG}^i(s,t)$ does have a pole at $s = 1$ with a residue independent of $t$:
 \es{M15Limit}{
   \lim_{s\to 1} \biggl[ (s-1) M_{\text{SG}, {{\bf 15}_s}}(s, t) \biggr] = - \frac{1}{\pi} \,.
 }
and thus $M_\text{SG}$ in each of the expansions presented above contributes to $\lambda_{1, 0, {\bf 15}_s}^2$ an amount equal to
 \es{Gotlam15Again}{
  \frac{1}{2\pi i }  \int_{-i \infty}^{ i \infty} dt\,  
   \Gamma^2 \left( 1 - \frac{t}{2} \right) \Gamma^2 \left( \frac{t-1}{2} \right)  =  2 \pi^2 \,.
 }

Note that although we have not yet discussed Mellin amplitudes for loop corrections, by suitably adding to them an appropriate multiple of $M_\text{SG}$ we can always define them such that they do not contribute to the $\sqrt{U}$ term, so that $\lambda_{1,0,{\bf 15}_s}^2$ is purely fixed by the coefficient of $M_\text{SG}$. Furthermore, because the three-point function of three stress tensor multiplets is uniquely determined up to an overall coefficient \cite{Liendo:2015cgi}, $\lambda_{1,0,{\bf 15}_s}^2$ must be proportional to the stress-tensor three-point function, which itself is proportional to $1/c_T$ according to the conformal Ward identity \cite{Osborn:1993cr}. We hence determine that
\es{B1}{
A_\text{SG}^1&=B_\text{SG}^1=C_\text{SG}^{1} =  \frac{c_T}{2\pi^2}   \lambda_{1,0,{\bf 15}_s}^2 \,, \qquad 
   \widetilde B_\text{SG}^1  = 0 \,, \\ 
A_\text{SG}^l &= B_\text{SG}^l = \widetilde B_\text{SG}^l = C_\text{SG}^l = 0 \,, \qquad \text{for $l>1$} \,.
}

Our final step is to determine the relationship between $\lambda^2_{1,0,{\bf15}_s}$ and $c_T^{-1}$. We can do so by considering the free $\mathcal{N}=6$ theory of four complex scalars and four 2-component complex fermions, where the scalars $\phi^a$ ($\bar\phi_a$) transform in the ${\bf4}$ (${\bar{\bf4}}$) of $SU(4)_R$.  (This is the same as the $U(1)_k \times U(1)_{-k}$ ABJM theory in the limit $k \to \infty$ considered in the previous section.)
We write $S$ in this case as
\es{Sfree}{
S_b{}^a = \phi^a\bar\phi_b-\frac{\delta_b^a}{4}\phi^c\bar\phi_c\,,
}
and then define $S(\vec x,X)$ as in \eqref{SNorm}. We then perform Wick contractions with the propagator $\langle \phi^a(\vec{x}) \bar\phi_b(0) \rangle = \frac{\delta_{b}^a}{ \abs{\vec{x}}}$ to find the 4-point function \eqref{SSSScor} with the crossing independent coefficients
\es{free4}{
 \cS^i_\text{free}(U,V)= \begin{pmatrix} 1 & U & \frac{U}{V} & \frac{U}{\sqrt{V}} & \frac{\sqrt{U}}{\sqrt{V}}& \sqrt{U} \end{pmatrix} \,,
}
so that by computing $S_{{\bf 15}_s}(U,1)$ and comparing to \eqref{S15Approx}, we find that $\lambda^2_{1,0,{\bf15}_s}=4$. This free theory has 8 real scalars and 8 Majorana fermions, so $c_T = 16$ according to \eqref{CanStress}. Because the relationship between $\lambda^2_{1,0,{\bf15}_s}$ and $c_T^{-1}$ is fixed by the superconformal Ward identity, we conclude that in general
\es{cTtoLam}{
\lambda^2_{1,0,{\bf15}_s}=\frac{64}{c_T}\,.
}
Combining \eqref{cTtoLam} with \eqref{B1} we conclude that
\es{K2}{
A_\text{SG}^1&=B_\text{SG}^1=C_\text{SG}^{1} =  \frac{32}{\pi^2}\,,
}
which is the same coefficient that was found for the $\mathcal{N}=8$ case in \cite{Zhou:2017zaw}.  This is the same coefficient we would obtain if we decomposed the known ${\cal N}=8$ answer from \cite{Zhou:2017zaw} into ${\cal N}=6$ language as we did in Section~\ref{secM4}.  Indeed, the supergravity term does not depend on $k$ when written in terms of $c_T$, because it is proportional to the effective 4d Newton constant $G_4 \propto 1/c_T$.

\subsection{Constraints from supersymmetric localization}
\label{Consusy}

Let us now explore the constraints on the coefficients $A_{i,j}^l$, $B_{i,j}^l$, $\widetilde B_{i,j}^l$, and $C_{i,j}^l$ coming from the supersymmetric localization constraints of Section~\ref{locloc}. 
To do so, we can compute the integrated constraints $I_{++}[\cS^i]$ in \eqref{Ipp} and $I_{+-}[\cS^i]$ in \eqref{Ipm} using the explicit Mellin amplitudes for $M_\text{SG}^i$, $M_4^i$, and $M_3^i$ given in \eqref{SugMellin}, \eqref{R4Mellin}, and \eqref{degree2}, respectively.  We have:  
\begin{equation}\begin{aligned}\label{ints34}
I_{++}[M_\text{SG}^i] &= 12\,,\quad &I_{+-}[M_\text{SG}^i] &= -\pi^2\,,\\
I_{++}[M_3^i] &= \frac83 \,,\quad &I_{+-}[M_3^i] &= \frac 23 \pi^2\,,\\
I_{++}[M_4^i] &= \frac{288}{35}\,,\quad &I_{+-}[M_4^i] &= \frac {8} 7 \pi^2 \,.
\end{aligned}\end{equation}
(For the details of the computation that gives \eqref{ints34}, see Appendix \ref{INTEGRALS}.)

Plugging \eqref{ints34} into  \eqref{DerSimpOneMassFinal} and \eqref{DerSimpTwoMassesFinal} and using Eqs.~\eqref{finitekupppp}, \eqref{largeLam}, and \eqref{finiteMupppp}, we can obtain the following results.  First, without using the constraints from the flat space limit or the constraints \eqref{B1} coming from the superconformal block expansion, the supersymmetric localization constraints \eqref{DerSimpOneMassFinal} and \eqref{DerSimpTwoMassesFinal} reproduce the coefficients in the first line of \eqref{B1}.  This is a stringent consistency check on the accuracy of our computations.

Second, using the constraints \eqref{B1} coming from the superconformal block expansion as an input, the supersymmetric localization constraints allow us to fix the coefficients at the next two orders in each of the expansions \eqref{finitekM}, \eqref{thooftM}, and \eqref{finitemuM}.  The result is
%
\es{coeffFix}{
\text{finite $k$:}\qquad A_4^4&=\frac{2240 }{(6\pi^4 k)^{2/3}}\,, \qquad
 A_3^3 = A_3^4 = 0 \,,\\
\text{'t Hooft:}\qquad B_3^4&=-\frac{54 \sqrt{2} \zeta (3)}{\pi ^5}\,,\qquad B_4^4=\frac{105 \zeta (3)}{2 \sqrt{2} \pi ^5}\,,\qquad \widetilde B_4^4=\frac{4480 \sqrt{2}}{3 \pi ^3}\,, \\
 B_3^3  &= \widetilde{B}_3^3 =  \widetilde{B}_3^4  = 0 \,, \\
\text{finite $\mu$:}\qquad C_3^4&=-\frac{576\ 2^{3/8} 3^{\frac14}\zeta (3)}{\pi ^{23/4} \mu^{3/8}}\,,\qquad C_4^4=\frac{2^{\frac38}280}{3^{3/4} \pi ^{23/4} } \left(4\sqrt{2} \pi ^3\mu^{\frac18}+3 \zeta
   (3)\mu^{-\frac38}\right)\,, \\
   C_3^3 &=0 \,.
}
These equations agree with the constraints from the flat space limit, thus providing a very non-trivial precision test of AdS/CFT\@.

Third, using both the constraints \eqref{B1} as well as the constraints coming from the flat space limit as input, the constraints from supersymmetric localization allow us to conclude that
 \es{Constraints5}{
  A_3^5 = A_4^5 = B_3^5 = B_4^5 = 0 \,.
 }

We can then plug these values back into \eqref{finitekM}, \eqref{thooftM}, and \eqref{finitemuM} to get the final answers \eqref{MellinSummary} as advertised in the Introduction.

\section{Discussion}
\label{disc}

In this paper we used superconformal symmetry, the flat space limit, and most importantly supersymmetric localization results for the mass deformed sphere free energy to compute the $R^4$ correction to the stress tensor multiplet bottom component four point function $\langle SSSS\rangle$ in $\mathcal{N}=6$ $U(N)_k\times U(N)_{-k}$ ABJM theory in the large $N$ finite $\mu=N/k^5$ limit. After taking the flat space limit we matched the known type IIA string theory S-matrix for finite $g_s$, which is the first check of AdS/CFT of this type for local operators. This finite $\mu$ result interpolates between the large $N$ finite $k$ limit at $\mu\to\infty$ and the large 't Hooft coupling $\lambda\sim N/k$ limit at $\mu\to0$, which in the flat space limit are related to the S-matrix of M-theory and weakly coupled type IIA string theory, respectively.

There were several technical innovations in this work relative to similar studies of $\mathcal{N}=8$ ABJM theory in \cite{Binder:2018yvd} and $\mathcal{N}=4$ SYM in \cite{Binder:2019jwn}, which all stem from the fact that our theory is not maximally supersymmetric like these other theories. One implication is that the stress tensor multiplet is $\frac13$-BPS, not $\frac12$-BPS as in the other cases, so the Ward identities that we derived for various four point functions in this multiplet are the first such derivation for operators annihilated by less than half the supercharges. Another novelty of this calculation was that demanding bulk locality, i.e.~that higher derivative corrections to supergravity correspond to polynomial Mellin amplitudes, in stress tensor correlators other than $\langle SSSS\rangle$ gave additional constraints, unlike the maximally supersymmetric cases were only $\langle SSSS\rangle$ gave such constraints. Finally, in the flat space limit, stress tensor multiplet correlators in holographic theories are dual to supergraviton multiplet amplitudes in one more dimension. For maximally supersymmetric supergravity there is just one such amplitude supermultiplet, but for our sub-maximal case two amplitudes exist, which is related to the fact that we found an extra subleading term in the large $N$ expansion of $\langle SSSS\rangle$ relative to the analogous expressions in $\mathcal{N}=8$ ABJM and $\mathcal{N}=4$ SYM.

A crucial ingredient in our finite $g_s$ check of AdS/CFT was the conjecture that the all orders in large $N$ localization expression for derivatives $\partial^4_{m_\pm}F\big\vert_{m_\pm=0}$ and $\partial^2_{m_+}\partial^2_{m_-}F\big\vert_{m_\pm=0}$ of the mass deformed sphere partition function $F(m_\pm)$ in \cite{Nosaka:2015iiw} only receive non-perturbative corrections of form $e^{-\sqrt{N/k}}$ and $e^{-\sqrt{Nk}}$.  When $m_\pm = 0$, these corrections can be interpreted as instanton effects in string theory, and it was proven in \cite{Marino:2016new,Hatsuda:2013gj,Hatsuda:2012dt,Calvo:2012du,Marino:2011eh,Drukker:2010nc,Marino:2009jd,Hatsuda:2013oxa} that for $F(0)$ they do take the form mentioned above.  Since a small mass deformation changes the geometry only slightly, we expect that for sufficiently small masses these instanton effects have the same $N$ and $k$ scaling as for $m_\pm = 0$.  It would be interesting to find a more rigorous justification of this fact in the future.

Looking ahead, there are more localization constraints that can be used to fix $\langle SSSS\rangle$. As discussed in Section \ref{locloc}, the $\mathcal{N}=6$ ABJM free energy can be computed using localization as a function of not only the two masses $m_\pm$ considered in this work, but also of a third mass $\widetilde m$.  The reason why there are three mass parameters is that, as an ${\cal N} = 2$ SCFT, any ${\cal N} = 6$ SCFT has $SU(2) \times SU(2) \times U(1)$ flavor symmetry, and the Cartan of the flavor symmetry algebra is three-dimensional.  In addition to the three mass parameters, one can also consider placing the theory on a squashed sphere parameterized by squashing parameter $b$ \cite{Hama:2011ea} (with $b=1$ corresponding to the round case).  There are then seven potentially independent combinations of four derivatives of these parameters that can be related to integrated 4-point functions of the stress tensor multiplet:
\es{seven}{
 \partial^4_{m_\pm}F\,,\quad
\partial^2_{m_+}\partial^2_{m_-}F\,,\quad
 \partial^4_{b}F\,,\quad
 \partial^2_{b}\partial^2_{ m_\pm}F\,,\quad
 \partial^4_{\widetilde m}F\,,\quad
 \partial^2_{m_\pm}\partial^2_{\widetilde m}F\,,\quad
 \partial^2_{b}\partial^2_{\widetilde m}F\,,
}
all evaluated at $m_\pm = \widetilde{m} = 0$ and $b=1$.  Only the first two were considered in this work. In Section~\ref{4POINT}, we showed that there are seven polynomial Mellin amplitudes of maximal degree six,\footnote{As shown in Table \ref{lowOrderTab}, we have one degree 3, one degree 4, two degree 5, and three degree 6.} as well as the supergravity Mellin amplitude that is already fixed by the conformal Ward identity. This means that we could potentially use localization to fix the coefficients of all these Mellin amplitudes, which would thus allow us to determine the $D^4R^4$ term in the large $N$ finite $\mu$ limit, that could be checked in the flat space limit against the known \cite{Green:1999pu} finite $g_s$ term in the type IIA S-matrix. These are the highest order terms we would expect to be able to fix with $\mathcal{N}=6$ supersymmetry.

For $\mathcal{N}=8$ ABJM theory, the $U(1)$ flavor symmetry combines with $SU(4)_R$ to form the larger R-symmetry $SO(8)_R$, so the dependence on $\widetilde m$ is now related to that on $m_\pm$. As discussed in \cite{Binder:2018yvd}, there are only two quartic Casimir invariants for $SO(8)$, so only the first four constraints in \eqref{seven} would be linearly independent. On the other hand, for $\mathcal{N}=8$ there are only three polynomial Mellin amplitudes of maximal degree 7, so we could fix the tree level $D^6R^4$ term, which is the highest order term that is protected by supersymmetry. In fact, there is only one additional Mellin amplitude at maximal degree 8, so four constraints would seem sufficient to fix tree level $D^8R^4$, but this term is not expected to be fixed by supersymmetry, so it is likely that one of these constraints becomes redundant for $\mathcal{N}=8$ ABJM when we take the large $N$ limit.

To go beyond these protected coefficients, we need a more general method such as the numerical conformal bootstrap. Our computation of the $\mathcal{N}=6$ Ward identities for $\langle SSSS\rangle$ opens the door to a numerical bootstrap study of $\mathcal{N}=6$ ABJM theory, which would generalize the $\mathcal{N}=8$ studies of \cite{Chester:2014fya,Chester:2014mea,Agmon:2017xes}. In the $\mathcal{N}=8$ case, the bootstrap bounds were found to be conjecturally saturated by CFT data in ABJM theory, so that all low-lying CFT data, both protected and unprotected, could be read off up to numerical error. If a similar thing occurs for $\mathcal{N}=6$ ABJM theory, then we can use this unprotected CFT data to extend the derivation in this work to higher order, and perhaps even interpolate between M-theory at finite $k$ and type IIA string theory at weak and strong coupling in the 't Hooft limit of ABJM theory.

\section*{Acknowledgments} 

We thank Yifan Wang for collaboration at early stages of this project, and we also thank him as well as Ofer Aharony and Igor Klebanov for useful discussions.   DJB and SSP are supported in part by the Simons Foundation Grant No.~488653, and by the US NSF under Grant No.~1820651\@.  DJB is also supported in part by the General Sir John Monash Foundation.  SMC is supported by the Zuckerman STEM Leadership Fellowship. SSP is also supported in part by an Alfred P.~Sloan Research Fellowship.  

\appendix

\section{Useful details on the conformal block expansion}

\subsection{Derivation of the $SU(4)$ invariants}
\label{SU4STRUCTURES}

The $SU(4)$ invariants presented in \eqref{TR} can be derived as follows.  The $T_R(X_i)$ are eigenfunctions of the $SU(4)$ quadratic Casimir $C_2$ acting on $X_1$ and $X_2$, namely
\es{casEig}{
C_2T_R(X_i)=c_RT_R(X_i)\,,
}
where 
 \es{CasX12}{
  C_2 T (X_i) &= \sum_{a=1}^{15} \biggl( T([t^a, [t^a, X_1]], X_2, X_3, X_4)
   + T(X_1, [t^a, [t^a, X_2]], X_2, X_3, X_4) \\
   &{}+ 2 T([t^a, X_1], [t^a,  X_2], X_3, X_4) \biggr) \,.
 }
Here, $t^a$, $a = 1, \ldots, 15$, are the (hermitian traceless) $SU(4)$ generators.  In the normalization where $\tr (t^a t^b) = \frac{\delta^{ab}}{2}$, the eigenvalues $c_R$ are
\es{eigs}{
c_{\bf1}=0\,,\quad c_{\bf15}=4\,,\quad c_{{\bf20}'}=6\,,\quad c_{\bf45}=c_{\overline{\bf45}}=8\,,\quad c_{\bf84}=10\,.
}
In the basis given in \eqref{SSSScor}, the tensor structures $T_R(X_i)$ obeying \eqref{casEig} are then those given in \eqref{TR}. (See also Eq.~(B.25) of~\cite{Liendo:2015cgi}.)

In terms of $\cS^i$, the functions of $(U, V)$ corresponding to the various representations are
 \es{SRFromSi}{
  \cS_{\bf 1} &= \cS^1 +  \frac{1}{30} \left( 2\cS^2 + 2\cS^3 -\cS^4  \right)  + \frac{1}{2} \left( \cS^5 + \cS^6 \right) \,, \\
  \cS_{{\bf 15}_a} &= \frac{1}{8} \left( -\cS^2 + \cS^3 \right) + \frac{1}{2} \left( \cS^5 - \cS^6 \right) \,, \\
  \cS_{{\bf 15}_s} &= \frac{1}{6} \left( \cS^2 + \cS^3 - \cS^4 \right) + \frac{1}{2} \left( \cS^5 + \cS^6 \right)  \,, \\
  \cS_{{\bf 20}'} &= \frac{1}{24} \left( \cS^2 + \cS^3 - 2 \cS^4 \right) \,, \\
  \cS_{{\bf 45} \oplus \overline{\bf 45}} &= \frac{1}{8} \left( \cS^2 - \cS^3 \right) \,, \\
  \cS_{{\bf 84}} &= \frac{1}{16}  \left( \cS^2 + \cS^3  +2 \cS^4 \right) \,.
 }

\subsection{Extracting OPE coefficients}
\label{OPEEXTRACTION}

We will be interested in extracting\footnote{See \cite{Zhou:2017zaw,Chester:2018lbz} for similar calculations in $\mathcal{N}=8$ SCFTs.} two OPE coefficients of protected ($1/3$-BPS) scalar operators in the $S \times S$ OPE:  the OPE coefficient of an operator with $\Delta = 1$ in the ${\bf 15}_s$ irrep of $SU(4)$ (this is the same as the external operator $S_a{}^b$), and that of an operator with $\Delta = 2$ in the ${\bf 84}$.  In the theories of interest to us, both of these operators are the lowest dimension operators in their corresponding R-symmetry channels.

Let us start with $\lambda_{1, 0, {\bf 15}_s}^2$, and let us take $U \to 0$ while setting $V = 1$.  In this limit, $G_{1, 0}(U, V) \approx \sqrt{U}/4$, so we must have 
 \es{S15Approx}{
  \cS_{{\bf 15}_s}(U, 1)  = \frac{\lambda_{1, 0, {\bf 15}_s}^2}{4} \sqrt{U} + \cdots \,.
 }
Thus, in order to extract $\lambda_{1, 0, {\bf 15}_s}^2$, all we need to do is extract the coefficient of $\sqrt{U}$ in the small $U$ expansion of $\cS_{{\bf 15}_s}(U, 1)$.   Note that the disconnected piece $\cS_{\text{disc}, {\bf 15}_s}(U, 1) = O(U)$ in this limit, so the $\sqrt{U}$ term in the small $U$ expansion of $\cS_{{\bf 15}_s}(U, 1)$ must come from a pole at $s=1$ in the Mellin amplitude $M_{{\bf 15}_s}(s, t)$ corresponding to $\cS_{{\bf 15}_s}(U, V)$, namely
 \es{Mellin15}{
  M_{{{\bf 15}_s}} &\equiv  \frac 16 \left( M^2 + M^3 - M^4 \right) + \frac 12 (M^5 + M^6)
 }
(see \eqref{SRFromSi}).  Performing the $s$ integral in \eqref{melDef} and picking up the residue at $s=1$, we obtain
 \es{S15ApproxMellin}{
  \cS_{{\bf 15}_s}(U, 1)  = -\frac{\sqrt{U}}{8i }  \int_{ -i \infty}^{ i \infty} dt\,  
   \Gamma^2 \left( 1 - \frac{t}{2} \right) \Gamma^2 \left( \frac{t-1}{2} \right)
    \lim_{s\to 1} \biggl[ (s-1) M_{{{\bf 15}_s}}(s, t) \biggr] + \cdots \,,
 }
where the integration contour can be chosen such that $\Re t < 2$.  Comparing with \eqref{S15Approx}, we have
 \es{Gotlam15}{
  \lambda_{1, 0, {\bf 15}_s}^2 = -\frac{1}{2i }  \int_{ -i \infty}^{ i \infty} dt\,  
   \Gamma^2 \left( 1 - \frac{t}{2} \right) \Gamma^2 \left( \frac{t-1}{2} \right)
    \lim_{s\to 1} \biggl[ (s-1) M_{{{\bf 15}_s}}(s, t) \biggr] \,.
 }

Next, let us consider extracting $\lambda_{2, 0, {\bf 84}}^2$ by considering $\cS_{\bf 84}(U, 1)$ in the limit $U \to 0$.  Because $G_{2, 0}(U, 1) = \frac{U}{16} + \cdots$ in this limit, we have
 \es{S84Approx}{
  \cS_{{\bf 84}}(U, 1)  = \frac{\lambda_{2, 0, {\bf 84}}^2}{16} U + \cdots \,.
 }
So, in this case, we should evaluate the coefficient multiplying $U$ in the small $U$ expansion of $\cS_{{\bf 84}}(U, 1) $.  This coefficient receives contributions from the disconnected piece, $\cS_{\text{disc}, {\bf 84}}(U, V) = (U + U/V) / 16$, which gives
 \es{SDiscApprox}{
  \cS_{\text{disc}, {\bf 84}}(U, 1) = \frac{U}{8} \,,
 }
as well as from the connected piece from the $s=2$ pole in the Mellin integral.  The Gamma functions in the definition \eqref{melDef} of the Mellin transform have a double pole at $s=2$, so 
\es{Mellin84}{
  M_{\bf 84} = \frac{1}{16}  \left( M^2 + M^3  +2 M^4 \right)
 }
must vanish at least linearly as $s \to 2$.  Combining the contribution of this pole with \eqref{SDiscApprox}, we have
 \es{S84Approx2Tmp}{
 \cS_{{\bf 84}}(U, 1)  = U \biggl[ \frac 18 +   \frac{i}{2 \pi} \int_{-i \infty}^{i \infty} dt\,  
   \Gamma^2 \left( 1 - \frac{t}{2} \right) \Gamma^2 \left( \frac{t}{2} \right)
    \lim_{s\to 2} \frac{M_{{{\bf 84}}}(s, t) }{s-2} \biggr] + \cdots \,.
 } 
The integration contour here must be such that $\Re t$ is smaller than the minimum between $2$ and the pole in $t$ of $M_{{{\bf 84}}}(s, t) $ with the smallest real part, and such that $2 - \Re t$  is smaller than the minimum between $2$ and the pole in $u$ of $M_{{{\bf 84}}}(s, t) $ with the smallest real part.  Such a condition is obeyed by $0< \Re t < 2$ for polynomial $M_{{{\bf 84}}}(s, t) $, but it is tricky to impose it when $M_{{{\bf 84}}}(s, t) $ has both a pole at $t=1$ and a pole at $u=1$, as is the case for the SUGRA amplitude.  In the case that both of these poles are present, let us use $0 < \Re t < 1$.  Because if we closed the $t$ contour on the right we would pick up both the pole at $t=1$ and that at $u=1$, we should subtract by hand the contribution from the pole at $u=1$.  Thus, the correct formula is
  \es{S84Approx2}{
 \cS_{{\bf 84}}(U, 1)  &= U \biggl[ \frac 18 + \pi^2 \lim_{s \to 2}  \lim_{t \to 3-s} \frac{(u-1) M_{{{\bf 84}}}(s, t) }{s-2}  \\
 &{}+\frac{i}{2 \pi} \int_{-i \infty}^{i \infty} dt\,  
   \Gamma^2 \left( 1 - \frac{t}{2} \right) \Gamma^2 \left( \frac{t}{2} \right)
    \lim_{s\to 2} \frac{M_{{{\bf 84}}}(s, t) }{s-2} \biggr] + \cdots \,.
 } 

Comparing with \eqref{S84Approx}, we extract
 \es{lam84}{
  \lambda_{2, 0, {\bf 84}}^2
   &= 2  + 16 \pi^2 \lim_{s \to 2}  \lim_{t \to 3-s} \frac{(u-1) M_{{{\bf 84}}}(s, t) }{s-2} \\
    &{}+   \frac{8i}{\pi} \int_{-i \infty}^{ i \infty} dt\,  
   \Gamma^2 \left( 1 - \frac{t}{2} \right) \Gamma^2 \left( \frac{t}{2} \right)
    \lim_{s\to 2} \frac{M_{{{\bf 84}}}(s, t) }{s-2} \,,
 }
with the $t$ contour obeying $0< \Re t < 1$.

\section{Discrete symmetries of $\cN = 6$ theories}
\label{DISCRETE}

Both $\cN = 6$ SCFTs and flat space scattering amplitudes may posses various discrete symmetries that can be used to impose selection rules.  The symmetries we will focus on here are parity $\cP$, time reversal combined with charge conjugation, $\cC \cT$, and a discrete R-symmetry we will call $\cZ$.  Even for theories that break these symmetries, organizing the SCFT correlators and scattering amplitudes in terms of them will prove very useful.


\subsection{Review of spinor helicity formalism}
\label{REVIEWSPINOR}

For massless fermions, the Dirac equation for the wavefunction of 4-component spinors implies
 \es{uvEqs}{
   \slashed v_{\pm}(p) = 0\,,\qquad \overline u_\pm(p)\slashed p = 0\,.
 }
Here $\pm$ indicated the helicity $h = \pm \frac 12$ of the wavefunction. If we take our Dirac matrices to be in the Weyl basis, namely
 \es{Weyl}{
  \gamma^0 = \begin{pmatrix}
   0 & 1 \\
   1 & 0 
   \end{pmatrix} \,, \qquad
    \gamma^i = \begin{pmatrix}
    0 & \sigma^i \\
    -\sigma^i & 0 
    \end{pmatrix} \,, \qquad
     \gamma^5 = \begin{pmatrix} - 1 & 0 \\
     0 & 1 \end{pmatrix}   \,,
 }
where $1$ stands for the $2\times 2$ identity matrix and $\sigma^i$, $i = 1, 2, 3$ are the standard Pauli matrices, then the top two components of the Dirac spinor transform in the $(1/2,0)$ and bottom two in the $(0,1/2)$ of $SO(3,1)$.  For a given momentum $p^\mu = (E, E \sin \theta \cos \phi, E \sin \theta \sin \phi, E \cos \theta)$, we can then define the angle and square brackets as 
 \es{AngleSquare}{
  | p \rangle^{\dot a}
   &= \sqrt{2E} \begin{pmatrix}
    \cos \frac{\theta}{2} \\
    \sin \frac{\theta}{2} e^{i \phi}
   \end{pmatrix} \,, \qquad
    | p ]_a = \sqrt{2E} \begin{pmatrix}
    \sin \frac{\theta}{2} \\
    -\cos \frac{\theta}{2} e^{i \phi}
   \end{pmatrix}  \,, \\
  [p|^a &=  \sqrt{2E} \begin{pmatrix}
    \cos \frac{\theta}{2} \\
    \sin \frac{\theta}{2} e^{-i \phi}
   \end{pmatrix} \,, \qquad
    \langle p |_{\dot a} = \sqrt{2E} \begin{pmatrix}
    \sin \frac{\theta}{2} \\
    -\cos \frac{\theta}{2} e^{-i \phi}
   \end{pmatrix}  
 }
such that 
 \es{FourComp}{
v_+(p) &= \begin{pmatrix} |p]_a \\ 0\end{pmatrix}\,, \qquad v_-(p) = \begin{pmatrix} 0 \\ |p\rangle^{\dot a} \end{pmatrix}\,, \\
\overline u_+(p) &= \begin{pmatrix} [p|^a & 0\end{pmatrix}\,, \qquad \overline u_-(p) = \begin{pmatrix} 0  & \langle p |_{\dot a} \end{pmatrix}
 }
are solutions to \eqref{uvEqs}.

Let us consider the scattering of massless particles $b_i^\pm$ for $i = 1, 2, \ldots$. We define the scattering amplitude to be:
\begin{equation}
A[b_1^\pm b_2^\pm \ldots] \delta^{(4)}(p_1+p_2+\ldots) = \langle a_1^\pm(p_1) a_2^\pm(p_2)\ldots \rangle\end{equation}
where $a_i^\pm(p)$ is the annihilation operator of the $i^{\text{th}}$ particle, annihilating a particle of helicity $\pm$ and momentum $p_i$.

\subsection{Discrete symmetries for scattering amplitudes}
\label{DISCRETEAMPLITUDES}

We will begin by discussing the discrete symmetries of the 4d amplitudes, motivated by two reasons:  1)  given that in $\cN = 6$ supergravity, we have two $\cC \cP \cT$ conjugate multiplets, we should understand how $\cC \cP \cT$ relates the scattering amplitudes;  and 2) we can use discrete symmetries in order to classify the structures that appear in the super-amplitude.   As mentioned above, we will discuss parity $\cP$, the product $\cC \cT$, as well as a discrete R-symmetry we denote by $\cZ$.

Under parity $\cP$, we reverse the spatial components of the momentum of a particle, while leaving the spin unchanged. Flipping the direction of $\vec{p}$ is equivalent to sending $\theta \to \pi - \theta$ and $\phi \to \phi \pm \pi$ in \eqref{AngleSquare}.  Under this transformation, the spinors in the first line of \eqref{AngleSquare} get interchanged and so do the spinors on the bottom line.  Thus, parity acts\footnote{In terms of the four-component spinors \eqref{FourComp}, the action of parity takes the usual form: 
$$ v_\pm(p^0,-\vec p) = \gamma^0 v_{\pm}(p^0,\vec p)  = \begin{pmatrix}0&1\\1&0\end{pmatrix}  v_{\pm}(p^0,\vec p) \,.$$ 
} as either $\cP_{a \dot a}$ or $\cP^{\dot a a}$
 \es{ParitySpinorHelicity}{
  \cP_{a \dot a} | p \rangle^{\dot a} = | p ]_a \,, \qquad 
   \cP^{\dot a a} | p ]_a = | p \rangle^a \,,  \qquad
    [p|^a \cP_{a \dot a} = \langle p |_{\dot a} \,, \qquad
     \langle p |_{\dot a} \cP^{\dot a a} = [p |^a \,.
 }
Hence the effect of parity is to swap all angle brackets with square brackets and vice versa, while leaving all coefficients unchanged.  For instance, $\cP (c \langle 12 \rangle ) = c [12]$ for any constant $c$.

The second discrete symmetry we consider is $\cC \cT$.  Under $\cC \cT$, the spatial components of momentum also flip sign, just like for $\cP$, but in addition $\cC \cT$ also implements complex conjugation.  Thus, from \eqref{AngleSquare}, we see that $\cC \cT$ acts as either $(\cC \cT)_{\dot a \dot b}$ or $(\cC \cT)^{ab}$ as follows:
 \es{CTAction}{
  (\cC \cT)_{\dot a \dot b} |p \rangle^{\dot b} = \langle p|_{\dot a} \,, \qquad
   (\cC \cT)^{ab} |p]_b = [p|^a \,, \qquad
    \langle p|_{\dot a}  (\cC \cT)_{\dot a \dot b} = |p \rangle^{\dot b} \,, \qquad
     [p|^a (\cC \cT)^{ab}  =  |p]_b \,.
 }
Thus, the effect of $\cC \cT$ is to flip all the brackets and perform complex conjugation on the coefficients---for instance $ \cC \cT (c \langle 12  \rangle) = c^* \langle 2 1 \rangle$ for any constant $c$.  

The combined transformation of the two symmetries above, $\cC \cP \cT$, is a symmetry of all unitary QFTs.  On amplitudes, it acts by exchanging angle brackets with flipped square brackets and vice versa, and it complex conjugates the coefficients.  For instance, $\cC \cP \cT ( c \langle 12 \rangle ) = c^* [21]$.  Using $\cC \cP \cT $, we can relate a given amplitude to the amplitude of the $\cC \cP \cT $ conjugate particles.  For particles $b_1$, $b_2$, etc.~with $\cC \cP \cT$ conjugate particles $\overline b_1$, $\overline b_2$, etc., we have
 \es{AmpCPT}{
  \cC \cP \cT  \left( A[b_1^\pm b_2^\pm \ldots] \right) = A[\overline b_1^\mp \overline b_2^\mp \ldots] \,.
 }

Because $\cC \cT$ does not change the helicity of the particles, it relates a given amplitude to itself.  Thus, we can classify the various scattering amplitudes based on whether they are $\cC \cT$-even or $\cC \cT$-odd.  (In a $\cC \cT$-preserving theory, such as pure $\cN = 6$ supergravity, all amplitudes should be $\cC \cT$-even.  But the $\cC \cT$ symmetry may be broken by higher derivative corrections.)  For instance, if we consider the amplitude
 \es{Ahphpamam}{
 A[h^+h^+a^-a^-] = [12]^4\langle 34\rangle^2 f_3(s,t) \,,
 }
we see that 
\begin{equation}\begin{split}\cC \cT \left(A[h^+h^+a^-a^-]\right) &= [12]^4\langle 34\rangle^2 f_3^*(s,t) 
\end{split}\end{equation}
and so the amplitude \eqref{Ahphpamam} is $\cC \cT$ even / odd if $f_3(s, t)$ is real / pure imaginary.  From this we conclude that $\cA[\Phi\Phi\Phi\Phi]$ can be thought of as containing two distinct superstructures, one of which is $\cC \cT$ even and the other $\cC \cT$ odd. Similar manipulations show that $\cA[\Phi\Phi\Phi\Psi]$ also contains a $\cC \cT$ even and $\cC \cT$ odd structure, corresponding to $f_2(s,t)$ purely real and purely imaginary respectively.

On the other hand, one can show that $\cA[\Phi\Phi\Psi\Psi]$ is always $\cC \cT$ even, even in a theory in which $\cC \cT$ is not a symmetry. We can see this by considering the graviton scattering amplitude:
 \es{GravScatt}{
  A[h^+h^+h^-h^-] &= [12]^4\langle 34\rangle^4f_1(s,t) \,, \qquad
   A[h^-h^-h^+h^+] = \langle 12\rangle^4 [34]^4 f_1(s,t)\,,
 }
where $A[h^-h^-h^+h^+]$ is related to $A[h^+h^+h^-h^-]$ under crossing both $1\leftrightarrow 3$ and $2\leftrightarrow4$.  But the two amplitudes in \eqref{GravScatt} are also related by $\cC \cP \cT$, 
 \es{GravScattCPT}{
  \cC \cP \cT \left( A[h^+h^+h^-h^-] \right) 
   = A[h^-h^-h^+h^+] = \langle 12\rangle^4 [34]^4 f_1^*(s,t) \,,
 }
and from comparing this expression with \eqref{GravScatt} we conclude that $f_1(s, t)$ must be real.   Then 
 \es{CTGravScatt}{
  \cC \cT \left( A[h^+h^+h^-h^-] \right) = [12]^4\langle 34\rangle^4f_1^*(s,t) = A[h^+h^+h^-h^-] \,,
 }
and so  $A[h^+h^+h^-h^-]$ is always $\cC \cT$-even.  This relation extends to the full multiplet thus showing that $A[\Phi \Phi \Psi \Psi]$ is $\cC \cT$-even.

Let us now consider all possible discrete R-symmetries of $\cN = 6$ supergravity and its higher derivative corrections.   Before doing so, let us recall that, as discussed in the main text, the various particles in the $\Phi$ and $\Psi$ multiplets transform under an $SU(6)_R$ R-symmetry that is a symmetry of pure supergravity and of the higher derivative corrections considered here.  Under $SU(6)_R$, the supercharges transform contravariantly
\begin{equation}
\eta^I\rightarrow M^I{}_{J}\eta^J \,,
\end{equation}
where $M^I{}_{J}$ is a unitary matrix with determinant $1$. The supergraviton fields $h^\pm\,,\ \psi_I^\pm\,,\ g_{IJ}^\pm\,,\ldots$ transform covariantly, so that overall the superfields $\Phi$ and $\Psi$ are invariant. 

To see what discrete R-symmetries might be possible, let us first focus on the pure supergravity case and consider relaxing the condition that $M^I{}_{J}$ has determinant $1$.  Instead, let us consider a more general element of $U(6)$. Without loss of generality let us consider a transformation:
\begin{equation}
\eta^I\rightarrow e^{i\theta}\eta^I\end{equation}
in the center of $U(6)$.  We can also allow the superfields $\Phi$ and $\Psi$ to pick up an overall phase:
\begin{equation}
\Phi\rightarrow e^{i\alpha}\Phi\,,\qquad \Psi\rightarrow e^{i\beta}\Psi.
\end{equation}
The supergravitons will then transform as:
\begin{equation}\begin{aligned}
h^+ &\rightarrow e^{i\alpha} h^+\,,\quad &a^+ &\rightarrow e^{i\beta} a^+\,, \\
\psi^+ &\rightarrow e^{i\alpha-i\theta} \psi^+\,,\quad &\chi^+ &\rightarrow e^{i\beta-i\theta} \chi^+\,, \\
g^+ &\rightarrow e^{i\alpha-2i\theta} g^+\,,\quad &\overline \phi &\rightarrow e^{i\beta-2i\theta} \overline \phi\,, \\
&\vdots & &\vdots \\
a^+ &\rightarrow e^{i\alpha-6i\theta} a^+\,,\quad & h^- &\rightarrow e^{i\beta-6i\theta} h^-\,.
\end{aligned}\end{equation}
We cannot however choose $\alpha\,, \beta$, and $\theta$ arbitrarily. The graviton and gauge fields are real, and so we can only transform them by a factor of $\pm 1$. This restricts us to the cases $e^{i\theta} = \pm i$ or $e^{i\theta} = \pm 1$, as well as $e^{i \alpha} = \pm 1$ and $e^{i \beta} = \pm 1$.  The case $e^{i \theta} = \pm 1$ is already in $SU(6)$, so let us focus on the possibility $e^{i\theta} = \pm i$.  To determine $e^{i\alpha}$ and $e^{i\beta}$, let us make use of the $\cC \cT$-invariance of supergravity in order to write the scattering amplitudes for three gravitons as
\begin{equation}
  A[h^+h^+h^-] = g\frac{[12]^6}{[13]^2[23]^2}\,,\qquad A[h^-h^-h^+] = g\frac{\langle 12\rangle ^6}{\langle13\rangle^2\langle23\rangle^2} \,, 
  \end{equation}
with real $g$.  Since the right-hand sides of these equations are invariant under the transformation considered above, we deduce that $e^{i\alpha} = 1$ and $e^{i\beta} = -1$.

We can now check that the transformation:
\begin{equation}\cZ: \Phi\rightarrow\Phi\,,\qquad\Psi\rightarrow -\Psi\,,\qquad \eta^I\rightarrow  i\eta^I\end{equation}
is in fact a symmetry of pure supergravity, as is the symmetry $-\cZ$ which sends $\eta^I\rightarrow -i\eta^I$. Under both $\cZ$ and $-\cZ$ the gauge fields flip sign
\begin{equation}
\cZ:\ a^\pm \rightarrow -a^\mp\,,\qquad g^\pm\rightarrow -g^\mp
\end{equation}
while the gravitons $h^\pm$ and the graviscalar $\phi$ are left invariant. The fermions will transform with additional factors of $i$:
\begin{equation}\cZ:\ \Psi^\pm\rightarrow\pm i\Psi^\mp\,,\qquad F^\pm\rightarrow \mp i F^\mp\,,\qquad \chi^\pm\rightarrow \pm i \chi^\mp\,.
\end{equation}
The full symmetry group is now $(\mathbb Z_4\times SU(6))/\mathbb Z_2$, the subgroup of $U(6)$ of matrices with determinant $\pm 1$. 

Note however that only fermion bilinears are physical. As a result, the transformation $\eta^I\rightarrow -\eta^I$ acts trivially on all amplitudes. After quotienting the $SU(6)$ by this $\mathbb Z_2$ symmetry, we find that the symmetry group acting on the amplitudes is $\mathbb Z_2\times (SU(6)/\mathbb Z_2)$, with $\cZ^2 = I.$

While $\cZ$ is a discrete R-symmetry of pure supergravity, it may or may not be a symmetry of the corrections to supergravity, so we can classify the various amplitude structures as $\cZ$-even or $\cZ$-odd.   Since $\delta^{(12)}(Q)$ contains twelve $\eta$'s, it is even under $\cZ$, and so we conclude that $\cA[\Phi\Phi\Psi\Psi]$ and $\cA[\Phi\Phi\Phi\Phi]$ are even under $\cZ$ and that $\cA[\Phi\Phi\Phi\Psi]$ is odd. We can alternatively deduce this from \eqref{topcomp}, since $\cA[\Phi\Phi\Phi\Psi]$ contains an amplitude with an odd number of gauge fields, while the other two amplitudes contain an even number.

\begin{table}
\begin{center}
\hspace{-.4in}
{\renewcommand{\arraystretch}{1.2}
\begin{tabular}{|r|c|c|c|c|}\hline
Amplitude & $\cC \cT$ & $\cZ$ & First counterterm & \# derivatives \\ \hline
$\cA[\Phi\Phi\Phi\Phi]\,,\quad \cA[\Psi\Psi\Psi\Psi]$ & $\pm$ & $+$ & $F^2R^2$  & $6$ \\
$\cA[\Phi\Phi\Psi\Psi]$ & $+$ & $+$ & $R^4$ & $8$ \\
$\cA[\Phi\Phi\Phi\Psi]\,,\quad \cA[\Psi\Psi\Psi\Phi]$ & $\pm$ & $-$ & $D^8FR^3$ & $15$ \\ \hline
\end{tabular}}
\caption{Four particle scattering in $\cN = 6$ supergravity. The dimension is the mass dimension of the lowest bulk counterterm contributing to the amplitude, and $\cC \cT$ and $\cZ$ are the discrete symmetries defined in the main text.}
\label{4ptAmps}
\end{center}
\end{table}

We can summarize these results in Table~\ref{4ptAmps}. In total, the scattering of four supergravitons is fixed up to five arbitrary functions of $s$ and $t$. To determine the $\cZ$ and $\cC \cT$ even part of the amplitude there are two functions, while for each of the other combinations there is a single function. Because the only superamplitude contribution to scalar scattering, $\cA[\Phi\Phi\Psi\Psi]$, is automatically $\cZ$ and $\cC \cT$ invariant, it is impossible to know whether these symmetries are present or not in the full theory just by considering scalar scattering, without any additional information.

\subsection{Discrete symmetries for $\cN = 6$ SCFTs}
\label{DISCRETESCFT}

Analogous $\cP$, $\cC \cT$, and $\cZ$ symmetries exist for $\cN = 6$ superconformal theories, with $\cC \cT \cP$ always being a symmetry.  Individually, $\cP$, $\cC \cT$, and $\cZ$ may not be symmetries of a given theory, as we will see, but they are symmetries of the free theory (or more generally of the $U(1)_k \times U(1)_{-k}$ ABJM theory for all $k$) and of the leading order large $c_T$ holographic correlators.

Under $\cP$ and $\cC \cT$, the $\Delta = 1$ operators $S$ are even, while the $\Delta = 2$ operators $P$ are odd.    Just as for amplitudes, we expect that three out of the five superconformal structures given in Table~\ref{SCFTDet} are $\cP$ or $\cC \cT$ even, while the other two are $\cP$ and $\cC \cT$ odd.

The $\cZ$ R-symmetry is trickier in the case of SCFTs than for scattering amplitudes, because while in the case of scattering amplitudes it commutes with the $SU(6)_R$ R-symmetry, for SCFTs it does not commute with the $SO(6)_R$ R-symmetry.  Instead, it extends $SO(6)_R$ to $O(6)_R$.  Let us define the $\cZ$ generator so that it corresponds to the $O(6)$ matrix 
 \es{ZDefFund}{
   \cZ^{IJ} = \diag \{-1, -1, -1, 1, 1, 1 \} 
 } 
that is not part of $SO(6)$.  The group $O(6)$ has two 6-dimensional representations:  the vector representation ${\bf 6}^+$ under which a vector $v^I$ transforms as $v^I \to \cZ^{IJ} v^J$, and the pseudovector representation under which $v^I \to - \cZ^{IJ} v^J$.  By convention, we take the supercharges to transform as the ${\bf 6}^+$.\footnote{We could've considered the supercharges to transform as a pseudovector, but this choice is related to the first choice by an $SO(6)$ rotation.}   The representations of $O(6)$ appearing in the stress tensor multiplet are all antisymmetric products of the ${\bf 6}^+$, because we can start with the stress-energy tensor, which is a singlet, and obtain all the other operators by acting with anti-symmetric products of the superconformal generators.  Thus:  the rank-0 tensor is the singlet ${\bf 1}^+$ that is invariant under $\cZ$;  the rank-1 anti-symmetric tensor is the ${\bf 6}^+$;  the rank-2 anti-symmetric tensor is the adjoint representation ${\bf 15}^+$;  the rank-3 anti-symmetric tensor, the ${\bf 20}$ is irreducible under $O(6)$ but would've been reducible to ${\bf 10} + \overline{\bf 10}$ under $SO(6)$;  the rank-4 anti-symmetric tensor is the ${\bf 15}^-$ and can also be represented as a rank-2 anti-symmetric tensor with the same $SO(6)$ transformation properties as the ${\bf 15}^+$ except for an additional minus sign under $\cZ$;  the rank-5 anti-symmetric tensor is the ${\bf 6}^-$ and can also be represented as a pseudovector;  and lastly, the rank-6 anti-symmetric tensor ${\bf 1}^-$ is invariant under $SO(6)$ but it gets multiplied by $(-1)$ under $\cZ$.  See Table~\ref{chargeConjugation} for a list of conformal primaries of the stress tensor multiplet and the $O(6)$ representations under which they transform.  In particular, note that the superconformal primary $S$ is an $O(6)$ antisymmetric rank-2 pseudotensor.

\begin{table}
\begin{center}
\hspace{-.4in}
{\renewcommand{\arraystretch}{1.2}
\begin{tabular}{| c |c | c | c | c | c | c | c|}\hline 
Operators &$T^{\mu\nu}$ & $\psi^{\alpha\mu}$ & $J^\mu$ & $F^\alpha$ & $S, P$ & $\chi^\alpha$ &  $j^\mu$ \\ \hline
$O(6)$    &$\mathbf 1^+$  & $\mathbf 6^+$ & $\mathbf {15}^+$ & $\mathbf{20}$ & $\mathbf{15}^-$ & $\mathbf{6}^-$ & $\mathbf 1^-$ \\
$SO(6)$   &$\mathbf 1$ & $\mathbf 6$ & $\mathbf {15}$ & ${\mathbf {\overline {10}}} + {\mathbf {10}}$ & $\mathbf {15}$ & $\mathbf{6}$ & $\mathbf 1$ \\ \hline
\end{tabular}}
\caption{$O(6)$ and $SO(6)$ assignments for operators in the stress-tensor multiplet.}
\label{chargeConjugation}
\end{center}
\end{table}



To gain intuition about the $\cZ$ transformation, let us describe how it acts in the free $\cN = 6$ theory of $4$ complex fields $C^a$ and $4$ complex two-component fermions $\psi^a$ where it is actually a symmetry.  Both $\phi^a$ and $\psi^a$ transform in the ${\bf 4}$ of $SU(4)_R$, and their conjugates $\phi^\dagger_a$ and $\psi^\dagger_a$ transform in the $\overline{\bf 4}$ of $SU(4)_R$.  In this case, one can show that the $\cZ$ symmetry acts as charge conjugation
 \es{ZActionFree}{
  \phi^{a\prime} = \phi_a^\dagger \,, \qquad 
   \phi_a^{\dagger \prime} = \phi^a \,,
 }
and similarly on $\psi^a$ and $\psi^\dagger_a$.    Indeed, from $\phi^a$ and $\psi^a$, we can construct the various operators in the stress-tensor multiplet.  For example, 
 \es{StressOps}{
  S_a{}^b &= \phi_a^\dagger \phi^b - \frac{1}{4} \delta_a^b \phi_c^\dagger \phi^c \,, \\
  P_a{}^b &= \psi_a^\dagger \psi^b - \frac{1}{4} \delta_a^b \psi_c^\dagger \psi^c \,, \\
  j^\mu &= -i \left( \phi_a^\dagger \partial_\mu \phi^a - (\partial_\mu \phi_a^\dagger)  \phi^a \right) + \psi^\dagger_a \gamma^\mu \psi^a \,, \\
  (J^\mu)_a{}^b &=-i \left( \phi_a^\dagger \partial_\mu \phi^b - (\partial_\mu \phi_a^\dagger)  \phi^b \right) 
   + \psi^\dagger_a \gamma^\mu \psi^b - \frac{1}{4} \delta_a^b \psi^\dagger_c \gamma^\mu \psi^c \,, \\
  \text{etc.}& 
 }
It is easy to see that under \eqref{ZActionFree}, $j^\mu$ acquires a $-1$ factor, as implied by Table~\ref{chargeConjugation}.  To see whether $S_a{}^b$, $P_a{}^b$, $(J^\mu)_a{}^b$ transform in the expected way, we should represent these operators as rank-two anti-symmetric tensors of $SO(6)$.  This is done by defining 
 \es{AntisymmDef}{
   S^{IJ} = i S_a{}^{b} C^{[I}_{bc} \overline{C}^{J]ac} \,, 
 }
and similarly for $P$ and $J^\mu$, with the $C$ matrices given in \eqref{Cmat} and $\overline{C}$ being their complex conjugates.  (The $C$ and $\overline{C}$ matrices are the Clebsch-Gordan coefficients for the ${\bf 6}$ of $SU(4)$ in the products $\overline{\bf 4} \otimes \overline{\bf 4}$ and ${\bf 4} \otimes {\bf 4}$, respectively.)  One can check that \eqref{ZActionFree} implies
 \es{ZActiononSPJ}{
  S^{IJ} &\to -\cZ^{IK} \cZ^{JL} S^{KL} \,, \\
  P^{IJ} &\to -\cZ^{IK} \cZ^{JL} P^{KL} \,, \\
  (J^\mu)^{IJ} &\to \cZ^{IK} \cZ^{JL} (J^\mu)^{KL} \,, 
 }
as expected from Table~\ref{chargeConjugation}.  One can make similar checks for the other operators in the stress tensor multiplet.

One can ask whether $\cZ$ is a symmetry in ABJM theory as well, where the scalars $\phi^a$ and fermions $\psi^a$ are bifundamental fields transforming in the $({\bf N}, \overline{\bf N})$ of the $U(N)_k \times U(N)_{-k}$ gauge symmetry.  If the two gauge fields corresponding to the $U(N)$ factors are $A_{1\mu}$ and $A_{2 \mu}$, the action is invariant under $\cZ$ provided that $A_{1\mu} - A_{2 \mu}$ change sign under $\cZ$.  In the $N=1$ case, this can be accomplished by requiring $A_{1\mu} \to -A_{1\mu} $ and $A_{2\mu} \to - A_{2\mu}$ under $\cZ$, and one can check that the action (including the Chern-Simons terms) is invariant under this transformation.  Thus, $\cZ$ is a symmetry of the $U(1)_k \times U(1)_{-k}$ ABJM theory.  Such a transformation of $A_{1\mu}$ and $A_{2 \mu}$ does not leave the action invariant in the non-Abelian case due to the cubic terms in the Chern-Simons action.  In the non-Abelian case, however, one can consider sending $A_{1\mu} \leftrightarrow A_{2 \mu}$ under $\cZ$, which also has the effect of flipping the sign of $A_{1\mu} - A_{2\mu}$.  Under this transformation, the action stays unchanged with the only exception that $k \to -k$.  Thus, the $\cZ$ transformation is not a symmetry of the $U(N)_k \times U(N)_{-k}$ for $N>1$.

Note that $\cC \cT$ and $\cP$ are not separately symmetries either of ABJM theory with $k>1$, because they also send $k \to -k$.  However, the combination $\cP \cZ$ where $\cZ$ is assumed to interchange the two gauge fields in addition to acting as in \eqref{ZActionFree} becomes a new parity symmetry of ABJM theory \cite{Aharony:2008ug}.   To summarize, ABJM theory with $k=1$ preserves $\cC \cT$, $\cP$, $\cZ$ separately, while ABJM theory with $k>1$ preserves only $\cC \cP \cT$ and $\cP \cZ$ (or $\cC \cT \cZ$).

Having discussed $\cC \cT$, $\cP$, and $\cZ$, let us now argue that the 4-point superconformal invariants (i.e.~invariants under $OSp(6|4)$) can be classified as even or odd under $\cP$ (or $\cC \cT$) and $\cZ$.  This fact may not be immediately obvious, because it may happen that the superconformal Ward identities mix together $\cZ$-odd with $\cZ$-even structures.  The argument that this mixing does not occur is as follows.   As we have seen, the stress tensor multiplet naturally forms a representation of $(\mathbb Z_2\times \mathbb Z_2)\ltimes OSp(6|4)$, that is, the superconformal group extended by the action of parity $\cP$ (or $\cC \cT$) and also $\cZ$. As we shall see, this means that we can classify correlation functions of the stress tensor multiplet by their $\cP$ (or $\cC \cT$) and $\cZ$ transformation properties. It is a consequence of the following proposition:
\\

\noindent{\bf Proposition 1:} Let $H$ be a group, $G$ a normal subgroup of $H$, and $\bf a$ be a representation of $H$. Then the space $\mathcal V$ of $G$-invariant maps from $\bf a\rightarrow\bf 1$ forms a representation of $H/G$.

\noindent{\bf Proof:} The space of functionals from $\cF: \bf a\rightarrow 1$ naturally forms a representation of $H$. Since $\mathcal V$ is the space of $G$-invariant maps, a functional $V$ is in $\mathcal V$ if and only if
\begin{equation}\label{VCon}
gV = V \text{ for all } g\in G.
\end{equation}
Next we prove that for any $h\in H$, then $hV \in \mathcal V$. To do so, let us check that $hV$ satisfies \eqref{VCon}:
\begin{equation}
g(hV) = (hh^{-1}) ghV = h (h^{-1}gh) V = h g'V = hV,
\end{equation}
where we have used the fact that $G$ is a normal subgroup of $H$ to write $h^{-1}gh = g'$ for some $g'\in G.$ Hence $\mathcal V$ is a representation of $H$ for which $G$ acts trivially, and so we conclude that $\mathcal V$ is a representation of $H/G$.
\\

To apply this to $OSp(6|4)$, take $H$ to be the group $(\mathbb Z_2\times \mathbb Z_2)\ltimes OSp(6|4)$ group. We can think of the stress tensor multiplet as a superfield $\cS^a_{\ b}\left(x^\mu,\theta^\alpha_c\right)$ which forms a representation $\bf t$ of the group $H$. The $OSp(6|4)$ invariant structures in
$$\bigg\langle \cS^{a_1}_{\ b_1}\big(x_1^\mu,\theta_1^{\alpha_1 A_1}\big)\cS^{a_2}_{\ b_2}\big(x_2^\nu,\theta_2^{\alpha_2 A_2}\big)\ldots\cS^{a_n}_{\ b_n}\big(x_n^\nu,\theta_n^{\alpha_n A_n}\big)\bigg\rangle$$ 
are then maps from ${\bf t}^{\otimes n}\rightarrow \bf 1$, and so by proposition 1 can be classified by their representations under $H/G \approx \mathbb Z_2\times \mathbb Z_2$.

Analogous to the amplitudes case, the correlator $\langle SSSS\rangle$ is always invariant under $\cP$, $\cC \cT$, and $\cZ$ separately. This is also true for $\langle SSPP\rangle$ (and also $\langle PPPP\rangle$), so if we are only interested in the Ward identities relating these correlators, we can restrict to structures that are even under all of these transformations without loss of generality.  This also means it is impossible to check whether a theory is $\cP$ or $\cZ$ invariant from just $\langle SSSS\rangle$ without having more information about the theory.  If, however, we had some information about the spectrum of the theory, then we could potentially determine whether a theory is parity-preserving or not based on the conformal block expansion of $\langle SSSS \rangle$.  Without such extra assumptions, in order to see whether a theory is invariant under $\cP$ or $\cZ$, we would need to see whether the $\cP$-odd or $\cZ$-odd part of a correlator such as $\langle SSSJ^\mu\rangle$ vanishes.  The amplitudes calculation furthermore suggests that together $\langle SSSS\rangle$ and $\langle SSSJ^\mu\rangle$ should suffice to fix all four-point functions of the stress tensor multiplet operators. 

\section{Relating $\cA[\Phi\Phi\Psi\Psi]$ to $\langle SSSS\rangle$}
\label{AMPTOCFT}
In this appendix, we shall explain how to relate the superamplitude $\cA[\Phi\Phi\Psi\Psi]$ to the large $s, t$ behaviour of the Mellin amplitudes in $\langle SSSS\rangle$. The calculation proceeds in two steps. First we compute the amplitude $A[\phi_{ABCD}\phi_{EFGH}\overline\phi_{IJ}\overline\phi_{KL}]$, where we have made explicit the $SU(6)$ indices on $\phi$ and $\overline\phi$. We then relate $\phi$ and $\overline \phi$ to the CFT operator $S^{\ a}_{b}$, requiring us to convert the $SU(6)$ structures to $SO(6)$ structures.

To compute $A[\phi_{ABCD}\phi_{EFGH}\overline\phi_{IJ}\overline\phi_{KL}]$, we must differentiate $\cA[\Phi\Phi\Psi\Psi]$ with respect to the Grassmannian variables:
\begin{equation}\begin{split}
A[\phi_{ABCD}\phi_{EFGH}\overline\phi_{IJ}\overline\phi_{KL}] &= \frac{\partial}{\partial\eta_1^A} \cdots \frac{\partial}{\partial\eta_4^L}\cA[\Phi\Phi\Psi\Psi] \\
&= \frac{\partial}{\partial\eta_1^A} \cdots \frac{\partial}{\partial\eta_4^L} \delta^{12}(Q)\frac{[12]^4}{\langle34\rangle^2}f_1(s,t) \\
&= \frac{[12]^4}{2^4\langle34\rangle^2}f_1(s,t)\frac{\partial}{\partial\eta_1^A} \cdots \frac{\partial}{\partial\eta_4^L} \prod_{M = 1}^6\sum_{i,j=1}^4 \langle ij\rangle\eta_i^M\eta_j^M\\
\end{split}\end{equation}
To simplify the process of differentiating $\delta^{(12)}(Q)$, we can use $SU(6)$ invariance to expand
\begin{equation}\begin{split}
A&[\phi_{ABCD}\phi_{EFGH}\overline\phi_{IJ}\overline\phi_{KL}] \\
&= \epsilon_{ABCDIJ}\epsilon_{EFGHKL}F_1(s,t)+\epsilon_{ABCDKL}\epsilon_{EFGHIJ}F_2(s,t)+\epsilon_{ABEFIK}\epsilon_{CDGHJL}F_3(s,t).
\end{split}\end{equation}
for some functions $F_i(s,t)$. We can then choose specific numbers for each index $A$ through $L$ to isolate each structure, and hence to find that
\begin{equation}
F_1(s,t) = 2s^2u(4t-u)f_1(s,t)\,,\quad F_2(s,t) = 2s^2t(4u-t)f_1(s,t)\,\quad F_3(s,t) = -s^2tuf_1(s,t)\,.
\end{equation}

Now we must relate $A[\phi\phi\overline\phi\overline\phi]$ to $\langle SSSS\rangle.$ To do so, we can rewrite $S_a^{\ b}$ as an antisymmetric $6\times 6$ matrix:
\begin{equation}
\check S^{IJ} = S_a^{\ b} C^{[I}_{bc} \overline{C}^{J]ac} \,,
\end{equation}
where $C^{I}_{ac}$ are $SO(6)$ gamma matrices. Explicit expressions for these matrices are given in Appendix \ref{WARDAP}. Up to normalization, we then find that
\begin{equation}
\check S^{IJ} \underset{\text{flat space}}\longrightarrow \phi_{ABCD}\epsilon^{ABCDIJ} + \delta^{IA}\delta^{JB}\overline \phi_{AB}.
\end{equation}
This expression for $\check S^{IJ}$ breaks the $SU(6)$ symmetry down to $SO(6)$ due to the presence of the $\delta^{IA}$ symbol. Applying this to the four-point function, we find that
\begin{equation}\label{SCORtoA}
\langle \check S^{I_1J_1}\dots\check S^{I_4J_4}\rangle \underset{\text{flat space}}\longrightarrow \text{sum of contracted permutations of } A[\phi\phi\overline\phi\overline\phi].
\end{equation}
We must now expand our final answer in terms of the $SO(6)$ structures appearing in \eqref{SSSScor}. To do so we choose a series of polarization matrices $(X_i)^a_{\ b}$ and then define
\begin{equation}
\check X_i^{IJ} = (X_i)^a_{\ b} C^{[I}_{ac} \overline{C}^{J]bc} \,.
\end{equation}
Contracting both sides of  \eqref{SCORtoA} with matrices $X_i^{IJ}$, on the left-hand side we find that
\begin{equation}\begin{split}
\langle \check S^{I_1J_1}(\vec x_1)\dots\check S^{I_4J_4}(\vec x_4)\rangle \check X^{I_1J_1}_1 \cdots\check X^{I_4J_4}_4 &\propto \langle S(\vec x_1,X_1) \cdots S(\vec x_4,X_4)\rangle \\
&= \frac 1{x_{12}^2x_{34}^2} \left[\cS^1(U,V)A_{12}A_{34} +\dots+\cS^6(U,V)B_{1342}\right]\,.
\end{split}\end{equation}
We then Mellin transform and take the flat-space limit \eqref{FlatLimit} to find that
\begin{equation}
\langle \check S^{I_1J_1}(\vec x_1)\dots\check S^{I_4J_4}(\vec x_4)\rangle \check X^{I_1J_1}_1 \cdots \check X^{I_4J_4}_4\underset{\text{flat space}}\longrightarrow \frac {\cN}{x_{12}^2x_{34}^2} \left[\cA^1(s,t)A_{12}A_{34} +\dots+\cA^6(s,t)B_{1342}\right]
\end{equation}
for some overall normalization constant $\cN$. Computing the right-hand side of \eqref{SCORtoA} is more straightforward; we simply contract the $X_i^{I_iJ_i}$ matrices with the various permutations of $A[\phi\phi\overline\phi\overline\phi]$. By imposing \eqref{SCORtoA} for many differents matrices $(X_i)^a_{\ b}$ we can completely determine $\cA^i(s,t)$ in terms of $f_1(s,t)$, and upon choosing a suitable value for $\cN$ we can reproduce \eqref{flatSSSS}.

\section{Supersymmetric Ward identities}
\label{WARDAP}

\subsection{Stress tensor multiplet four-point functions}

To describe the supersymmetric variations which relate operators in the stress tensor multiplet, it will be convenient to introduce index-free notation to encode the $\mathfrak{so}(6)\approx \mathfrak{su}(4)$ representations which appear. We will use indices $I,J,\ldots$ for the $\bf 6$; and raised and lowered $a,b,\ldots$ indices for the $\bf 4$ and $\overline{\bf 4}$ as in section \ref{4POINT}. The gamma matrices $C^I_{ab}$ and ${\overline C}^{I a b}$ convert antisymmetric tensors of the $\bf 4$ and $\overline {\bf 4}$ into the $\bf 6$; a convenient basis for these matrices is:
\es{Cmat}{
C_1=&\begin{pmatrix} 0 & \sigma_1 \\ -\sigma_1 & 0 \\\end{pmatrix}\,,\qquad\quad\;\; C_2=\begin{pmatrix} 0 & -\sigma_3 \\ \sigma_3 & 0 \\\end{pmatrix}\,,\qquad\quad C_3=\begin{pmatrix} i\sigma_2 & 0 \\ 0 & i\sigma_2 \\\end{pmatrix}\,,\\ 
C_4=&-i\begin{pmatrix} 0 & i\sigma_2 \\ i\sigma_2 & 0 \\\end{pmatrix}\,,\qquad C_5=-i\begin{pmatrix} 0 & I_2 \\ -I_2 & 0 \\\end{pmatrix}\,,\qquad C_6=-i\begin{pmatrix} -i\sigma_2 & 0 \\ 0 & i\sigma_2 \\\end{pmatrix}\,,
}
where $\sigma_i$ are the Pauli matrices.

We can now describe operators in index-free notation as:
\begin{equation}
S(\vec x,X) = X^{\ b}_{a}S_b^{\ a}(\vec x)\,,\qquad F(\vec x,Y) = Y^{ab}F_{ab}(\vec x)\,,\qquad \chi_I(\vec x,Z) = Z^I\chi_I(\vec x)\,.
\end{equation}
with analogous notation for other operators in the stress tensor multiplet. To implement tracelessness of $S_b^{\ a}$ we impose the condition $X^{\ a}_{a} = 0$, and similarly we impose that the matrix $Y^{ab}$ is symmetric. We can alternative think of the matrix $X^{\ b}_{a}$ as an antisymmetric tensor $\check X^{IJ}$ via the mapping
\begin{equation}\check X^{IJ} = X^a_{\ b} C^{[I}_{ac} \overline{C}^{J]bc} \,.\end{equation}
Similarly, the $Z^I$ can also be written as antisymmetric tensors $\slashed Z_{ab} = C^I_{ab}Z_I$ and $\overline{\slashed Z}^{a b} = \overline C_I^{a b}Z^I$.

We can normalize our operators by defining their two-point functions, as we did for $S$ in \eqref{sNorm}:
\begin{equation}\begin{split}
\langle \chi^\alpha(\vec x_1,Z_1)\chi^\beta(\vec x_2,Z_2) \rangle &= (Z_1\cdot Z_2)\frac{i\slashed x_{12}}{x_{12}^4} \,, \\
\langle F^\alpha(\vec x_1,Y)\overline F^\beta(\vec x_2,\overline Y) \rangle &=  Y^{ab}\overline Y_{ab}\frac{i\slashed x_{12}}{x_{12}^4} \,, \\
\langle P(\vec x_1,X_1)P(\vec x_2,X_2)\rangle &= \frac{\text{Tr}(X_1X_2)}{x_{12}^4}  \,.
\end{split}\end{equation}

We can expand four point correlators as a sum over conformally invariant and $\mathfrak{so}(6)$ invariant structures. As explained in section \ref{WARDSMAIN} we restrict to those structures which are parity preserving and $\cC$ invariant. For instance,
\begin{equation}\begin{aligned}
\langle S(\vec x_1,X_1)S(\vec x_2,X_2)P(\vec x_3,X_3)&P(\vec x_4,X_4)\rangle\\
= \frac 1{x_{12}^2 x_{34}^4}&\Bigg[\cR^1(U,V)A_{12}A_{34}+\cR^2(U,V)A_{13}A_{24}+\cR^3(U,V)A_{14}A_{34} \\
&+ \cR^4(U,V)B_{1423} + \cR^5(U,V)B_{1234}+\cR^6(U,V)B_{1342}\Bigg] \label{SSPP} \,, 
\end{aligned}\end{equation}
where we define as in \eqref{ABRDef} the structures
\begin{equation}
A_{ij} = \tr(X_iX_j) \,, \qquad  B_{ijkl} = \tr(X_i X_j X_k X_l) + \tr(X_l X_k X_j X_i) \,.
 \end{equation}
Similarly, for the fermionic correlators we can expand:
\begin{equation}\begin{split}\label{sscc4}
\langle &S(\vec x_1,X_1)S(\vec x_2,X_2)\chi^\alpha(\vec x_3,Z_3)\chi^\beta(\vec x_4,Z_4)\rangle \\
&= \frac{i\slashed x_{34}^{\alpha\beta}}{x_{12}^2x_{34}^4}\bigg[\text{Tr}(X_1X_2)(Z_3\cdot Z_4)\cC^{1,1} + (Z_3\check X_1\check X_2Z_4) \cC^{2,1}+ (Z_3\check X_1\check X_2Z_4)\cC^{3,1}\bigg]\\
&+\frac{i(\slashed x_{13}\slashed x_{24}\slashed x_{12})^{\alpha\beta}}{2x_{12}^4x_{34}^4}\bigg[\text{Tr}(X_1X_2)(Z_3\cdot Z_4)\cC^{1,2} + (Z_3\check X_1\check X_2Z_4)\cC^{2,2}+ (Z_3\check X_1\check X_2Z_4)\cC^{3,2}\bigg]\,,\\
\\\langle &S(\vec x_1,X_1)S(\vec x_2,X_2)\chi^\alpha(\vec x_3,Z_3)F^\beta(\vec x_4,Y_4)\rangle\\
&=\frac{i\slashed x_{34}^{\alpha\beta}}{x_{12}^2x_{34}^4}\bigg[\text{Tr}(X_1X_2Y_4\slashed Z_3)\cE^{1,1}+\text{Tr}(X_2X_1Y_4\slashed Z_3)\cE^{2,1}+\text{Tr}(X_2Y_4X_1^T\slashed Z_3)\cE^{3,1}\bigg]\\
&+\frac{i(\slashed x_{13}\slashed x_{24}\slashed x_{12})^{\alpha\beta}}{2x_{12}^4x_{34}^4}\bigg[\text{Tr}(X_1X_2Y_4\slashed Z_3)\cE^{1,2}+\text{Tr}(X_2X_1Y_4\slashed Z_3)\cE^{2,2}+\text{Tr}(X_2Y_4X_1^T\slashed Z_3)\cE^{3,2}\bigg]\,,\\
\\\langle &S(\vec x_1,X_1)S(\vec x_2,X_2)F^\alpha(\vec x_3,Z_3) F^\beta(\vec x_4,Y_4)\rangle\\
&=\frac{i\slashed x_{34}^{\alpha\beta}}{x_{12}^2x_{34}^4}\bigg[\big(\epsilon_{abcd}(X_1)^a_{\ e}Y_1^{eb}(X_2)^c_{\ f}Y_2^{fd}\big)\cF^{1,1}+\big(\epsilon_{abcd}(X_1)^a_{\ e}Y_2^{eb}(X_2)^c_{\ f}Y_1^{fd}\big)\cF^{2,1}\bigg]\\
&+\frac{i(\slashed x_{13}\slashed x_{24}\slashed x_{12})^{\alpha\beta}}{2x_{12}^4x_{34}^4}\bigg[\big(\epsilon_{abcd}(X_1)^a_{\ e}Y_1^{eb}(X_2)^c_{\ f}Y_2^{fd}\big)\cF^{1,2}+\big(\epsilon_{abcd}(X_1)^a_{\ e}Y_2^{eb}(X_2)^c_{\ f}Y_1^{fd}\big)\cF^{2,2}\bigg]\,,\\
\\\langle &S(\vec x_1,X_1)S(\vec x_2,X_2)\overline F^\beta(\vec x_3,\overline Y_3)F^\alpha(\vec x_4,Y_4)\rangle\\
&=\frac{i\slashed x_{34}^{\alpha\beta}}{x_{12}^2x_{34}^4}\bigg[\text{Tr}(X_1X_2)\text{Tr}(Y_4\overline Y_3)\cG^{1,1}+\text{Tr}(Y_4\overline Y_3X_2X_1)\cG^{2,1}\\
&\hspace{70pt}+\text{Tr}(Y_4\overline Y_3X_2X_1)\cG^{3,1}+\text{Tr}(Y_4X_2^T\overline Y_3X_1)\cG^{4,1}\bigg]\\
&+\frac{i(\slashed x_{13}\slashed x_{24}\slashed x_{12})^{\alpha\beta}}{2x_{12}^4x_{34}^4}\bigg[\text{Tr}(X_1X_2)\text{Tr}(Y_4\overline Y_3)\cG^{1,2}+\text{Tr}(Y_4\overline Y_3X_2X_1)\cG^{2,2}\\
&\hspace{70pt}+\text{Tr}(Y_4\overline Y_3X_2X_1)\cG^{3,2}+\text{Tr}(Y_4X_2^T\overline Y_3X_1)\cG^{4,2}\bigg]\,.
\end{split}\end{equation}

\subsection{Ward identities}
As discussed in section \ref{WARDSMAIN}, to compute the supersymmetric Ward identities we need only the action of the Poincar\'e supercharges $Q^{\alpha I}$ on the operators in the stress tensor multiplet. Using the index free notation of the prevoius section, these variations can be written as
\begin{equation}\label{SUSYtrans}\begin{aligned}
\delta^{\alpha}(Z)S(\vec x,X) =&\ \frac 1 4\left[F^\alpha(\vec x, X\cdot \slashed Z) +\overline F^\alpha(\vec x, \overline{\slashed Z} \cdot X)\right] + \frac 14\chi^\alpha(\vec x,\check X\cdot Z) \,, \\
\delta^{\alpha}(Z)F^\beta(\vec x,Y) =&\ \frac 12\epsilon^{\alpha\beta}P(\vec x,Y\cdot \overline{\slashed Z}) +  \sigma^{\alpha\beta}_\mu J^\mu (\vec x,\slashed Z_1 \cdot \overline{\slashed Z}_2 - \slashed Z_2 \cdot \overline{\slashed Z}_1)  \\
&\ {}-\frac{i}{2}\sigma_\mu^{\alpha\beta}\partial^\mu S(\vec x, Y\cdot \overline{\slashed Z}) \\
\delta^{\alpha}(Z_1)\chi^\beta(\vec x,Z_2) =&\ \frac 12 \epsilon^{\alpha\beta}P(\vec x,\slashed Z_1 \cdot \overline{\slashed Z}_2 - \slashed Z_2 \cdot \overline{\slashed Z}_1) + Z_1\cdot Z_2\ i\sigma_\mu^{\alpha\beta}j^\mu(\vec x) \,, \\
&\ {}+\frac i 8\sigma_\mu^{\alpha\beta}\partial^\mu S(\vec x,\slashed Z_1 \cdot \overline{\slashed Z}_2 - \slashed Z_2 \cdot \overline{\slashed Z}_1) \,, \\
\delta^\alpha (Z) P(\vec x,X) =&\ \frac i 6\left(\sigma_\mu^{\alpha\beta}\partial^\mu F_\beta(\vec x,X\cdot \slashed Z)+\sigma_\mu^{\alpha\beta}\partial^\mu F_\beta(\vec x,\overline{\slashed Z} \cdot X)\right)\\
&\ {}- \frac i 6 \sigma_\mu^{\alpha\beta}\partial^\mu\chi_\beta(\vec x,\check X\cdot Z) \,, \\
&\text{etc.}
\end{aligned}\end{equation}
Here, $\delta^{\alpha}(Z)$ represents the action of $Z_I Q^{\alpha I}$ on the various operators, and $\sigma_\mu$ are the 3d gamma matrices, which 
we can take to be the Pauli matrices. We have omitted the supersymmetric variations of $J, j, \psi$, and $T$ as they are not needed in this work.
\allowdisplaybreaks

We will now give the Ward identities for two scalars and two fermions derived in Section~\ref{BULKLOC}. We will begin with $\langle SS\chi\chi\rangle$ and $\langle SS\chi F\rangle$, which can be derived from $\delta\langle SSS\chi\rangle$. We will omit those functions of the cross-ratios that are related to these under crossing.

The expressions for $\langle SS\chi\chi\rangle$ are:
\begin{align}\begin{autobreak}\label{Cward}
\cC^{1,1} = 
-\frac{1}{2 U}\Big(U^2 \partial_V\mathcal{S}^1(U,V)
+4 U^2 \partial_U\mathcal{S}^1(U,V)
+4 U^2 \partial_U\mathcal{S}^5(U,V)
+U (V-U) \partial_V\mathcal{S}^2(U,V)
+U (-U+V-1) \partial_U\mathcal{S}^2(U,V)
+U V \partial_V\mathcal{S}^3(U,V)
+U (U+V-1) \partial_U\mathcal{S}^3(U,V)
+2 U V \partial_V\mathcal{S}^4(U,V)
+2 U (V-1) \partial_U\mathcal{S}^4(U,V)
-4 U \mathcal{S}^1(U,V)
-3 U \mathcal{S}^5(U,V)
-U \mathcal{S}^6(U,V)
+(U-V+1) \mathcal{S}^2(U,V)
-(V-1) \mathcal{S}^3(U,V)
+(U-2 V+2) \mathcal{S}^4(U,V)\Big)
\end{autobreak}\,,\\
\begin{autobreak}
\cC^{2,1} =
 -\frac{1}{32 U}\Big(U^2 \partial_U\mathcal{S}^2(U,V)
 +U^2 \partial_U\mathcal{S}^3(U,V)
 -U^2 \partial_V\mathcal{S}^1(U,V)
 +U (U+V) \partial_V\mathcal{S}^2(U,V)
 +U V \partial_U\mathcal{S}^2(U,V)
 -U \partial_U\mathcal{S}^2(U,V)
 +U V \partial_V\mathcal{S}^3(U,V)
 +U V \partial_U\mathcal{S}^3(U,V)
 -U \partial_U\mathcal{S}^3(U,V)
 +2 U V \partial_V\mathcal{S}^4(U,V)
 +2 U V \partial_U\mathcal{S}^4(U,V)
 -2 U \partial_U\mathcal{S}^4(U,V)
 -U \mathcal{S}^2(U,V)
 +U \mathcal{S}^4(U,V)
 +U \mathcal{S}^5(U,V)
 -U \mathcal{S}^6(U,V)
 -V \mathcal{S}^2(U,V)
 +\mathcal{S}^2(U,V)
 -V \mathcal{S}^3(U,V)
 +\mathcal{S}^3(U,V)
 -2 V \mathcal{S}^4(U,V)
 +2 \mathcal{S}^4(U,V)\Big)
\end{autobreak}\,,\\\begin{autobreak}
\cC^{1,2} =
\frac{1}{2} (U ((3 V+1) \partial_V\mathcal{S}^1(U,V)
+3 U \partial_U\mathcal{S}^1(U,V)
-\partial_V\mathcal{S}^2(U,V)
-\partial_U\mathcal{S}^2(U,V)
+(U-1) \partial_U\mathcal{S}^3(U,V)
-2 \partial_U\mathcal{S}^4(U,V)
+V (\partial_V\mathcal{S}^3(U,V)
+4 \partial_V\mathcal{S}^5(U,V))
+4 U \partial_U\mathcal{S}^5(U,V))
+\mathcal{S}^2(U,V)
+\mathcal{S}^3(U,V)
+2 \mathcal{S}^4(U,V))
\end{autobreak}\,, \\ 
\begin{autobreak}
\cC^{2,2} =
\frac{1}{32} (U ((V-1) \partial_V\mathcal{S}^1(U,V)
+U \partial_U\mathcal{S}^1(U,V)
+\partial_V\mathcal{S}^2(U,V)
-\partial_U\mathcal{S}^2(U,V)
+V \partial_V\mathcal{S}^3(U,V)
+(U-1) \partial_U\mathcal{S}^3(U,V)
-2 \partial_U\mathcal{S}^4(U,V))
+\mathcal{S}^2(U,V)
+\mathcal{S}^3(U,V)
+2 \mathcal{S}^4(U,V))
\end{autobreak}\,.\end{align}

The expressions for $\langle SSF\chi\rangle$ are:
\begin{align}\begin{autobreak}
\cE^{1,1} = 
-V \partial_V\mathcal{S}^2(U,V)
-(V-1) \partial_U\mathcal{S}^2(U,V)
-V \partial_V\mathcal{S}^3(U,V)
-(U+V-1) \partial_U\mathcal{S}^3(U,V)
-2 V \partial_V\mathcal{S}^4(U,V)
-2 (V-1) \partial_U\mathcal{S}^4(U,V)
-2 U \partial_U\mathcal{S}^5(U,V)
+\frac{(V-1) \mathcal{S}^2(U,V)}{U}+\frac{(V-1) \mathcal{S}^3(U,V)}{U}+\mathcal{S}^5(U,V)
+\mathcal{S}^6(U,V)
-\frac{(U-2 V+2) \mathcal{S}^4(U,V)}{U}
\end{autobreak}\,,\\
\begin{autobreak}
\cE^{3,1} = 
-U (\partial_V\mathcal{S}^2(U,V)
+\partial_U\mathcal{S}^2(U,V)
-\partial_U\mathcal{S}^3(U,V))
+\mathcal{S}^2(U,V)
-\mathcal{S}^3(U,V)
\end{autobreak}\,,\\
\begin{autobreak}
\cE^{1,2} =
U (-\partial_U\mathcal{S}^2(U,V)
+V \partial_V\mathcal{S}^3(U,V)
+(U-1) \partial_U\mathcal{S}^3(U,V)
-2 \partial_U\mathcal{S}^4(U,V)
+2 V \partial_V\mathcal{S}^5(U,V)
+2 U \partial_U\mathcal{S}^5(U,V))
+\mathcal{S}^2(U,V)
+\mathcal{S}^3(U,V)
+2 \mathcal{S}^4(U,V)
\end{autobreak}\,,\\
\begin{autobreak}
\cE^{3,2} = 
U (\partial_V\mathcal{S}^2(U,V)
-V \partial_V\mathcal{S}^3(U,V)
-U \partial_U\mathcal{S}^3(U,V))
\end{autobreak}\,.\end{align}

Next we shall give expressions for $\langle SSFF\rangle$ and $\langle SS\overline FF\rangle$, which can be computed from $\delta\langle SSSF\rangle$. Unlike the previous correlators, we cannot completely fix these in terms of $\langle SSSS\rangle$. We will instead also leave $\cF^{1,1}(U,V)$ and $\cF^{2,1}(U,V)$ undetermined. We then find that the other components of $\langle SSFF\rangle$ are:
\begin{align}\begin{autobreak}\label{SSFFward}
\cF^{2,1}(U,V) =
\frac{1}{V}\Big(-4 U V \partial_V\mathcal{S}^4(U,V)
-4 U V \partial_U\mathcal{S}^4(U,V)
-2 (U-2 V) \mathcal{S}^4(U,V)
+(U-V) \mathcal{F}^{1,1}(U,V)
+\mathcal{F}^{1,2}(U,V)\Big)
\end{autobreak}\,,\\
\begin{autobreak}
\cF^{2,2}(U,V)=
-\frac{1}{V}\Big(U\big(-4 V \partial_V\mathcal{S}^4(U,V)
-2 \mathcal{S}^4(U,V)
+\mathcal{F}^{1,1}(U,V)\big)
+\mathcal{F}^{1,2}(U,V)\Big)
\end{autobreak}\,.\end{align}

Furthermore, by imposing conservation on $\langle SSSJ\rangle$, we find that $\cF^{1,1}(U,V)$ and $\cF^{2,1}(U,V)$ are constrained by the Ward identities:
\begin{align}\begin{autobreak}\label{SSFFward2}
\cF^{1,1}(U,V) =
\frac{1}{3 U}2 U^3 (U+2 V-2) \partial^2_U\mathcal{S}^1(U,V)
+2 U^2 V (U+2 V-2) \partial^2_V\mathcal{S}^1(U,V)
+U^2 (U+2 V-2) \partial_U\mathcal{S}^1(U,V)
+2 U^2 (U+V-1) (U+2 V-2) \partial_U\partial_V\mathcal{S}^1(U,V)
-2 U^2 (U+2 V-2) \partial^2_U\mathcal{S}^2(U,V)
-2 U^2 V (U+2 V-2) \partial^2_U\mathcal{S}^3(U,V)
+8 U^2 V (U-V+1) \partial^2_U\mathcal{S}^4(U,V)
-2 U V^2 (U+2 V-2) \partial^2_V\mathcal{S}^3(U,V)
+8 U V^2 (U-V+1) \partial^2_V\mathcal{S}^4(U,V)
+U (2 U-V+1) (U+2 V-2) \partial_V\mathcal{S}^1(U,V)
-2 U V (U+2 V-2) \partial^2_V\mathcal{S}^2(U,V)
+U (U+2 V-2) \partial_U\mathcal{S}^2(U,V)
-2 U (U+V-1) (U+2 V-2) \partial_U\partial_V\mathcal{S}^2(U,V)
-(U-1) U (U+2 V-2) \partial_U\mathcal{S}^3(U,V)
-2 U V (U+V-1) (U+2 V-2) \partial_U\partial_V\mathcal{S}^3(U,V)
+4 (U-1) U (U-V+1) \partial_U\mathcal{S}^4(U,V)
+8 U V (U-V+1) (U+V-1) \partial_U\partial_V\mathcal{S}^4(U,V)
-(U-2 V+2) (U+2 V-2) \partial_V\mathcal{S}^2(U,V)
+V (U+2 V-2) (-3 U+2 V-2) \partial_V\mathcal{S}^3(U,V)
+4 V (U-V+1) (3 U-2 V+2) \partial_V\mathcal{S}^4(U,V)
-2 U \Big(U^2-U (2 V+1)
+(V-1)^2\Big) \partial_U\mathcal{F}^{1,1}(U,V)
+\Big(U^2 (1-2 V)
+U (4 V+3) (V-1)
-2 (V-1)^3\Big) \partial_V\mathcal{F}^{1,1}(U,V)
+\Big(U^2-3 U (V+1)
+2 (V-1)^2\Big) \partial_V\mathcal{F}^{1,2}(U,V)
+U (-U+V-1) \partial_U\mathcal{F}^{1,2}(U,V)
-(U+2 V-2) \mathcal{S}^2(U,V)
-(U+2 V-2) \mathcal{S}^3(U,V)
+4 (U-V+1) \mathcal{S}^4(U,V)
\end{autobreak}\,,\\
\begin{autobreak}
\cF^{1,2}(U,V) =
\frac{1}{3} \Big(2 U^3 \partial^2_U\mathcal{S}^1(U,V)
+2 U^2 V \partial^2_V\mathcal{S}^1(U,V)
+U^2 \partial_U\mathcal{S}^1(U,V)
+2 U^2 (U+V-1) \partial_U\partial_V\mathcal{S}^1(U,V)
-2 U^2 \partial^2_U\mathcal{S}^2(U,V)
-2 U^2 V \partial^2_U\mathcal{S}^3(U,V)
-4 U^2 V \partial^2_U\mathcal{S}^4(U,V)
-2 U V^2 \partial^2_V\mathcal{S}^3(U,V)
-4 U V^2 \partial^2_V\mathcal{S}^4(U,V)
+U (2 U-V+1) \partial_V\mathcal{S}^1(U,V)
-2 U V \partial^2_V\mathcal{S}^2(U,V)
+U \partial_U\mathcal{S}^2(U,V)
-2 U (U+V-1) \partial_U\partial_V\mathcal{S}^2(U,V)
-(U-1) U \partial_U\mathcal{S}^3(U,V)
-2 U V (U+V-1) \partial_U\partial_V\mathcal{S}^3(U,V)
-2 (U-1) U \partial_U\mathcal{S}^4(U,V)
-4 U V (U+V-1) \partial_U\partial_V\mathcal{S}^4(U,V)
-(U-2 V+2) \partial_V\mathcal{S}^2(U,V)
+V (-3 U+2 V-2) \partial_V\mathcal{S}^3(U,V)
+2 V (-3 U+2 V-2) \partial_V\mathcal{S}^4(U,V)
+U (U-V+1) \partial_U\mathcal{F}^{1,1}(U,V)
+2 U \partial_U\mathcal{F}^{1,2}(U,V)
+\Big(U (V+1)
-(V-1)^2\Big) \partial_V\mathcal{F}^{1,1}(U,V)
+(U+V-1) \partial_V\mathcal{F}^{1,2}(U,V)
-\mathcal{S}^2(U,V)
-\mathcal{S}^3(U,V)
-2 \mathcal{S}^4(U,V)\Big)
\end{autobreak}\,.\end{align}

We also find the following expressions for $\langle SS\overline FF\rangle$:
\begin{align}\begin{autobreak}\label{SSFGward}
\cG^{1,1}(U,V) = 
\frac{1}{U}\Big(-2 U^2 \partial_U\mathcal{S}^1(U,V)
-4 U^2 \partial_U\mathcal{S}^5(U,V)
-2 U V \partial_V\mathcal{S}^3(U,V)
-2 U (U+V-1) \partial_U\mathcal{S}^3(U,V)
-4 U V \partial_V\mathcal{S}^4(U,V)
-4 U (V-1) \partial_U\mathcal{S}^4(U,V)
+2 U \mathcal{S}^1(U,V)
+2 U \mathcal{S}^5(U,V)
+2 (V-1) \mathcal{S}^3(U,V)
-2 (U-2 V+2) \mathcal{S}^4(U,V)
+(U-V+1) \mathcal{F}^{1,1}(U,V)
+\mathcal{F}^{1,2}(U,V)\Big)
\end{autobreak}\,,\\
\begin{autobreak}
\cG^{2,1}(U,V) =
\frac{1}{U}\Big(4 U^2 \partial_U\mathcal{S}^5(U,V)
-2 U V \partial_V\mathcal{S}^2(U,V)
-2 U (V-1) \partial_U\mathcal{S}^2(U,V)
+2 U V \partial_V\mathcal{S}^3(U,V)
+2 U (U+V-1) \partial_U\mathcal{S}^3(U,V)
+4 U V \partial_V\mathcal{S}^4(U,V)
+4 U (V-1) \partial_U\mathcal{S}^4(U,V)
-2 U \mathcal{S}^5(U,V)
+2 U \mathcal{S}^6(U,V)
+2 (V-1) \mathcal{S}^2(U,V)
-2 (V-1) \mathcal{S}^3(U,V)
+2 (U-2 V+2) \mathcal{S}^4(U,V)
-(U-2 V+2) \mathcal{F}^{1,1}(U,V)
-2 \mathcal{F}^{1,2}(U,V)\Big)
\end{autobreak}\,,\\
\begin{autobreak}
\cG^{4,1}(U,V) = 
\frac{1}{V}U\Big(2 V \Big(\partial_V\mathcal{S}^2(U,V)
+\partial_U\mathcal{S}^2(U,V)
+\partial_U\mathcal{S}^3(U,V)
+2 (\partial_V\mathcal{S}^4(U,V)
+\partial_U\mathcal{S}^4(U,V)))
+2 \mathcal{S}^4(U,V)
-\mathcal{F}^{1,1}(U,V)\Big)
-\mathcal{F}^{1,2}(U,V)\Big)-2 \Big(\mathcal{S}^2(U,V)
+\mathcal{S}^3(U,V)
+2 \mathcal{S}^4(U,V)\Big)
\end{autobreak}\,,\\
\begin{autobreak}
\cG^{1,2}(U,V) = 
2 U \Big(U \partial_U\mathcal{S}^1(U,V)
+(U-1) \partial_U\mathcal{S}^3(U,V)
-2 \partial_U\mathcal{S}^4(U,V)
+V (\partial_V\mathcal{S}^1(U,V)
+\partial_V\mathcal{S}^3(U,V)
+2 \partial_V\mathcal{S}^5(U,V))
+2 U \partial_U\mathcal{S}^5(U,V)\Big)
+2 \mathcal{S}^3(U,V)
+4 \mathcal{S}^4(U,V)
-\mathcal{F}^{1,1}(U,V)
\end{autobreak}\,,\\
\begin{autobreak}
\cG^{2,2}(U,V) =
-2 U \Big(\partial_U\mathcal{S}^2(U,V)
+V \partial_V\mathcal{S}^3(U,V)
+(U-1) \partial_U\mathcal{S}^3(U,V)
-2 \partial_U\mathcal{S}^4(U,V)
+2 V \partial_V\mathcal{S}^5(U,V)
+2 U \partial_U\mathcal{S}^5(U,V)\Big)
+2 \mathcal{S}^2(U,V)
-2 \mathcal{S}^3(U,V)
-4 \mathcal{S}^4(U,V)
+2 \mathcal{F}^{1,1}(U,V)
+\mathcal{F}^{1,2}(U,V)
\end{autobreak}\,,\\
\begin{autobreak}
\cG^{4,2}(U,V) = 
\frac{1}{V}\Big(U (-2 V (\partial_V\mathcal{S}^2(U,V)
+V \partial_V\mathcal{S}^3(U,V)
+U \partial_U\mathcal{S}^3(U,V)
+2 \partial_V\mathcal{S}^4(U,V))
-2 \mathcal{S}^4(U,V)
+\mathcal{F}^{1,1}(U,V))
+\mathcal{F}^{1,2}(U,V)\Big)
-\mathcal{F}^{1,2}(U,V)
\end{autobreak}\,.
\end{align}

Finally, in section \ref{INTCOR} we need Ward identities relating $\langle SSPP\rangle$ to $\langle SSSS\rangle$. These expressions can be derived by considering the supersymmetric variation $\delta\langle SSP\chi\rangle$:
\begin{align}\begin{autobreak}
\cR^1(U,V) =
2 V^2 \partial^2_V\mathcal{S}^2(U,V)
+2 V^2 \partial^2_V\mathcal{S}^3(U,V)
+4 V^2 \partial^2_V\mathcal{S}^4(U,V)
+2 V (U+V-1) \partial_U\partial_V\mathcal{S}^2(U,V)
+2 U V \partial^2_U\mathcal{S}^2(U,V)
-\frac{V (-3 U+2 V-2) \partial_V\mathcal{S}^3(U,V)}{U}
+2 V (U+V-1) \partial_U\partial_V\mathcal{S}^3(U,V)
+2 U V \partial^2_U\mathcal{S}^3(U,V)
+4 V (U+V-1) \partial_U\partial_V\mathcal{S}^4(U,V)
+4 U V \partial^2_U\mathcal{S}^4(U,V)
+2 V \partial_V\mathcal{S}^5(U,V)
-2 V \partial_V\mathcal{S}^6(U,V)
+\frac{V (U-2 V+2) \partial_V\mathcal{S}^2(U,V)}{U}
+\frac{4 V (U-V+1) \partial_V\mathcal{S}^4(U,V)}{U}
-U \partial_U\mathcal{S}^1(U,V)
-(V+1) \partial_U\mathcal{S}^2(U,V)
-(-U+V+1) \partial_U\mathcal{S}^3(U,V)
-2 (-U+V+1) \partial_U\mathcal{S}^4(U,V)
-2 U \partial_U\mathcal{S}^6(U,V)
+\mathcal{S}^1(U,V)
-\frac{(U-2 (V+1)) \mathcal{S}^4(U,V)}{U}
+\mathcal{S}^5(U,V)
+\mathcal{S}^6(U,V)
+\frac{(V+1) \mathcal{S}^2(U,V)}{U}
+\frac{(V+1) \mathcal{S}^3(U,V)}{U}
\end{autobreak} \label{RFirst}\,, \\
\begin{autobreak}
\cR^2(U,V) =
-U^2 \partial_U\mathcal{S}^1(U,V)
-2 U^2 V \partial^2_U\mathcal{S}^1(U,V)
-4 U^2 V \partial^2_U\mathcal{S}^5(U,V)
-2 U V^2 \partial^2_V\mathcal{S}^1(U,V)
-4 U V^2 \partial^2_V\mathcal{S}^5(U,V)
+2 V^2 \partial^2_V\mathcal{S}^2(U,V)
+2 V^2 \partial^2_V\mathcal{S}^3(U,V)
+4 V^2 \partial^2_V\mathcal{S}^4(U,V)
-2 U V (U+V-1) \partial_U\partial_V\mathcal{S}^1(U,V)
+2 U V \partial^2_U\mathcal{S}^2(U,V)
+2 U V \partial^2_U\mathcal{S}^3(U,V)
+4 U V \partial^2_U\mathcal{S}^4(U,V)
-2 (U-1) U \partial_U\mathcal{S}^5(U,V)
-4 U V (U+V-1) \partial_U\partial_V\mathcal{S}^5(U,V)
-2 U \partial_U\mathcal{S}^6(U,V)
-V (3 U-2 V+2) \partial_V\mathcal{S}^1(U,V)
-(V+1) \partial_U\mathcal{S}^2(U,V)
+2 V (U+V-1) \partial_U\partial_V\mathcal{S}^2(U,V)
-(-U+V+1) \partial_U\mathcal{S}^3(U,V)
+2 V (U+V-1) \partial_U\partial_V\mathcal{S}^3(U,V)
-2 (-U+V+1) \partial_U\mathcal{S}^4(U,V)
+4 V (U+V-1) \partial_U\partial_V\mathcal{S}^4(U,V)
-2 V (3 U-2 V+1) \partial_V\mathcal{S}^5(U,V)
-2 V \partial_V\mathcal{S}^6(U,V)
+\frac{V (U-2 V+2) \partial_V\mathcal{S}^2(U,V)}{U}
-\frac{V (-3 U+2 V-2) \partial_V\mathcal{S}^3(U,V)}{U}
+\frac{4 V (U-V+1) \partial_V\mathcal{S}^4(U,V)}{U}
-\mathcal{S}^5(U,V)
+\mathcal{S}^6(U,V)
+\frac{(V+1) \mathcal{S}^2(U,V)}{U}
+\frac{(V+1) \mathcal{S}^3(U,V)}{U}
-\frac{(U-2 (V+1)) \mathcal{S}^4(U,V)}{U}
\end{autobreak}\,,\\
\begin{autobreak}
\cR^4(U,V) =
-\frac{1}{2} \Big(2 \Big(-2 U^2-(U+3) V+U+2 V^2+1\Big) \partial_V\mathcal{S}^5(U,V)
+2 U^2 \partial_U\mathcal{S}^1(U,V)
+2 U^2 (2 U+V-1) \partial^2_U\mathcal{S}^1(U,V)
+4 U^2 (U+V-1) \partial^2_U\mathcal{S}^5(U,V)
+\Big(4 U^2+U-2 (V-1)^2\Big) \partial_V\mathcal{S}^1(U,V)
-\frac{\Big(U^2+U (4-3 V)
+2 (V-1)^2\Big) \partial_V\mathcal{S}^2(U,V)}{U}
-\frac{2 \Big(U^2+U-2 (V-1)^2\Big) \partial_V\mathcal{S}^4(U,V)}{U}
-\frac{\Big((U (2 U-1)
-4) V-2 (U+1) (U-1)^2-2 V^2\Big) \partial_V\mathcal{S}^3(U,V)}{U}
+2 U V (2 U+V-1) \partial^2_V\mathcal{S}^1(U,V)
+2 U (U+V-1) (2 U+V-1) \partial_U\partial_V\mathcal{S}^1(U,V)
-2 U (2 U+V-1) \partial^2_U\mathcal{S}^2(U,V)
-2 U \Big(V-(U-1)^2\Big) \partial^2_U\mathcal{S}^3(U,V)
-4 U (U+V-1) \partial^2_U\mathcal{S}^4(U,V)
+4 U V (U+V-1) \partial^2_V\mathcal{S}^5(U,V)
+2 (U-1) U \partial_U\mathcal{S}^5(U,V)
+4 U (U+V-1)^2 \partial_U\partial_V\mathcal{S}^5(U,V)
+2 U \partial_U\mathcal{S}^6(U,V)
-2 V (2 U+V-1) \partial^2_V\mathcal{S}^2(U,V)
+(3 U+V-1) \partial_U\mathcal{S}^2(U,V)
-2 (U+V-1) (2 U+V-1) \partial_U\partial_V\mathcal{S}^2(U,V)
-2 V \Big(V-(U-1)^2\Big) \partial^2_V\mathcal{S}^3(U,V)
+(U+V-1) \partial_U\mathcal{S}^3(U,V)
+2 \Big((U-1)^2-V\Big) (U+V-1) \partial_U\partial_V\mathcal{S}^3(U,V)
-4 V (U+V-1) \partial^2_V\mathcal{S}^4(U,V)
+2 (U+V-1) \partial_U\mathcal{S}^4(U,V)
-4 (U+V-1)^2 \partial_U\partial_V\mathcal{S}^4(U,V)
+2 (U+V-1) \partial_V\mathcal{S}^6(U,V)
+\mathcal{S}^5(U,V)
-\mathcal{S}^6(U,V)
-\frac{(3 U+V-1) \mathcal{S}^2(U,V)}{U}
-\frac{(2 U+V-1) \mathcal{S}^3(U,V)}{U}
-\frac{(3 U+2 V-2) \mathcal{S}^4(U,V)}{U}
\Big)
\end{autobreak} \,,\\ 
\begin{autobreak}
\cR^5(U,V) =
-\frac{1}{2} \Big(-2 U^2 \partial^2_U\mathcal{S}^1(U,V)
-2 V^2 \partial^2_V\mathcal{S}^3(U,V)
-4 V^2 \partial^2_V\mathcal{S}^4(U,V)
+2 U V \partial^2_V\mathcal{S}^1(U,V)
+2 U (U+V-1) \partial_U\partial_V\mathcal{S}^1(U,V)
-2 U (U+V) \partial^2_U\mathcal{S}^2(U,V)
-2 U V \partial^2_U\mathcal{S}^3(U,V)
-4 U V \partial^2_U\mathcal{S}^4(U,V)
+2 U \partial_U\mathcal{S}^6(U,V)
+(U-2 V+2) \partial_V\mathcal{S}^1(U,V)
-2 V (U+V) \partial^2_V\mathcal{S}^2(U,V)
+(U+V+1) \partial_U\mathcal{S}^2(U,V)
-2 (U+V-1) (U+V) \partial_U\partial_V\mathcal{S}^2(U,V)
-(U-V-1) \partial_U\mathcal{S}^3(U,V)
-2 V (U+V-1) \partial_U\partial_V\mathcal{S}^3(U,V)
-2 (U-V-1) \partial_U\mathcal{S}^4(U,V)
-4 V (U+V-1) \partial_U\partial_V\mathcal{S}^4(U,V)
-2 V \partial_V\mathcal{S}^5(U,V)
+2 V \partial_V\mathcal{S}^6(U,V)
-\frac{(U+V) (U-2 V+2) \partial_V\mathcal{S}^2(U,V)}{U}
-\frac{V (3 U-2 V+2) \partial_V\mathcal{S}^3(U,V)}{U}
-\frac{4 V (U-V+1) \partial_V\mathcal{S}^4(U,V)}{U}
-\mathcal{S}^5(U,V)
-\mathcal{S}^6(U,V)
-\frac{(U+V+1) \mathcal{S}^2(U,V)}{U}
-\frac{(V+1) \mathcal{S}^3(U,V)}{U}
-\frac{(-U+2 V+2) \mathcal{S}^4(U,V)}{U}
\Big)
\end{autobreak} \label{RLast}\,.
\end{align}

\section{Mellin amplitudes}
\label{MellinApp}

In this appendix we will first review how to convert supersymmetric Ward identities to position space. We will then describe the Mellin space formulation for four point functions of two scalars and two fermions, following \cite{Faller:2017hyt}.


\subsection{Ward identities in Mellin space}
\label{scalscalscalscal}

From the definition of the Mellin transform $M^i(s,t)$ of $\langle SSSS\rangle$ in \eqref{melDef}, we can derive the effect of multiplication by $U^mV^n$ and of differentiating with respect to $U$ and $V$:
\begin{equation}\begin{split}\label{3DMellin}
\widehat{U^mV^n}{ M}(s,t)=&{ M}(s-2m,t+2m+2n)\left(1-\frac{s}{2}\right)_m^2\left(1-\frac t 2\right)_{-m-n}^2\left(1-\frac u2\right)_{n}^2 \,, \\
\widehat{\partial^m_U}{ M}(s,t)=&\left(\frac s2+1-m\right)_m \widehat U^{-m} { M}(s,t)\,,\\
\widehat{\partial^m_V} { M}(s,t)=&{\left(\frac{u}{2}-m\right)_m}\widehat V^{-m} { M}(s,t)\,.
\end{split}\end{equation}

We can apply these rules to the position space Ward identities in \eqref{SSSSward}, so that they act on $M^i(s,t)$.

\subsection{Scalar-Scalar-Fermion-Fermion}
\label{scalscalfermferm}
Next, we consider the Mellin transform of the 4-point function $\langle SS\psi^\alpha \psi^\beta\rangle$ of dimension one scalar operators $S$ and dimension $\frac32$ spin half operators $\psi^\alpha$ and $\psi^\beta$ with spinor indices $\alpha,\beta=1,2$. We consider parity even four point functions, which contain two conformal structures:
\es{GsFerm}{
\langle S(\vec{x}_1)S(\vec{x}_2)\psi^\alpha(\vec{x}_3)\psi^\beta(\vec{x}_4)\rangle=\frac{i\slashed x_{34}^{\alpha\beta}}{x_{12}^2x_{34}^4}\mathcal{H}_1(U,V)+\frac{i(\slashed x_{13}\slashed x_{24}\slashed x_{12})^{\alpha\beta}}{2x_{12}^4x_{34}^4}\mathcal{H}_2(U,V)\,.
}
The Mellin transforms ${ M}_i^{SS\psi\psi}(s,t)$ of the connected parts of the correlators $\mathcal{H}^{SS\psi\psi}_{\text{conn},i}$ can then be defined by
\es{MellinDefFerm}{
\mathcal{H}_{\text{conn},1}^{SS\psi\psi}(U,V)=&\int_{-i\infty}^{i\infty}\frac{ds\, dt}{(4\pi i)^2} U^{\frac s2}V^{\frac u2-1}\Gamma\left[1-\frac s2\right]\Gamma\left[2-\frac s2\right]\Gamma^{2}\left[1-\frac t2\right]\Gamma^{2}\left[1-\frac u 2\right]{ M}^{SS\psi\psi}_1(s,t)\,,\\
\mathcal{H}_{\text{conn},2}^{SS\psi\psi}(U,V)=&\int_{-i\infty}^{i\infty}\frac{ds\, dt}{(4\pi i)^2} U^{\frac s2}V^{\frac u2-1}\Gamma^2\left[2-\frac s2\right]\Gamma^{2}\left[1 -\frac t2\right]\Gamma^{2}\left[1-\frac u2\right]{ M}^{SS\psi\psi}_2(s,t)\,,\\
}
where as previously we define $u = 4 - s - t$. These expression were derive in \cite{Faller:2017hyt} using $AdS_4$ Witten diagram calculations, where the arguments of the Gamma functions were chosen so that bulk contact Witten diagrams correspond to polynomial Mellin amplitudes.

Derivatives of $U$ and $V$ and powers of $U$ and $V$ in position space act on ${ M}^{SS\psi\psi}_i(s,t)$ according to the definition \eqref{MellinDefFerm} as 
\es{3DMellinFerm}{
\widehat{\partial^m_U} { M}^{SS\psi\psi}_i(s,t)=&{\left(\frac{s}{2}+1-m\right)_m}\widehat U^{-m} { M}^{SS\psi\psi}_i(s,t)\,,\\
\widehat{\partial^m_V} { M}^{SS\psi\psi}_i(s,t)=&{\left(\frac{u}{2}-m\right)_m}\widehat V^{-m} { M}^{SS\psi\psi}_i(s,t)\,,\\
\widehat{U^mV^n}{ M}^{SS\psi\psi}_1(s,t)=&{ M}^{SS\psi\psi}_1(s-2m,t+2m+2n)\left(1-\frac{s}{2}\right)_m\left(2-\frac{s}{2}\right)_m\left(1-\frac t2\right)_{-m-n}^2\left(1-\frac{u}{2}\right)_n^2\,,\\
\widehat{U^mV^n}{ M}^{SS\psi\psi}_2(s,t)=&{ M}^{SS\psi\psi}_2(s-2m,t+2m+2n)\left(2-\frac{s}{2}\right)^2_m\left(1-\frac{t}{2}\right)^2_{-m-n}\left(1-\frac{u}{2}\right)_n^2\,.
}

\section{Evaluating $I_{+-}[\cS^i]$ and $I_{++}[\cS^i]$}
\label{INTEGRALS}

In this appendix, we will describe how to evaluate $I_{++}[\cS^i(U,V)]$ and $I_{+-}[\cS^i(U,V)]$ using the Mellin transform $M^i(s,t)$ defined in \eqref{melDef}.  Each of these reduces to integrals over $s$ and $t$, which can be evaluated by summing all poles that appear in the contour defined in \eqref{contour}. In some cases, the pole summation can be easily done using the Barnes lemma
\begin{equation}\label{barnes}
\int_{-i\infty}^{i\infty}\frac{ds}{2\pi i}\Gamma(a+s)\Gamma(b+s)\Gamma(c-s)\Gamma(d-s) = \frac{\Gamma(a+c)\Gamma(b+d)\Gamma(b+c)\Gamma(b+d)}{\Gamma(a+b+c+d)} \,,
\end{equation}
which holds for contours for which the poles of each Gamma function lie either to the left or to the right of the contour.

\subsection{$I_{+-}[\cS^i]$}
\label{pmint}

We begin by writing $I_{+-}[{\cal S}^i]$ \eqref{Ipm} as an integral over $M^i(s,t)$:
\begin{equation}\begin{split}
I_{+-}[{\cal S}^i] & = \int_0^\infty dr\int_0^\pi d\theta\, \sin \theta \, \frac{{\cal S}^1 \left( 1 + r^2 - 2 r \cos \theta ,  r^2 \right) }{1 + r^2 - 2 r \cos \theta}\\
&= \int \frac{ds\ dt}{(4\pi i)^2}\Bigg(\Gamma^2\left[1-\frac s2\right]\Gamma^2\left[1-\frac t2\right]\Gamma^2\left[\frac {s+t-2}{2}\right] M^1(s,t) \\
& \hspace{50pt} \times \int_0^\infty dr\int_0^\pi d\theta\, \sin \theta \, \left( 1 + r^2 - 2 r \cos \theta \right)^{s/2-1} r^{2-s-t}\Bigg).
\end{split}\end{equation}
The integral of $r$ and $\theta$ can now be explicitly performed
to get
\begin{equation}\label{stInt}
\begin{split}
I_{+-}[{\cal S}^i] & = \int \frac{ds\ dt}{(4\pi i)^2} \frac{2\sqrt{\pi}}{(2-t)(s+t-2)}M^1(s,t) \\
&\times \Gamma \left[1-\frac{s}{2}\right] \Gamma \left[\frac{s+1}{2}\right] \Gamma \left[1-\frac{t}{2}\right] \Gamma \left[\frac{t-1}{2}\right] \Gamma \left[\frac {s+t-2}{2}\right] \Gamma \left[\frac{3-s-t}2\right]\,.
\end{split}\end{equation}

The polynomial Mellin amplitudes $M^1_3(s,t)$ \eqref{degree2} and $M^1_4(s,t)$ \eqref{R4Mellin} both equal $(t-2)(2-s-t)$ times a polynomial in $s,t$, so $I_{+-}[{\cal S}_n^i]$ can be evaluated for $n=3,4$ by writing the integrand as a sum of products of six Gamma functions in $s,t$ and then applying the Barnes lemma twice. For example, for $M^1_3(s,t)=(t-2)(2-s-t)$ we compute
\es{Ipm3}{
I_{+-}[{\cal S}^i_3] & = \int \frac{ds dt}{(4\pi i)^2} {2\sqrt{\pi}} \Gamma \left[1-\frac{s}{2}\right] \Gamma \left[\frac{s+1}{2}\right] \Gamma \left[1-\frac{t}{2}\right] \Gamma \left[\frac{t-1}{2}\right] \Gamma \left[\frac{2-s-t}{2}\right] \Gamma \left[\frac{3-s-t}2\right]\\
&=\int\frac{dt}{4\pi i}\pi ^{3/2} \Gamma \left[1-\frac{t}{2}\right] \Gamma
   \left[2-\frac{t}{2}\right] \Gamma \left[\frac{t-1}{2}\right]
   \Gamma \left[\frac{t}{2}\right]\\
   &=\frac{2\pi^2}{3}\,,
}
where the last two equalities followed from the Barnes lemma. We can evaluate $I_{+-}[{\cal S}^i_4] $ similarly to get the result in \eqref{ints34}.

The supergravity Mellin amplitude $M^1_1(s,t)$ \eqref{SugMellin} is also proportional to $(t-2)(2-s-t)$, but the remaining function is not a polynomial in $s,t$ and so we must work harder. We compute
\es{Ipm4}{
I_{+-}[{\cal S}^i_1] & = \int \frac{ds dt}{(4\pi i)^2}\frac{1}{4\pi^{2} s(2+s)}\left[\sqrt\pi(4+s)  \Gamma \left[1-\frac{s}{2}\right]-{4\Gamma\left[\frac{1-s}2\right]} \right] \\
&\qquad\times \Gamma \left[\frac{s+1}{2}\right] \Gamma \left[1-\frac{t}{2}\right] \Gamma \left[\frac{t-1}{2}\right] \Gamma \left[\frac {s+t-2}{2}\right] \Gamma \left[\frac{3-s-t}2\right]\\
 &= \int \frac{ds }{4\pi i}\frac{  \Gamma\left[1-\frac s2\right]\Gamma\left[\frac s2\right]  \Gamma \left[\frac{s+1}{2}\right]}{4\pi s(2+s)}\left[\sqrt\pi(4+s)  \Gamma \left[1-\frac{s}{2}\right]-{4\Gamma\left[\frac{1-s}2\right]} \right]\\
 &=-\pi^2\,,
}
where in the first equality we used the Barnes lemma, and in the second equality we summed over poles with the contour $0<\Re(s)<1$. Note that this contour is different from the range $0<\Re(s)<2$ that would follow form \eqref{contour}, since the supergravity term includes the stress tensor multiplet superblock, which contain extra poles that require a more constraining contour \cite{Dolan:2011dv}.

\subsection{$I_{++}[\cS^i]$}
\label{ppint}

$I_{++}[\cS^i]$ can be easily evaluated using Eq.~\eqref{IppRewrite3}.  For the polynomial Mellin amplitudes $M_3^i$ and $M_4^i$, the first term in \eqref{IppRewrite3} vanishes, and in the second term we have
 \es{M3484}{
  \lim_{s \to 2} \frac{M_{3, {\bf 84}}}{s-2} &= -\frac{1}{24} \,, \\
  \lim_{s \to 2} \frac{M_{3, {\bf 84}}}{s-2} &= -\frac{1}{5}  - \frac{3 t (t-2)}{56} \,.
 }
For the supergravity term, the first term in \eqref{IppRewrite3} gives $8 \pi/3$ and in the integrand of the second term we have
 \es{MSG84}{
  \lim_{s \to 2} \frac{M_{\text{SG}, {\bf 84}}}{s-2} &=
   \frac{(t-2) \Gamma \left( \frac{1-t}{2} \right)}{8 \sqrt{\pi} t (t + 2) \Gamma\left( 1 - \frac{t}{2} \right) } 
    - \frac{t^2 \Gamma \left( \frac{t-1}{2} \right)}{16 \sqrt{\pi} (t-2)(t-4) 
     \Gamma \left( 1 + \frac{t}{2} \right) } \\
      &{}- \frac{t^4 - 4 t^3 - 12 t^2 + 32 t - 32}{16 t (t-4)(t^2 - 4) } \,.
 }
Using \eqref{IppRewrite3}, we then obtain the results given in \eqref{ints34}.

\bibliographystyle{ssg}
\bibliography{N6draft}

\begingroup\raggedright\begin{thebibliography}{100}

\bibitem{Maldacena:1997re}
J.~M. Maldacena, ``{The Large $N$ limit of superconformal field theories and
  supergravity},'' {\em Int. J. Theor. Phys.} {\bf 38} (1999) 1113--1133,
  \href{http://xxx.lanl.gov/abs/hep-th/9711200}{{\tt hep-th/9711200}}. [Adv.
  Theor. Math. Phys.2,231(1998)].

\bibitem{Witten:1998qj}
E.~Witten, ``{Anti-de Sitter space and holography},'' {\em Adv. Theor. Math.
  Phys.} {\bf 2} (1998) 253--291,
  \href{http://xxx.lanl.gov/abs/hep-th/9802150}{{\tt hep-th/9802150}}.

\bibitem{Gubser:1998bc}
S.~S. Gubser, I.~R. Klebanov, and A.~M. Polyakov, ``{Gauge theory correlators
  from noncritical string theory},'' {\em Phys. Lett.} {\bf B428} (1998)
  105--114, \href{http://xxx.lanl.gov/abs/hep-th/9802109}{{\tt
  hep-th/9802109}}.

\bibitem{DHoker:1998vkc}
E.~D'Hoker, D.~Z. Freedman, and W.~Skiba, ``{Field theory tests for correlators
  in the AdS / CFT correspondence},'' {\em Phys. Rev.} {\bf D59} (1999) 045008,
  \href{http://xxx.lanl.gov/abs/hep-th/9807098}{{\tt hep-th/9807098}}.

\bibitem{Freedman:1998bj}
D.~Z. Freedman, S.~D. Mathur, A.~Matusis, and L.~Rastelli, ``{Comments on 4
  point functions in the CFT / AdS correspondence},'' {\em Phys. Lett.} {\bf
  B452} (1999) 61--68, \href{http://xxx.lanl.gov/abs/hep-th/9808006}{{\tt
  hep-th/9808006}}.

\bibitem{DHoker:1998bqu}
E.~D'Hoker and D.~Z. Freedman, ``{Gauge boson exchange in $AdS_{d+1}$},'' {\em
  Nucl. Phys.} {\bf B544} (1999) 612--632,
  \href{http://xxx.lanl.gov/abs/hep-th/9809179}{{\tt hep-th/9809179}}.

\bibitem{DHoker:1998ecp}
E.~D'Hoker and D.~Z. Freedman, ``{General scalar exchange in $AdS_{d+1}$},''
  {\em Nucl. Phys.} {\bf B550} (1999) 261--288,
  \href{http://xxx.lanl.gov/abs/hep-th/9811257}{{\tt hep-th/9811257}}.

\bibitem{DHoker:1999bve}
E.~D'Hoker, D.~Z. Freedman, S.~D. Mathur, A.~Matusis, and L.~Rastelli,
  ``{Graviton and gauge boson propagators in AdS(d+1)},'' {\em Nucl. Phys.}
  {\bf B562} (1999) 330--352,
  \href{http://xxx.lanl.gov/abs/hep-th/9902042}{{\tt hep-th/9902042}}.

\bibitem{DHoker:1999kzh}
E.~D'Hoker, D.~Z. Freedman, S.~D. Mathur, A.~Matusis, and L.~Rastelli,
  ``{Graviton exchange and complete four point functions in the AdS / CFT
  correspondence},'' {\em Nucl. Phys.} {\bf B562} (1999) 353--394,
  \href{http://xxx.lanl.gov/abs/hep-th/9903196}{{\tt hep-th/9903196}}.

\bibitem{DHoker:1999mqo}
E.~D'Hoker, D.~Z. Freedman, and L.~Rastelli, ``{AdS/CFT four point functions:
  How to succeed at $z$ integrals without really trying},'' {\em Nucl. Phys.}
  {\bf B562} (1999) 395--411,
  \href{http://xxx.lanl.gov/abs/hep-th/9905049}{{\tt hep-th/9905049}}.

\bibitem{DHoker:1999jke}
E.~D'Hoker, D.~Z. Freedman, S.~D. Mathur, A.~Matusis, and L.~Rastelli,
  ``{Extremal correlators in the AdS / CFT correspondence},''
  \href{http://xxx.lanl.gov/abs/hep-th/9908160}{{\tt hep-th/9908160}}.

\bibitem{Freedman:1998tz}
D.~Z. Freedman, S.~D. Mathur, A.~Matusis, and L.~Rastelli, ``{Correlation
  functions in the CFT(d)/AdS(d+1) correspondence},'' {\em Nucl. Phys.} {\bf
  B546} (1999) 96--118, \href{http://xxx.lanl.gov/abs/hep-th/9804058}{{\tt
  hep-th/9804058}}.

\bibitem{deHaro:2002vk}
S.~de~Haro, A.~Sinkovics, and K.~Skenderis, ``{On a supersymmetric completion
  of the R4 term in 2B supergravity},'' {\em Phys. Rev.} {\bf D67} (2003)
  084010, \href{http://xxx.lanl.gov/abs/hep-th/0210080}{{\tt hep-th/0210080}}.

\bibitem{Policastro:2006vt}
G.~Policastro and D.~Tsimpis, ``{$R^4$, purified},'' {\em Class. Quant. Grav.}
  {\bf 23} (2006) 4753--4780,
  \href{http://xxx.lanl.gov/abs/hep-th/0603165}{{\tt hep-th/0603165}}.

\bibitem{Paulos:2008tn}
M.~F. Paulos, ``{Higher derivative terms including the Ramond-Ramond
  five-form},'' {\em JHEP} {\bf 10} (2008) 047,
  \href{http://xxx.lanl.gov/abs/0804.0763}{{\tt 0804.0763}}.

\bibitem{Liu:2013dna}
J.~T. Liu and R.~Minasian, ``{Higher-derivative couplings in string theory:
  dualities and the B-field},'' {\em Nucl. Phys.} {\bf B874} (2013) 413--470,
  \href{http://xxx.lanl.gov/abs/1304.3137}{{\tt 1304.3137}}.

\bibitem{Goncalves:2014ffa}
V.~Gon{\c c}alves, ``{Four point function of $\mathcal{N}=4$ stress-tensor
  multiplet at strong coupling},'' {\em JHEP} {\bf 04} (2015) 150,
  \href{http://xxx.lanl.gov/abs/1411.1675}{{\tt 1411.1675}}.

\bibitem{Rastelli:2017ymc}
L.~Rastelli and X.~Zhou, ``{Holographic Four-Point Functions in the $(2, 0)$
  Theory},'' \href{http://xxx.lanl.gov/abs/1712.02788}{{\tt 1712.02788}}.

\bibitem{Rastelli:2017udc}
L.~Rastelli and X.~Zhou, ``{How to Succeed at Holographic Correlators Without
  Really Trying},'' \href{http://xxx.lanl.gov/abs/1710.05923}{{\tt
  1710.05923}}.

\bibitem{Rastelli:2016nze}
L.~Rastelli and X.~Zhou, ``{Mellin amplitudes for $AdS_5\times S^5$},'' {\em
  Phys. Rev. Lett.} {\bf 118} (2017), no.~9 091602,
  \href{http://xxx.lanl.gov/abs/1608.06624}{{\tt 1608.06624}}.

\bibitem{Binder:2019jwn}
D.~J. Binder, S.~M. Chester, S.~S. Pufu, and Y.~Wang, ``{$\mathcal{N}=4$
  Super-Yang-Mills Correlators at Strong Coupling from String Theory and
  Localization},'' \href{http://xxx.lanl.gov/abs/1902.06263}{{\tt 1902.06263}}.

\bibitem{Zhou:2017zaw}
X.~Zhou, ``{On Superconformal Four-Point Mellin Amplitudes in Dimension
  $d>2$},'' \href{http://xxx.lanl.gov/abs/1712.02800}{{\tt 1712.02800}}.

\bibitem{Chester:2018aca}
S.~M. Chester, S.~S. Pufu, and X.~Yin, ``{The M-Theory S-Matrix From ABJM:
  Beyond 11D Supergravity},'' \href{http://xxx.lanl.gov/abs/1804.00949}{{\tt
  1804.00949}}.

\bibitem{Binder:2018yvd}
D.~J. Binder, S.~M. Chester, and S.~S. Pufu, ``{Absence of $D^4 R^4$ in
  M-Theory From ABJM},'' \href{http://xxx.lanl.gov/abs/1808.10554}{{\tt
  1808.10554}}.

\bibitem{Chester:2018dga}
S.~M. Chester and E.~Perlmutter, ``{M-Theory Reconstruction from (2,0) CFT and
  the Chiral Algebra Conjecture},''
  \href{http://xxx.lanl.gov/abs/1805.00892}{{\tt 1805.00892}}.

\bibitem{Giusto:2019pxc}
S.~Giusto, R.~Russo, A.~Tyukov, and C.~Wen, ``{Holographic correlators in
  AdS$_3$ without Witten diagrams},''
  \href{http://xxx.lanl.gov/abs/1905.12314}{{\tt 1905.12314}}.

\bibitem{Rastelli:2019gtj}
L.~Rastelli, K.~Roumpedakis, and X.~Zhou, ``{$AdS_3\times S^3$ Tree-Level
  Correlators: Hidden Six-Dimensional Conformal Symmetry},''
  \href{http://xxx.lanl.gov/abs/1905.11983}{{\tt 1905.11983}}.

\bibitem{Giusto:2018ovt}
S.~Giusto, R.~Russo, and C.~Wen, ``{Holographic correlators in AdS$_{3}$},''
  {\em JHEP} {\bf 03} (2019) 096,
  \href{http://xxx.lanl.gov/abs/1812.06479}{{\tt 1812.06479}}.

\bibitem{Zhou:2018ofp}
X.~Zhou, ``{On Mellin Amplitudes in SCFTs with Eight Supercharges},'' {\em
  JHEP} {\bf 07} (2018) 147, \href{http://xxx.lanl.gov/abs/1804.02397}{{\tt
  1804.02397}}.

\bibitem{Goncalves:2019znr}
V.~Gon{\c c}alves, R.~Pereira, and X.~Zhou, ``{${\bf 20}'$ Five-Point Function
  from $AdS_5\times S^5$ Supergravity},''
  \href{http://xxx.lanl.gov/abs/1906.05305}{{\tt 1906.05305}}.

\bibitem{Pestun:2007rz}
V.~Pestun, ``{Localization of gauge theory on a four-sphere and supersymmetric
  Wilson loops},'' {\em Commun. Math. Phys.} {\bf 313} (2012) 71--129,
  \href{http://xxx.lanl.gov/abs/0712.2824}{{\tt 0712.2824}}.

\bibitem{Kapustin:2009kz}
A.~Kapustin, B.~Willett, and I.~Yaakov, ``{Exact Results for Wilson Loops in
  Superconformal Chern-Simons Theories with Matter},'' {\em JHEP} {\bf 1003}
  (2010) 089, \href{http://xxx.lanl.gov/abs/0909.4559}{{\tt 0909.4559}}.

\bibitem{Polchinski:1999ry}
J.~Polchinski, ``{S matrices from AdS space-time},''
  \href{http://xxx.lanl.gov/abs/hep-th/9901076}{{\tt hep-th/9901076}}.

\bibitem{Susskind:1998vk}
L.~Susskind, ``{Holography in the flat space limit},'' {\em AIP Conf. Proc.}
  {\bf 493} (1999), no.~1 98--112,
  \href{http://xxx.lanl.gov/abs/hep-th/9901079}{{\tt hep-th/9901079}}.

\bibitem{Giddings:1999jq}
S.~B. Giddings, ``{Flat space scattering and bulk locality in the AdS / CFT
  correspondence},'' {\em Phys. Rev.} {\bf D61} (2000) 106008,
  \href{http://xxx.lanl.gov/abs/hep-th/9907129}{{\tt hep-th/9907129}}.

\bibitem{Penedones:2010ue}
J.~Penedones, ``{Writing CFT correlation functions as AdS scattering
  amplitudes},'' {\em JHEP} {\bf 03} (2011) 025,
  \href{http://xxx.lanl.gov/abs/1011.1485}{{\tt 1011.1485}}.

\bibitem{Fitzpatrick:2011hu}
A.~L. Fitzpatrick and J.~Kaplan, ``{Analyticity and the Holographic
  S-Matrix},'' {\em JHEP} {\bf 10} (2012) 127,
  \href{http://xxx.lanl.gov/abs/1111.6972}{{\tt 1111.6972}}.

\bibitem{Fitzpatrick:2011jn}
A.~L. Fitzpatrick and J.~Kaplan, ``{Scattering States in AdS/CFT},''
  \href{http://xxx.lanl.gov/abs/1104.2597}{{\tt 1104.2597}}.

\bibitem{Aharony:2008ug}
O.~Aharony, O.~Bergman, D.~L. Jafferis, and J.~Maldacena, ``{${\cal N}=6$
  superconformal Chern-Simons-matter theories, M2-branes and their gravity
  duals},'' {\em JHEP} {\bf 10} (2008) 091,
  \href{http://xxx.lanl.gov/abs/0806.1218}{{\tt 0806.1218}}.

\bibitem{Aharony:2008gk}
O.~Aharony, O.~Bergman, and D.~L. Jafferis, ``{Fractional M2-branes},'' {\em
  JHEP} {\bf 0811} (2008) 043, \href{http://xxx.lanl.gov/abs/0807.4924}{{\tt
  0807.4924}}.

\bibitem{Green:1997di}
M.~B. Green and P.~Vanhove, ``{D instantons, strings and M theory},'' {\em
  Phys. Lett.} {\bf B408} (1997) 122--134,
  \href{http://xxx.lanl.gov/abs/hep-th/9704145}{{\tt hep-th/9704145}}.

\bibitem{Nekrasov:2002qd}
N.~A. Nekrasov, ``{Seiberg-Witten prepotential from instanton counting},'' {\em
  Adv. Theor. Math. Phys.} {\bf 7} (2003), no.~5 831--864,
  \href{http://xxx.lanl.gov/abs/hep-th/0206161}{{\tt hep-th/0206161}}.

\bibitem{Nekrasov:2003rj}
N.~Nekrasov and A.~Okounkov, ``{Seiberg-Witten theory and random partitions},''
  {\em Prog. Math.} {\bf 244} (2006) 525--596,
  \href{http://xxx.lanl.gov/abs/hep-th/0306238}{{\tt hep-th/0306238}}.

\bibitem{Losev:1997tp}
A.~Losev, N.~Nekrasov, and S.~L. Shatashvili, ``{Issues in topological gauge
  theory},'' {\em Nucl. Phys.} {\bf B534} (1998) 549--611,
  \href{http://xxx.lanl.gov/abs/hep-th/9711108}{{\tt hep-th/9711108}}.

\bibitem{Moore:1997dj}
G.~W. Moore, N.~Nekrasov, and S.~Shatashvili, ``{Integrating over Higgs
  branches},'' {\em Commun. Math. Phys.} {\bf 209} (2000) 97--121,
  \href{http://xxx.lanl.gov/abs/hep-th/9712241}{{\tt hep-th/9712241}}.

\bibitem{Dolan:2008vc}
F.~Dolan, ``{On Superconformal Characters and Partition Functions in Three
  Dimensions},'' {\em J.Math.Phys.} {\bf 51} (2010) 022301,
  \href{http://xxx.lanl.gov/abs/0811.2740}{{\tt 0811.2740}}.

\bibitem{Liendo:2015cgi}
P.~Liendo, C.~Meneghelli, and V.~Mitev, ``{On Correlation Functions of BPS
  Operators in 3d ${\mathcal{N}}$ = 6 Superconformal Theories},'' {\em Commun.
  Math. Phys.} {\bf 350} (2017), no.~1 387--419,
  \href{http://xxx.lanl.gov/abs/1512.06072}{{\tt 1512.06072}}.

\bibitem{Cordova:2016emh}
C.~Cordova, T.~T. Dumitrescu, and K.~Intriligator, ``{Multiplets of
  Superconformal Symmetry in Diverse Dimensions},'' {\em JHEP} {\bf 03} (2019)
  163, \href{http://xxx.lanl.gov/abs/1612.00809}{{\tt 1612.00809}}.

\bibitem{Chester:2014fya}
S.~M. Chester, J.~Lee, S.~S. Pufu, and R.~Yacoby, ``{The $ \mathcal{N}=8 $
  superconformal bootstrap in three dimensions},'' {\em JHEP} {\bf 09} (2014)
  143, \href{http://xxx.lanl.gov/abs/1406.4814}{{\tt 1406.4814}}.

\bibitem{Jafferis:2010un}
D.~L. Jafferis, ``{The Exact Superconformal R-Symmetry Extremizes $Z$},'' {\em
  JHEP} {\bf 1205} (2012) 159, \href{http://xxx.lanl.gov/abs/1012.3210}{{\tt
  1012.3210}}.

\bibitem{Closset:2012vg}
C.~Closset, T.~T. Dumitrescu, G.~Festuccia, Z.~Komargodski, and N.~Seiberg,
  ``{Contact Terms, Unitarity, and F-Maximization in Three-Dimensional
  Superconformal Theories},'' {\em JHEP} {\bf 1210} (2012) 053,
  \href{http://xxx.lanl.gov/abs/1205.4142}{{\tt 1205.4142}}.

\bibitem{Green:1997as}
M.~B. Green, M.~Gutperle, and P.~Vanhove, ``{One loop in eleven-dimensions},''
  {\em Phys. Lett.} {\bf B409} (1997) 177--184,
  \href{http://xxx.lanl.gov/abs/hep-th/9706175}{{\tt hep-th/9706175}}.
  [,164(1997)].

\bibitem{Russo:1997mk}
J.~G. Russo and A.~A. Tseytlin, ``{One loop four graviton amplitude in
  eleven-dimensional supergravity},'' {\em Nucl. Phys.} {\bf B508} (1997)
  245--259, \href{http://xxx.lanl.gov/abs/hep-th/9707134}{{\tt
  hep-th/9707134}}.

\bibitem{Green:1998by}
M.~B. Green and S.~Sethi, ``{Supersymmetry constraints on type IIB
  supergravity},'' {\em Phys. Rev.} {\bf D59} (1999) 046006,
  \href{http://xxx.lanl.gov/abs/hep-th/9808061}{{\tt hep-th/9808061}}.

\bibitem{Green:2005ba}
M.~B. Green and P.~Vanhove, ``{Duality and higher derivative terms in M
  theory},'' {\em JHEP} {\bf 01} (2006) 093,
  \href{http://xxx.lanl.gov/abs/hep-th/0510027}{{\tt hep-th/0510027}}.

\bibitem{Pioline:2015yea}
B.~Pioline, ``{$D^{6}R^{4}$ amplitudes in various dimensions},'' {\em JHEP}
  {\bf 04} (2015) 057, \href{http://xxx.lanl.gov/abs/1502.03377}{{\tt
  1502.03377}}.

\bibitem{Green:1997ud}
M.~B. Green, ``{Connections between M theory and superstrings},'' {\em Nucl.
  Phys. Proc. Suppl.} {\bf 68} (1998) 242--251,
  \href{http://xxx.lanl.gov/abs/hep-th/9712195}{{\tt hep-th/9712195}}.
  [,242(1997)].

\bibitem{Polchinski:1998rr}
J.~Polchinski, {\em {String theory. Vol. 2: Superstring theory and beyond}}.
\newblock Cambridge University Press, 2007.

\bibitem{Green:2008uj}
M.~B. Green, J.~G. Russo, and P.~Vanhove, ``{Low energy expansion of the
  four-particle genus-one amplitude in type II superstring theory},'' {\em
  JHEP} {\bf 02} (2008) 020, \href{http://xxx.lanl.gov/abs/0801.0322}{{\tt
  0801.0322}}.

\bibitem{Gomez:2010ad}
H.~Gomez and C.~R. Mafra, ``{The Overall Coefficient of the Two-loop
  Superstring Amplitude Using Pure Spinors},'' {\em JHEP} {\bf 05} (2010) 017,
  \href{http://xxx.lanl.gov/abs/1003.0678}{{\tt 1003.0678}}.

\bibitem{DHoker:2005jhf}
E.~D'Hoker, M.~Gutperle, and D.~H. Phong, ``{Two-loop superstrings and
  S-duality},'' {\em Nucl. Phys.} {\bf B722} (2005) 81--118,
  \href{http://xxx.lanl.gov/abs/hep-th/0503180}{{\tt hep-th/0503180}}.

\bibitem{Gomez:2013sla}
H.~Gomez and C.~R. Mafra, ``{The closed-string 3-loop amplitude and
  S-duality},'' {\em JHEP} {\bf 10} (2013) 217,
  \href{http://xxx.lanl.gov/abs/1308.6567}{{\tt 1308.6567}}.

\bibitem{Alday:2018kkw}
L.~F. Alday, ``{On Genus-one String Amplitudes on $AdS_5 \times S^5$},''
  \href{http://xxx.lanl.gov/abs/1812.11783}{{\tt 1812.11783}}.

\bibitem{Alday:2017vkk}
L.~F. Alday and S.~Caron-Huot, ``{Gravitational S-matrix from CFT dispersion
  relations},'' {\em JHEP} {\bf 12} (2018) 017,
  \href{http://xxx.lanl.gov/abs/1711.02031}{{\tt 1711.02031}}.

\bibitem{Alday:2018pdi}
L.~F. Alday, A.~Bissi, and E.~Perlmutter, ``{Genus-One String Amplitudes from
  Conformal Field Theory},'' \href{http://xxx.lanl.gov/abs/1809.10670}{{\tt
  1809.10670}}.

\bibitem{Bashkirov:2011fr}
D.~Bashkirov, ``{A Note on ${\cal N}\ge 6$ Superconformal Quantum Field
  Theories in three dimensions},''
  \href{http://xxx.lanl.gov/abs/1108.4081}{{\tt 1108.4081}}.

\bibitem{Nosaka:2015iiw}
T.~Nosaka, ``{Instanton effects in ABJM theory with general R-charge
  assignments},'' {\em JHEP} {\bf 03} (2016) 059,
  \href{http://xxx.lanl.gov/abs/1512.02862}{{\tt 1512.02862}}.

\bibitem{Marino:2011eh}
M.~Marino and P.~Putrov, ``{ABJM theory as a Fermi gas},'' {\em J. Stat. Mech.}
  {\bf 1203} (2012) P03001, \href{http://xxx.lanl.gov/abs/1110.4066}{{\tt
  1110.4066}}.

\bibitem{Kos:2013tga}
F.~Kos, D.~Poland, and D.~Simmons-Duffin, ``{Bootstrapping the $O(N)$ vector
  models},'' {\em JHEP} {\bf 06} (2014) 091,
  \href{http://xxx.lanl.gov/abs/1307.6856}{{\tt 1307.6856}}.

\bibitem{Mack:2009mi}
G.~Mack, ``{D-independent representation of Conformal Field Theories in D
  dimensions via transformation to auxiliary Dual Resonance Models. Scalar
  amplitudes},'' \href{http://xxx.lanl.gov/abs/0907.2407}{{\tt 0907.2407}}.

\bibitem{Elvang:2010jv}
H.~Elvang, D.~Z. Freedman, and M.~Kiermaier, ``{A simple approach to
  counterterms in ${\cal N}=8$ supergravity},'' {\em JHEP} {\bf 11} (2010) 016,
  \href{http://xxx.lanl.gov/abs/1003.5018}{{\tt 1003.5018}}.

\bibitem{Elvang:2010xn}
H.~Elvang, D.~Z. Freedman, and M.~Kiermaier, ``{SUSY Ward identities,
  Superamplitudes, and Counterterms},'' {\em J. Phys.} {\bf A44} (2011) 454009,
  \href{http://xxx.lanl.gov/abs/1012.3401}{{\tt 1012.3401}}.

\bibitem{Wang:2015jna}
Y.~Wang and X.~Yin, ``{Constraining Higher Derivative Supergravity with
  Scattering Amplitudes},'' {\em Phys. Rev.} {\bf D92} (2015), no.~4 041701,
  \href{http://xxx.lanl.gov/abs/1502.03810}{{\tt 1502.03810}}.

\bibitem{Wang:2015aua}
Y.~Wang and X.~Yin, ``{Supervertices and Non-renormalization Conditions in
  Maximal Supergravity Theories},''
  \href{http://xxx.lanl.gov/abs/1505.05861}{{\tt 1505.05861}}.

\bibitem{Elvang:2011fx}
H.~Elvang, Y.-t. Huang, and C.~Peng, ``{On-shell superamplitudes in N<4 SYM},''
  {\em JHEP} {\bf 09} (2011) 031, \href{http://xxx.lanl.gov/abs/1102.4843}{{\tt
  1102.4843}}.

\bibitem{Freedman:2018mrv}
D.~Z. Freedman, R.~Kallosh, and Y.~Yamada, ``{Duality Constraints on
  Counterterms in Supergravities},'' {\em Fortsch. Phys.} {\bf 66} (2018),
  no.~10 1800054, \href{http://xxx.lanl.gov/abs/1807.06704}{{\tt 1807.06704}}.

\bibitem{Elvang:2015rqa}
H.~Elvang and Y.-t. Huang, {\em {Scattering Amplitudes in Gauge Theory and
  Gravity}}.
\newblock Cambridge University Press, 2015.

\bibitem{McGady:2013sga}
D.~A. McGady and L.~Rodina, ``{Higher-spin massless $S$-matrices in
  four-dimensions},'' {\em Phys. Rev.} {\bf D90} (2014), no.~8 084048,
  \href{http://xxx.lanl.gov/abs/1311.2938}{{\tt 1311.2938}}.

\bibitem{Hama:2011ea}
N.~Hama, K.~Hosomichi, and S.~Lee, ``{SUSY Gauge Theories on Squashed
  Three-Spheres},'' {\em JHEP} {\bf 1105} (2011) 014,
  \href{http://xxx.lanl.gov/abs/1102.4716}{{\tt 1102.4716}}.

\bibitem{Beem:2013sza}
C.~Beem, M.~Lemos, P.~Liendo, W.~Peelaers, L.~Rastelli, and B.~C. van Rees,
  ``{Infinite Chiral Symmetry in Four Dimensions},'' {\em Commun. Math. Phys.}
  {\bf 336} (2015), no.~3 1359--1433,
  \href{http://xxx.lanl.gov/abs/1312.5344}{{\tt 1312.5344}}.

\bibitem{Chester:2014mea}
S.~M. Chester, J.~Lee, S.~S. Pufu, and R.~Yacoby, ``{Exact Correlators of BPS
  Operators from the 3d Superconformal Bootstrap},'' {\em JHEP} {\bf 03} (2015)
  130, \href{http://xxx.lanl.gov/abs/1412.0334}{{\tt 1412.0334}}.

\bibitem{Beem:2016cbd}
C.~Beem, W.~Peelaers, and L.~Rastelli, ``{Deformation quantization and
  superconformal symmetry in three dimensions},'' {\em Commun. Math. Phys.}
  {\bf 354} (2017), no.~1 345--392,
  \href{http://xxx.lanl.gov/abs/1601.05378}{{\tt 1601.05378}}.

\bibitem{Dedushenko:2016jxl}
M.~Dedushenko, S.~S. Pufu, and R.~Yacoby, ``{A one-dimensional theory for Higgs
  branch operators},'' \href{http://xxx.lanl.gov/abs/1610.00740}{{\tt
  1610.00740}}.

\bibitem{Dedushenko:2017avn}
M.~Dedushenko, Y.~Fan, S.~S. Pufu, and R.~Yacoby, ``{Coulomb Branch Operators
  and Mirror Symmetry in Three Dimensions},''
  \href{http://xxx.lanl.gov/abs/1712.09384}{{\tt 1712.09384}}.

\bibitem{Dedushenko:2018icp}
M.~Dedushenko, Y.~Fan, S.~S. Pufu, and R.~Yacoby, ``{Coulomb Branch
  Quantization and Abelianized Monopole Bubbling},''
  \href{http://xxx.lanl.gov/abs/1812.08788}{{\tt 1812.08788}}.

\bibitem{Agmon:2017xes}
N.~B. Agmon, S.~M. Chester, and S.~S. Pufu, ``{Solving M-theory with the
  Conformal Bootstrap},'' \href{http://xxx.lanl.gov/abs/1711.07343}{{\tt
  1711.07343}}.

\bibitem{Hanada:2012si}
M.~Hanada, M.~Honda, Y.~Honma, J.~Nishimura, S.~Shiba, and Y.~Yoshida,
  ``{Numerical studies of the ABJM theory for arbitrary N at arbitrary coupling
  constant},'' {\em JHEP} {\bf 05} (2012) 121,
  \href{http://xxx.lanl.gov/abs/1202.5300}{{\tt 1202.5300}}.

\bibitem{Marino:2016new}
M.~Marino, ``{Localization at large N in Chern--Simons-matter theories},'' {\em
  J. Phys.} {\bf A50} (2017), no.~44 443007,
  \href{http://xxx.lanl.gov/abs/1608.02959}{{\tt 1608.02959}}.

\bibitem{Hatsuda:2013gj}
Y.~Hatsuda, S.~Moriyama, and K.~Okuyama, ``{Instanton Bound States in ABJM
  Theory},'' {\em JHEP} {\bf 05} (2013) 054,
  \href{http://xxx.lanl.gov/abs/1301.5184}{{\tt 1301.5184}}.

\bibitem{Hatsuda:2012dt}
Y.~Hatsuda, S.~Moriyama, and K.~Okuyama, ``{Instanton Effects in ABJM Theory
  from Fermi Gas Approach},'' {\em JHEP} {\bf 01} (2013) 158,
  \href{http://xxx.lanl.gov/abs/1211.1251}{{\tt 1211.1251}}.

\bibitem{Calvo:2012du}
F.~Calvo and M.~Marino, ``{Membrane instantons from a semiclassical TBA},''
  {\em JHEP} {\bf 05} (2013) 006, \href{http://xxx.lanl.gov/abs/1212.5118}{{\tt
  1212.5118}}.

\bibitem{Drukker:2010nc}
N.~Drukker, M.~Marino, and P.~Putrov, ``{From weak to strong coupling in ABJM
  theory},'' {\em Commun.Math.Phys.} {\bf 306} (2011) 511--563,
  \href{http://xxx.lanl.gov/abs/1007.3837}{{\tt 1007.3837}}.

\bibitem{Marino:2009jd}
M.~Marino and P.~Putrov, ``{Exact Results in ABJM Theory from Topological
  Strings},'' {\em JHEP} {\bf 06} (2010) 011,
  \href{http://xxx.lanl.gov/abs/0912.3074}{{\tt 0912.3074}}.

\bibitem{Hatsuda:2013oxa}
Y.~Hatsuda, M.~Marino, S.~Moriyama, and K.~Okuyama, ``{Non-perturbative effects
  and the refined topological string},'' {\em JHEP} {\bf 09} (2014) 168,
  \href{http://xxx.lanl.gov/abs/1306.1734}{{\tt 1306.1734}}.

\bibitem{GREEN1982444}
M.~B. Green and J.~H. Schwarz, ``Supersymmetrical string theories,'' {\em
  Physics Letters B} {\bf 109} (1982), no.~6 444 -- 448.

\bibitem{Osborn:1993cr}
H.~Osborn and A.~Petkou, ``{Implications of conformal invariance in field
  theories for general dimensions},'' {\em Annals Phys.} {\bf 231} (1994)
  311--362, \href{http://xxx.lanl.gov/abs/hep-th/9307010}{{\tt
  hep-th/9307010}}.

\bibitem{Green:1999pu}
M.~B. Green, H.-h. Kwon, and P.~Vanhove, ``{Two loops in eleven-dimensions},''
  {\em Phys. Rev.} {\bf D61} (2000) 104010,
  \href{http://xxx.lanl.gov/abs/hep-th/9910055}{{\tt hep-th/9910055}}.

\bibitem{Chester:2018lbz}
S.~M. Chester, ``{AdS$_4$/CFT$_3$ for Unprotected Operators},''
  \href{http://xxx.lanl.gov/abs/1803.01379}{{\tt 1803.01379}}.

\bibitem{Faller:2017hyt}
J.~Faller, S.~Sarkar, and M.~Verma, ``{Mellin Amplitudes for Fermionic
  Conformal Correlators},'' {\em JHEP} {\bf 03} (2018) 106,
  \href{http://xxx.lanl.gov/abs/1711.07929}{{\tt 1711.07929}}.

\bibitem{Dolan:2011dv}
F.~Dolan and H.~Osborn, ``{Conformal Partial Waves: Further Mathematical
  Results},'' \href{http://xxx.lanl.gov/abs/1108.6194}{{\tt 1108.6194}}.

\end{thebibliography}\endgroup

\end{document}